\documentclass[lettersize,journal]{IEEEtran}
\usepackage{amsmath,amsfonts}
\usepackage[ruled,vlined]{algorithm2e}
\usepackage{array}
\usepackage{subcaption}

\usepackage{outlines}
\usepackage{caption}
\usepackage{multirow}
\usepackage{hyperref}
\usepackage{comment}
\usepackage{enumerate}
\usepackage{booktabs}

\usepackage{textcomp}
\usepackage{stfloats}
\usepackage{url}
\usepackage{verbatim}
\usepackage{graphicx}
\usepackage{cite}
\begin{document}

\title{Data Analytics for Improving Energy Efficiency in Short Sea Shipping}

\author{Mohamed Abuella,~\IEEEmembership{} Hadi Fanaee,~\IEEEmembership{}
M. Amine Atou,~\IEEEmembership{} 
Slawomir Nowaczyk,~\IEEEmembership{} 
Simon Johansson,~\IEEEmembership{}and Ethan Faghani.~\IEEEmembership{}

\thanks{Manuscript received March 1, 2024; revised March 16, 2024.}

\thanks{This work was supported by the Sweden's innovation agency (Vinnova).(\textit{Corresponding author: Mohamed Abuella}.)\\
Mohamed Abuella, Hadi Fanaee, Amine Atoui, and Slawomir Nowaczy are with Center for Applied Intelligent Systems Research (CAISR), Halmstad University, Kristian IV:s väg 80523, Halmstad, 30118 Sweden (e-mail: \{mohamed.abuella;
hadi.fanaee; amine.atoui; Slawomir.Nowaczyk\}@hh.se).
Simon Johansson and Ethan Faghani are with CetaSol AB, Gothenburg, 41251, Sweden (e-mail: \{simon.johansson and ethan.faghani\}@cetasol.com)}}

\markboth{Journal of \LaTeX\ Class Files,~Vol.~14, No.~8, August~2021}%
{Shell \MakeLowercase{\textit{et al.}}: A Sample Article Using IEEEtran.cls for IEEE Journals}


\maketitle
\begin{abstract}
To meet the urgent requirements for the climate change mitigation, several proactive measures of energy efficiency have been implemented in maritime industry. Many of these practices depend highly on the onboard data of vessel's operation and environmental conditions.
In this paper, a high resolution onboard data from passenger vessels in short-sea shipping (SSS) have been collected and preprocessed.
We first investigated the available data to deploy it effectively to model the physics of the vessel, and hence the vessel performance. Since in SSS, the weather measurements and forecasts might have not been in temporal and spatial resolutions that accurately representing the actual environmental conditions.
Then, We proposed a data-driven modeling approach for vessel energy efficiency. This approach addresses the challenges of data representation and energy modeling by combining and aggregating data from multiple sources and seamlessly integrates explainable artificial intelligence (XAI) to attain clear insights about the energy efficiency for a vessel in SSS.
After that, the developed model of energy efficiency has been utilized in developing a framework for optimizing the vessel voyage to minimize the fuel consumption and meeting the constraint of arrival time.
Moreover, we developed a spatial clustering approach for labeling the vessel paths to detect the paths for vessels with operating routes of repeatable and semi-repeatable paths. 
\end{abstract}

\begin{IEEEkeywords}
Short-sea shipping, energy efficiency, spatio-temporal aggregation, voyage speed optimization, Time-series analysis, spatial clustering, vessel path identification, maritime transportation,
\end{IEEEkeywords}

\section{Introduction}
\subsection{General Background}
Short-Sea Shipping (SSS) is a commercial transportation mode that does not involve intercontinental cross-ocean. The SSS provides a cost-efficient and environment-friendly alternative for transportation by utilizing inland and coastal waterways to transport the commercial freight \cite{sss_def}. 

On the other hand, the SSS produces some negative effects on the natural habitats and polluting the air along the coasts of populated cites~\cite{donner2018}. As a response to this, the International Maritime Organization (IMO) have conducted many studies and recommended standards and imposed policies for the maritime sector to reduce the carbon dioxide (CO$_2$) to 40\% by 2030 and cut 50\% of all GHGs by 2050, based on the emissions in 2008~\cite{Ampah2021}.

\begin{figure}[htb]
\centering
\includegraphics[width=\linewidth]{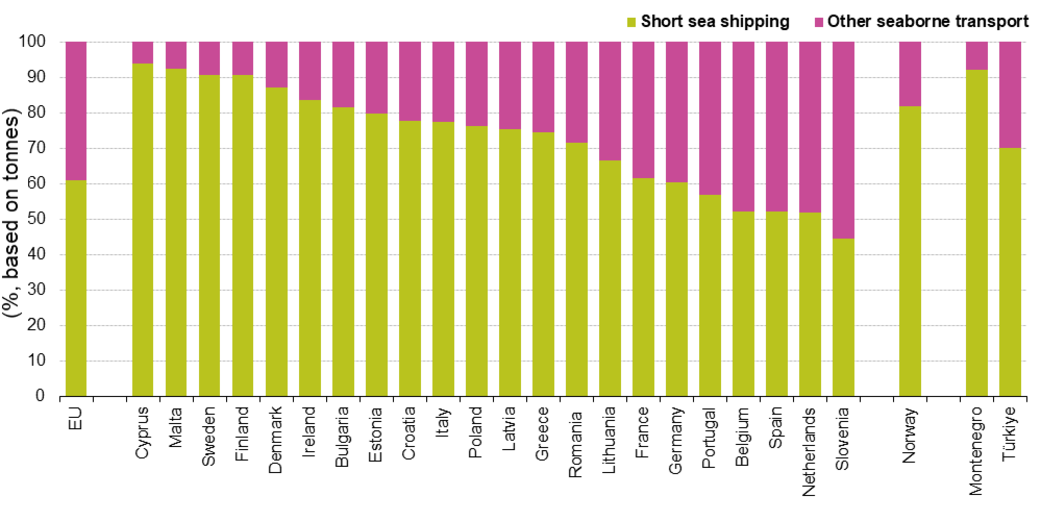}
\caption{European short-sea shipping of freight versus total sea transport, in 2021~\cite{Eurostat2021}.} 
\end{figure}

Furthermore, COVID-19 pandemic has accelerated the digitalization of the entire shipping industry globally, and hence attracted a  profound consideration to data collection and preparation stages~\cite{world2020accelerating}.
The operational and some environmental conditions can be accessed through an Automatic Identification System (AIS) messages, which is a service developed by the International Maritime Organization (IMO) in 2002 to record the sensor measurements and send the vessel position information for the traffic between other ships and neighboring shores~\cite{C2020}.

In a broader perspective, for improving the vessel's energy efficiency and harnessing more fuel savings and less GHG emissions, there are mainly two procedures. The first strategy is in the ship design stage, where the ship is built to obtain a body and equipped machinery that work efficiently. The second strategy is during the ship operation over the water or at the ports. This latter procedure can be achieved by adopting energy management plans that optimally enhance the energy efficiency and fuel consumption~\cite{zis2020ship}.

Maritime transportation is crucial for global trade, generating extensive vessel trajectory data that reveals intricate spatial and temporal navigation patterns. Understanding these patterns is vital for effective maritime traffic surveillance and management~\cite{tu2017exploiting}.


{It is crucial to distinguish between trajectory and path when studying movement data. \\
The term \textit{path} refers to the specific course taken by the object. 
While, a \textit{trajectory} refers to a sequence of consecutive geographical points, each representing a specific location at a given timestamp{~\cite{zheng2015trajectory}}. 
Thus, the trajectory typically denotes the movement of an object over time. In other words, a trajectory is a path with schedule and speed information.
For instance, if a vessel travels from origin port to destination port, its trajectory is the sequence of geographical points it passes through, while its path is the specific course taken, such as a fairway in a small river or through a specific canal. While the vessel trajectory is its navigation including time and space information, such as time schedule, speed, and location{~\cite{petry2019towards},~\cite{parent2013semantic}}.

Trajectories, observed in various scenarios such as pedestrian movements, vehicular routes, and natural events like wildlife migrations or hurricanes, involve time-evolving position data. Trajectory mining aims to uncover significant patterns within datasets, enabling tasks like path classification, anomaly detection for accidents or traffic congestions, surveillance for suspicious activities, and prediction of vessel trajectories in different landscapes{~\cite{lee2007trajectory}}.

Additionally, the term \textit{route} is synonymous with path or trajectory as long as it shares the same origin and destination. Conversely, any difference in either the origin or destination represents another route.
For example, when a vessel travels from an origin port to three destination ports, we say this vessel has three distinct routes to reach its three destinations.

Moreover, a \textit{Voyage} is generally means the period between a departure from a port to the departure from the next port. Voyage is commonly used in reference to sea travel, much like the term 'trip' is used in the context of air travel. }

{Path clustering - same as trajectory clustering, but we do not consider the time information. Our focus is primarily on spatial information of the moving object.
Path clustering, a versatile technique, involves grouping paths into clusters based on their similarity, demonstrating its effectiveness in a myriad of practical applications.}
In the realm of navigation, path identification empowers systems to generate clear and detailed instructions for users seeking their way. Traffic analysis benefits from path clustering as it facilitates the identification of diverse traffic patterns, such as the smooth flow of traffic on highways and the congestion often encountered on city streets. Path identification proves equally valuable in route planning, enabling the optimization of routes for transportation systems, including public and maritime transportation services~\cite{zheng2011computing}.

In the scope of the maritime industry, path identification from Automatic Identification System (AIS) data is a challenging task due to the high spatial freedom and, especially in coastal areas, the high frequency of ship's navigation maneuvers. Thus, it is imperative to develop a path identification tool that integrates with route planning systems for improving maritime safety and optimizing vessel routing. As data-driven approaches from AIS data continue to grow and evolve, path clustering will undoubtedly play an increasingly important role in understanding vessel behavior and supporting decision-making in maritime transportation~\cite{yan2021emerging}.

This paper proposes a data-driven frameworks for modeling, optimizing, and path identification approaches to improve energy efficiency in short-sea shipping. In addition, a framework for vessel path identification has been developed for route planning and management of resources in maritime industry. 

\subsection{Main Outcomes}
The main outcomes of this paper can be summarized as follow:
\begin{itemize}
    \item Modeling of energy efficiency: Develop a data-driven model for voyage energy efficiency, including:\\
    A spatiotemporal aggregation of operation and navigation data from onboard and external sources to capture the impact of both spatial and temporal factors on voyage energy efficiency. \\
    Introduce an efficiency score that considers both total fuel consumption and voyage duration to measure the voyage energy efficiency.
    \item Data clustering: Clustering the data of voyages and sorting them based on their efficiency scores. This clustering enables the voyage optimization algorithm to learn more insights for better actions, either by selecting the best voyages or by eliminating the worst voyages.
    \item {Time-series analysis models and comparative analysis: Four time-series based models are implemented as algorithms of voyage speed optimization. Then, a rigorous evaluation of their performance is conducted across different data clusters and using metrics that account for voyage efficiency.
    \item Practical implication: Demonstrate the significant effectiveness and practicality of the proposed approach for fixed-route vessels in short sea shipping, where the options for obtaining efficient voyages are limited. The approach also aligns with the guidance of domain experts, adhering to safety and traffic considerations.}
    \item The proposed clustering approach of vessel paths requires only position information, specifically longitude and latitude.
    \item The clustering approach has a proven added value for clustering challenging unseen or unknown paths.
    \item The approach is robust and interpretable by applying a similarity measure that reduces the influence of noise or outliers and offers a clear interpretation of path clustering.
    \item The approach has a customizable parameter to determine the number of path clusters, thereby enhancing the flexibility and adaptability of the framework and allowing users to tailor it to their specific needs.
    \item The approach also includes a method to study and analyze the patterns within specific segments of a path.
    \item It is a data-driven solution that can be used as a valuable asset for informed decision-making in route planning and optimization, traffic management, and resource allocation.
\end{itemize}
\section{Related Work}\label{Sec_relatedwork}
As digitalization and automation become increasingly prevalent in the maritime sector, the research addressing new challenges has been growing rapidly, particularly with regard to developing frameworks for energy efficiency and Maritime Situational Awareness (MSA) in cross-ocean shipping. On the other hand, the research progress has not kept pace for vessels operating in coastal areas. Thus, this literature review will focus on our primary area of interest, which is research that is related to short-sea shipping.
\subsection{Vessel Energy Efficiency Modeling}\label{Sec_XAIEE_relatedwork}
Recent research studies~\cite{Bellingmo2021, jorgensen2022ship} have explored energy-efficient routing for an electric ferry in Western Norway. They rely on operational data from onboard measurements and environmental conditions from the Norwegian Meteorological Institute, interpolated to the nearest temporal and spatial resolutions of the vessel's onboard data. 
Similarly, the researchers from Napa Ltd. in Finland conducted several studies on voyage optimization, including two cases~\cite{sugimotodigital, haranen2017role} where environmental conditions were collected from the weather forecasts. 
Other studies in the literature have also processed environmental data from different weather providers to match the vessel's operational onboard data, as reviewed in~\cite{zakaria2022instruments}. However, such approaches do not account for weather factors that influence both fuel consumption and the Estimated Time of Arrival (ETA), which is a crucial constraint when optimizing the vessel's voyage, especially in SSS. 

The maritime industry increasingly adopts digitization and Machine Learning (ML) techniques; however, their black-box nature remains a significant challenge. While ML can provide valuable insights, the reasoning behind the predictions made by such models is often difficult to comprehend due to their lack of explainability. 
To address this issue, Shapley additive explanations (SHAP)~\cite{Lundberg2017} were developed, providing a way to determine the contribution of each input feature toward the model's  output. SHAP is commonly used as a solution to the explainability issue in ML.
A recent study~\cite{lang2022comp} analyzed feature importance for the power consumption of a chemical tanker. The results indicate that the ship's speed through the water is the most influential feature, while ship heading and other weather features have relatively minor influences.
Kim et al.~\cite{kim2021explainable} utilized SHAP in combination with an anomaly detection algorithm to detect and interpret anomalies in onboard data from a cargo vessel. It allowed the identification of the specific sensor variable responsible for an anomaly, and SHAP-based clustering was used to interpret and group common anomaly patterns.
A validation study for explainability in the maritime time-series data~\cite{veerappa2022} compared two common model-agnostic XAI approaches, SHAP for a global method and LIME as a local method.
A literature review on XAI~\cite{arrieta2020explainable} discusses the importance of XAI as a key component in modern AI techniques. The authors present a taxonomy of existing contributions related to the explainability of different machine learning models.
Overall, the use and development of ML techniques in the maritime industry requires a careful balance between performance gain and explainability. 
\subsection{Vessel Voyage Optimization }\label{Sec_TSAEE_relatedwork}
Various techniques can be applied for the voyage optimization problem, as demonstrated in prior studies such as~\cite{chen2021art, Walther2016, fan2022review, moradi2022}. Some of these approaches have been implemented specifically for short-sea shipping, as indicated in the domain of~\cite{zakaria2022instruments} and~\cite{grifoll2018potential}.

Different modeling and optimization algorithms for ship weather routing have been investigated thoroughly in~\cite{walther2016modeling}. It was found that the effectiveness of these algorithms strongly depends on specific requirements concerning the objectives, control variables and constraints as well as the implementation.\\
The majority of these voyage optimization approaches are applied to ocean-crossing ships by controlling mainly the vessel speed and its route. Whereas, in the SSS, especially for fixed-route vessels like ours, there are fewer options available for voyage optimization.
Therefore, the scope of this paper is mainly focused on short-sea shipping and the pertinent research literature.

In the third IMO GHG study 2014~\cite{IMO2014third}, it was assumed that weather effects alone would be responsible for 15\% of additional power margin on top of the theoretical propulsion requirements of ocean-going ships, and a 10\% additional power requirement for coastal ships. \\
In a recent adaption of the Ship Traffic Emission Assessment Model (STEAM), propelling power is determined by wave height and directions, accounting for the environmental conditions in a highly detailed manner~\cite{C2020}.

Recent research studies~\cite{Bellingmo2021, jorgensen2022ship} have explored energy-efficient routing for an electric ferry in Western Norway. They rely on operational data from onboard measurements and environmental conditions from the Norwegian Meteorological Institute, and proposed a hybrid physics-guided machine learning model for optimizing the ship route. Based on their findings, the hybrid model was showing an energy reduction of 3.7\% compared to the actual consumption, simply by applying minor route and speed profile alterations as guided by the provided weather forecasts.

Researchers from Napa Ltd. in Finland conducted several studies on voyage optimization. In two of their studies~\cite{sugimotodigital} and~\cite{haranen2017role}.
These experts stated that their products of voyage optimization can achieve a fuel cost reduction of more than 10\% with 2\% to 4\% savings from trim optimization and 6\% to 8\% from speed and route optimization.~\cite{Wingrove2016}.

In the study conducted by Huotari et al.~\cite{huotari2021convex}, where they used a combined model with both dynamic programming and convex optimisation to obtain optimal speed profiles. The fuel savings were around 1.1\% and for voyages with substantial variance in environmental conditions, the fuel savings reached as high as 3.5\%.

In the review paper by Wang et al.~\cite{wang2022comprehensive} numerous studies are explored, including coastal and inland shipping, also revealing a diverse range of fuel savings outcomes.\\
Meanwhile, regarding weather routing, specifically through speed and route optimization, it has been demonstrated in~\cite{walther2021development} that the potential savings of carbon dioxide emissions and fuel costs are in range of 5\% to 10\%.
\subsection{Vessel Path Identification}\label{Sec_PathCluster_relatedwork}
Path clustering can be done using a variety of different methods~\cite{yuan2017review}. We will explore the related works of these various methods.

Clustering is gaining popularity for route extraction. Machine learning (ML) has recently been applied extensively for vessel path identification by learning patterns from historical data. 
{Lee et al., in{~\cite{lee2007trajectory}} introduced TRACLUS, a trajectory clustering algorithm employing a partition-and-group framework to discover common sub-trajectories. Demonstrating efficacy through formal trajectory partitioning and density-based clustering and efficiently identifies shared patterns in real trajectory data.}
The study presented in~\cite{pallotta2013vessel} introduced a framework, Traffic Route Extraction and Anomaly Detection (TREAD), which utilizes unsupervised learning for maritime route extraction. The primary emphasis is on anomaly detection and route prediction, highlighting the crucial role of AIS data in enhancing maritime situational awareness. The work specifically addresses challenges related to intermittency and persistence in AIS data.
Another method of route extraction was proposed in~\cite{Yan2020}, transforming ship trajectories into ship trip semantic objects (STSO) and utilizing graph theory for route extraction. The method proves robustness in extracting traffic routes for merchant ships but may have limitations for vessels with frequent navigation behavior changes, such as fishing vessels.
The approach in~\cite{zhao2019novel}, on the other hand, adopts a dynamic time warping (DTW) distance as a similarity measure and considers vessel course changes to analyze its trajectories. Experiments demonstrated its high accuracy in distinguishing and detecting similar vessel trajectories, outperforming existing methods in accuracy and cluster degree evaluation.
Moreover,~\cite{de2012machine} presents a machine learning framework for maritime vessel trajectory analysis, incorporating clustering, classification, and outlier detection. It employs piecewise linear segmentation for compression and alignment kernels to integrate geographical domain knowledge, enhancing task performance. Results show reduced computation time without compromising accuracy.

Capobianco et al.\cite{capobianco2021deep} proposed a deep learning approach using recurrent neural networks, employing a Bidirectional Long Short-Term Memory (BiLSTM) layer as an encoder and a Unidirectional Long Short-Term Memory (LSTM) layer as a decoder, for vessel trajectory prediction. Their model outperforms baseline approaches, showcasing the effectiveness of sequence-to-sequence neural networks.
In their study, Li et al.~\cite{li2023incorporation} present an AIS data-based machine learning method for feature extraction and unsupervised route planning for Maritime Autonomous Surface Ships (MASS). The method uses Automatic and Adaptive Dynamic Time Warping (AADTW), Spectral Clustering with Affinity Feature (SCAF), and a route optimization algorithm based on dynamic programming to extract features, obtain movement patterns, and plan routes. The proposed method outperforms existing methods by considering the impact of hidden factors and providing different routes for different types of MASS. 
The work in~\cite{li2023ais} systematically analyzes the performance of twelve ship trajectory prediction methods, including classical machine learning and emerging deep learning techniques. It compares twelve methods across three AIS datasets, representing different maritime traffic scenarios, and evaluates their effectiveness based on six indexes. The study concludes that traditional machine learning-based trajectory prediction methods struggle to meet the rising demands for accuracy and real-time performance, leading to increased interest in and promising results from deep learning-based approaches.
{EnvClus*, introduced in{~\cite{zygouras2024envclus}}, it is an innovative unsupervised data-driven framework for vessel trajectory forecasting, achieving a 33\% improvement over state-of-the-art methods. EnvClus* excels in accurately predicting vessel routes, particularly in long trips, showcasing its effectiveness in mobility analytics and trajectory prediction.}

A maritime traffic route extraction approach based on multi-dimensional density-based spatial clustering of applications with noise (MD-DBSCAN) was developed in~\cite{huang2023maritime}. The approach incorporates trajectory compression, similarity measures, and extraction of ship trajectory clusters. The approach demonstrates effectiveness in noise reduction and route extraction. The authors in~\cite{wang2021ship} proposed a trajectory clustering method based on Hierarchical Density-Based Spatial Clustering of Applications with Noise (HDBSCAN) and Hausdorff distance to generate a similarity matrix. The method adapts to shape characteristics and exhibits good clustering scalability and improved clustering results compared to DBSCAN, k-means, and spectral clustering algorithms.

Eljabu et al. proposed spatial clustering methods (SPTCLUST and SPTCLUST-II) in~\cite{eljabu2022spatial_I} and~\cite{eljabu2022spatial_II} respectively, for maritime traffic routes extraction from AIS data. The approach consists of data preprocessing, pathfinding, and route extraction without using traditional clustering algorithms. It achieved high F1-scores, 97\% and 99\%, for tankers and cargo maritime traffic routes. 

The study in~\cite{han2021modeling} enhanced the DBSCAN method by integrating the Mahalanobis distance metric for vessel behavior modeling. The proposed methodology includes clustering historical AIS data and detecting anomalies. The study showcases applicability to diverse water regions, contributing to situational awareness, collision prevention, and route planning.

Farahnakian et al.~\cite{farahnakian2023comprehensive} conducted a comprehensive examination of clustering-based techniques, including k-means, DBSCAN, Affinity Propagation (AP), and the Gaussian Mixtures Model (GMM), for detecting abnormal vessel behaviors from AIS data. Results indicate that k-means is particularly effective in detecting dark ships and spiral vessel movements, which is crucial for enhancing maritime safety.
Furthermore, the study~\cite{moavinis2023detection} proposed two methods for trajectory outlier detection, with the first utilizing DBSCAN clustering and Hausdorff distance, and the second employing Support Vector Machine (SVM) classifier and the Generalized Sequence Pattern algorithm. Both models outperform the baselines, with the SVM approach demonstrating superior performance in the identification of traffic patterns and outliers.
{The study{~\cite{petry2019towards}} developed MUITAS, a novel trajectory similarity measure addressing limitations in existing approaches for multiple-aspect trajectories enriched with heterogeneous semantic dimensions. Through evaluation on real datasets, MUITAS demonstrates robustness and outperforms current methods in precision at recall and clustering techniques for diverse mobility data.} 
{Moreover, the authors in{~\cite{varlamis2021building}} presented a methodology for extracting navigation network information from vessel trajectories, utilizing AIS data. The proposed model identifies key areas, speed, and course patterns, forming a network abstraction for optimizing ship routing and scheduling in the maritime industry, demonstrated through analysis in the eastern Mediterranean sea. This model is also useful in an outlier behavior detection.}

The research paper~\cite{liu2021visualization} offers a detailed survey of visual analytics for vessel trajectory data. The authors discuss a variety of methods, including map-based visualization, timeline-based visualization, and interactive visualization. 

{The survey{~\cite{parent2013semantic}} delves into the growing focus on semantically rich trajectories in movement data analysis, covering concepts, management issues, and techniques for constructing, enriching, and mining trajectories, with attention to emerging privacy challenges.}

The paper~\cite{zhang2022vessel} comprehensively reviews various approaches for vessel trajectory predicting, including clustering algorithms and machine learning algorithms. It also discusses the challenges and future research directions, such as the uncertainty in the data, the dynamic environment, and the computational complexity.

Among the identified challenges, which are subjects of ongoing research and require additional attention, three are worthy of specific mention:
navigating dynamic maritime environments poses a substantial challenge in accurately identifying vessel paths (I); ensuring stability, explainability, and managing the computational cost of the model add further complexity (II); finally, addressing the need for flexibility, scalability, and practical applicability is crucial for a comprehensive solution in the field of vessel path identification (III).
Motivated by these challenges, we aim to develop a framework that focuses on vessel path identification and potentially tackling such challenges faced in maritime transportation.
\section{Data Collection and Preparation}\label{Sec_Data_Col_Prep}

\subsection{Data Collection}\label{sec_data_col}
The ship’s onboard data have been received from our industry partner CetaSol AB in Gothenburg~\cite{CetaSol2022}.  The data has been gathered over a period of 15 months, between January 2020 and March 2021. It has a 3Hz frequency and records about the ship’s position, course direction, and speed. It is also including some of operational and meteorological data, such as fuel rate, engine speed, torque, acceleration, wind speed and direction.

Some information about the ship and its voyage can be found on Marine Traffic website~\cite{marinetrafffic}.\\
Other weather variables such as wave height and sea current speed and direction have been collected from external sources, Copernicus Marine Service~\cite{Copernicus2020} and Stormglass~\cite{StormGlass2022} APIs.

\begin{table}[htb]
\begin{center}
\caption{The navigational variables and their data sources.} \label{vars_abbrev_tab}
\begin{tabular}{cc||cc} 
\hline
Variable & Source & Variable & Source\\
\hline 
Latitude & Onboard & WindSpeed\_cps & Copernicus \\
Longitude & Onboard & WindDirection\_cps & Copernicus \\
SpeedOverGround & Onboard & WaveHeight & Copernicus \\
HeadingMagnetic & Onboard & WaveDirection & Copernicus \\
Pitch & Onboard & WindSpeed\_sg & Stormglass \\
Roll  & Onboard & WindDirection\_sg & Stormglass\\
WindSpeed\_onb & Onboard & CurrentSpeed & Stormglass \\
WindDirection\_onb & Onboard & CurrentDirection & Stormglass\\
\hline
\end{tabular}
\end{center}
\end{table}

\subsection{Data Preparation and Validation}\label{sec_data_valid}
The external weather data are past forecasts (hindcasts), which have reanalysed to become hourly in temporal resolution and with 0.25 to 0.5 degree as a spatial resolution.
Trilinear interpolation in time and space dimensions has been applied on external weather data to be more suitable for time and position frames of the given vessel routing. 
Therefore, the weather and onboard data are used in this analysis with a temporal resolution of 1-minute in average.

\begin{figure}[htb]
\centering
\begin{subfigure}[b]{\linewidth}
\centering
    \includegraphics[width=\linewidth]{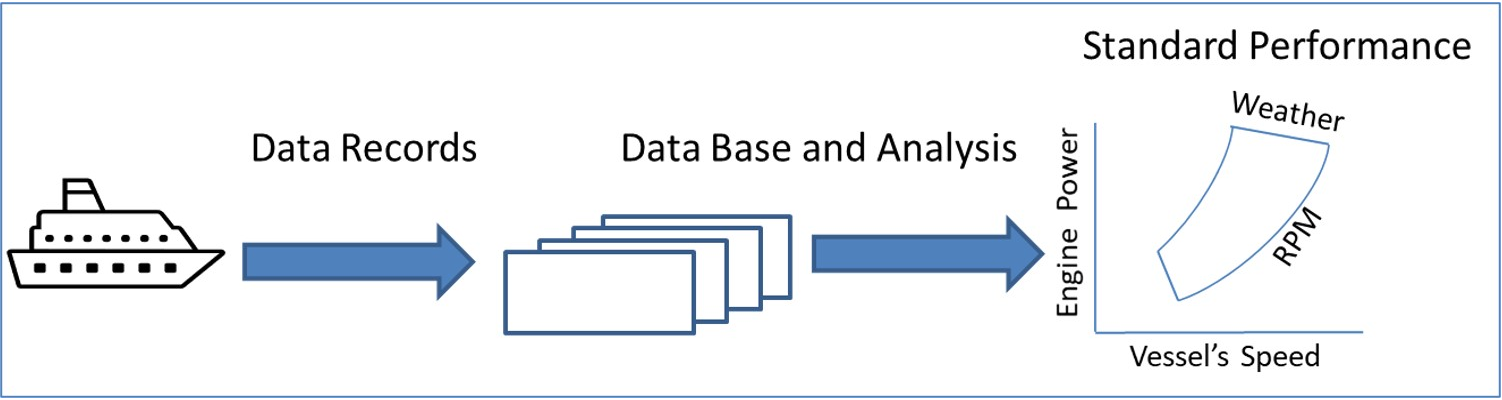}
    \caption{General data analytics for ship's standard performance~\cite{Carlton2018}.}
    \label{graph_std_perform}
\end{subfigure}
\hfill
\centering
\begin{subfigure}[b]{\linewidth}
\centering
    \includegraphics[width=\linewidth]{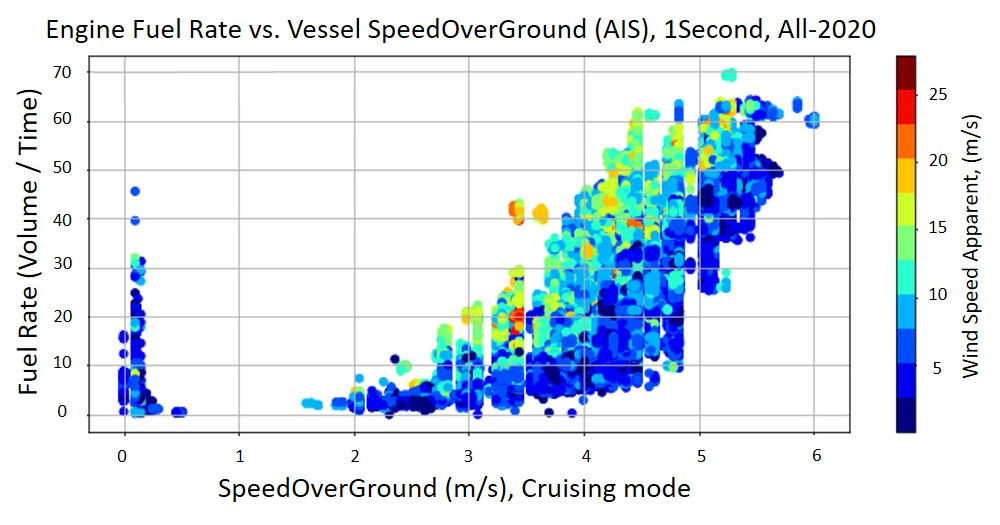}
    \caption{Vessel Buro's standard performance}
    \label{Buro_std}
\end{subfigure}
\caption{Vessel's data analytics and standard performance graph for the case study}
\label{graph_std}
\end{figure}

The data validation is conducted through the cruising-speeds mode is to reduce the other vessel effects on the fuel consumption, and thus, producing graphs that can be then compared with the general ship's standard performance.
The operational and weather data validation is carried out visually, as shown in Figure~\ref{graph_std}.

\section{Methodology}\label{Sec_method}

Energy efficiency modeling and data clustering is a vital component of our framework. 
The primary objective is to identify and sort the voyages based on their efficiency scores.
Then, train the models with the sorted data clusters iteratively, to distill insights from the voyages with different behaviours. 
Thus, the trained models will gain valuable insights into the performance and operational patterns of the vessel.
\subsection{Energy Efficiency Modeling}\label{sec_EE_Modeling}

This part presents a mathematical and visual overview of the fundamental theoretical background that forms the basis for modeling vessel energy efficiency---an indispensable element within our comprehensive framework. \\

To estimate the vessel's energy efficiency in this framework of voyage optimization, we employed a previously developed model equipped with artificial intelligence (XAI) and machine learning techniques. More details about this energy efficiency modeling approach can be found in~\cite{abuella2023xai}.

The efficiency score ($\mathrm{Eff_{score}}$) is calculated from the normalized total fuel and time for every voyage, as following:
\begin{equation}\label{eq_effsocre}
\mathrm{Eff_{Score}} = 1- \frac{2 \times  [Fuel_{Tl_{Nm}} \times Time_{Tl_{Nm}}]} {[Fuel_{Tl_{Nm}} + Time_{Tl_{Nm}}]}
\end{equation}

The efficiency score considers the proportional reduction in both fuel consumption and time, assessing the vessel's efficient use of resources during the voyage.

\begin{figure}[htb]
\centering
\begin{subfigure}{0.34\linewidth}
    \includegraphics[width=\linewidth]{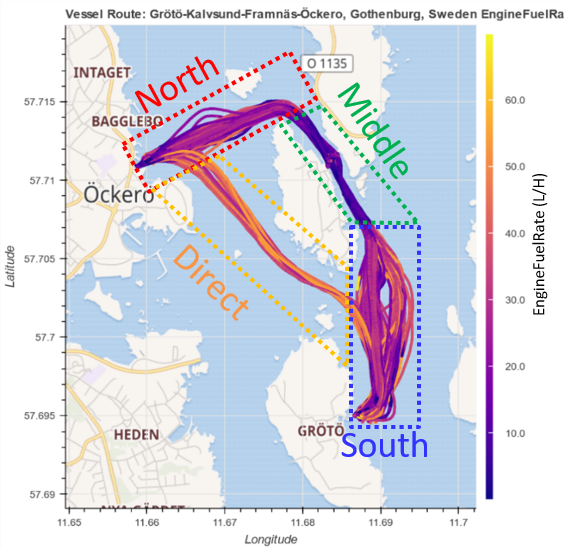}
    \caption{Vessel's route.}
    \label{Aggreg_rts}
\end{subfigure}
\hfill
\centering
\begin{subfigure}{0.64\linewidth}
\centering
   \includegraphics[width=\linewidth]{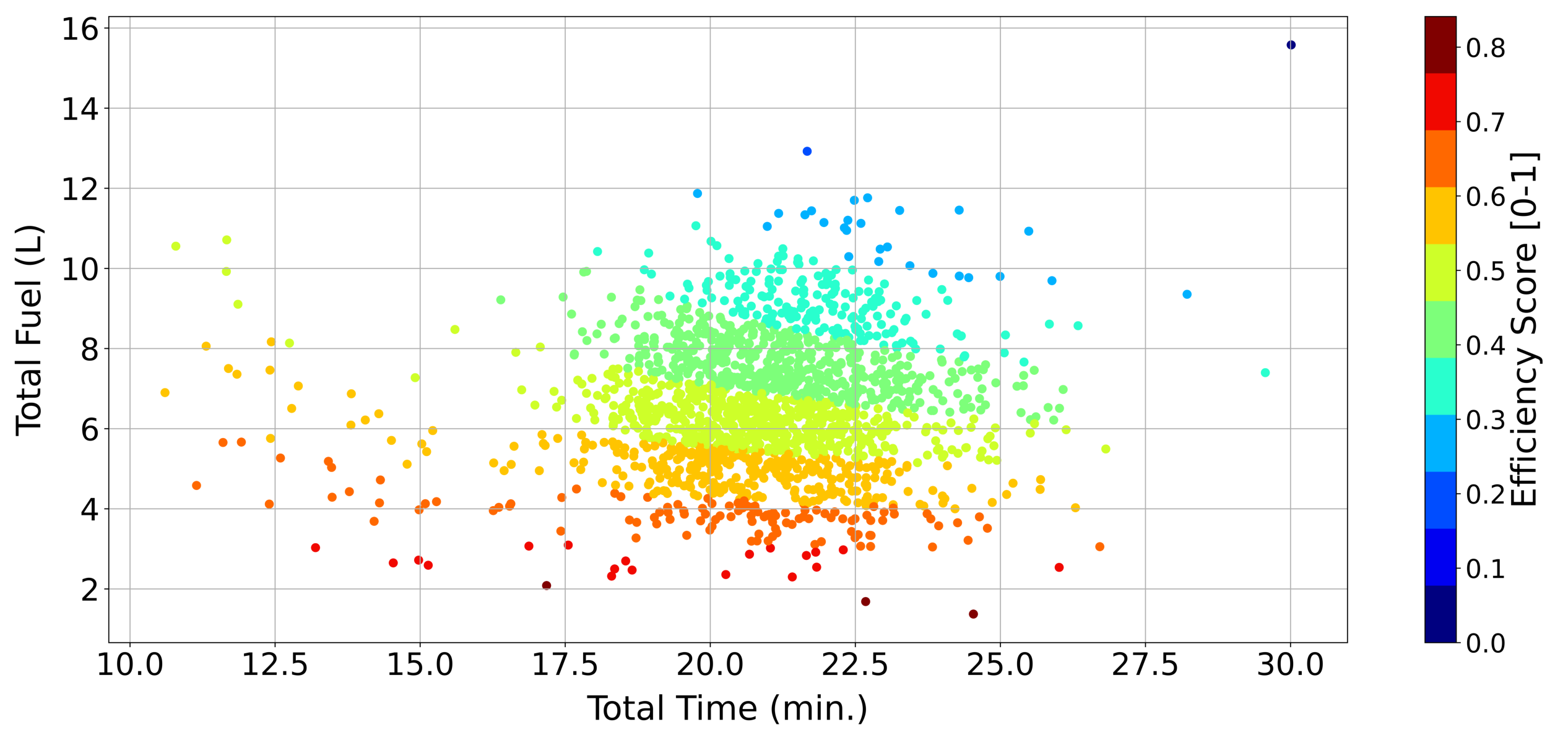}
\caption{Data are projected by Efficiency Score vs. fuel and time.}
\label{Global_Eff_vs_FL_Time}
\end{subfigure}
\caption{The vessel voyages and the aggregated data projected by the Efficiency Score.}
\label{Aggregation_Data}
\end{figure}

Figure~\ref{Aggregation_Data} facilitates to visualize the process of aggregation for the vessel's voyages. First by illustrating the voyages in terms of space (i.e., latitude and longitude) as shown in Figure~\ref{Aggreg_rts}, and second by representing the aggregated voyages as points in new dimensions of Efficiency Score versus total fuel and total time. The aggregated data and its new dimensions are projected as in Figure~\ref{Global_Eff_vs_FL_Time}.

The representation of voyages in terms of fuel and time is done by adopting the concept of the Efficiency Score (Eff-Score). The efficiency scores for all vessel's voyages are presented in Figure~\ref{Global_Eff_vs_FL_Time}.
It is evident that the voyage with lower fuel and shorter time have higher efficiency scores, and vice versa.

\begin{algorithm}[htb]
\caption{Modeling of Energy Efficiency and Clustering of Voyages Data Based on Their Energy Efficiency}
\label{data_proc_Alg}

\KwData{Voyages data of the vessel}
\KwResult{Clusters of Voyages}
{Load the operational and navigational data, including speed, course, fuel, position, distance, and weather\;
Tag the datapoints to its corresponding voyage, $V_{id}$\;}
\ForEach{voyage $V_{i}$ in voyages data}{
    Calculate total fuel consumption and time for $V_{i}$\;
    Normalize total fuel and time for $V_{i}$ based on their maximum values of all voyages\;
    Calculate the \textit{Eff-Score} as described in Eq.~\eqref{eq_effsocre}, and assign it to all datapoints of this voyage $V_{i}$\;}

Initialize four empty lists for each cluster: $Top75Pr$, $Top50Pr$, $Top25Pr$, $Top10Pr$ (Percentiles of Eff-Scores)\;

\ForEach{data point in all data}{
    Extract the Eff-Score of the data point\;
    \If{Eff-Score is in the top 75\%}{
        Append the data point to $Top75Pr$\;
    }
    \ElseIf{Eff-Score is in the top 50\%}{
        Append the data point to $Top50Pr$\;
    }
    \ElseIf{Eff-Score is in the top 25\%}{
        Append the data point to $Top25Pr$\;
    }
    \ElseIf{Eff-Score is in the top 10\%}{
        Append the data point to $Top10Pr$\;
    }
}
\end{algorithm}

\begin{figure}[htb]
\centering
\centering
\includegraphics[width=\linewidth]{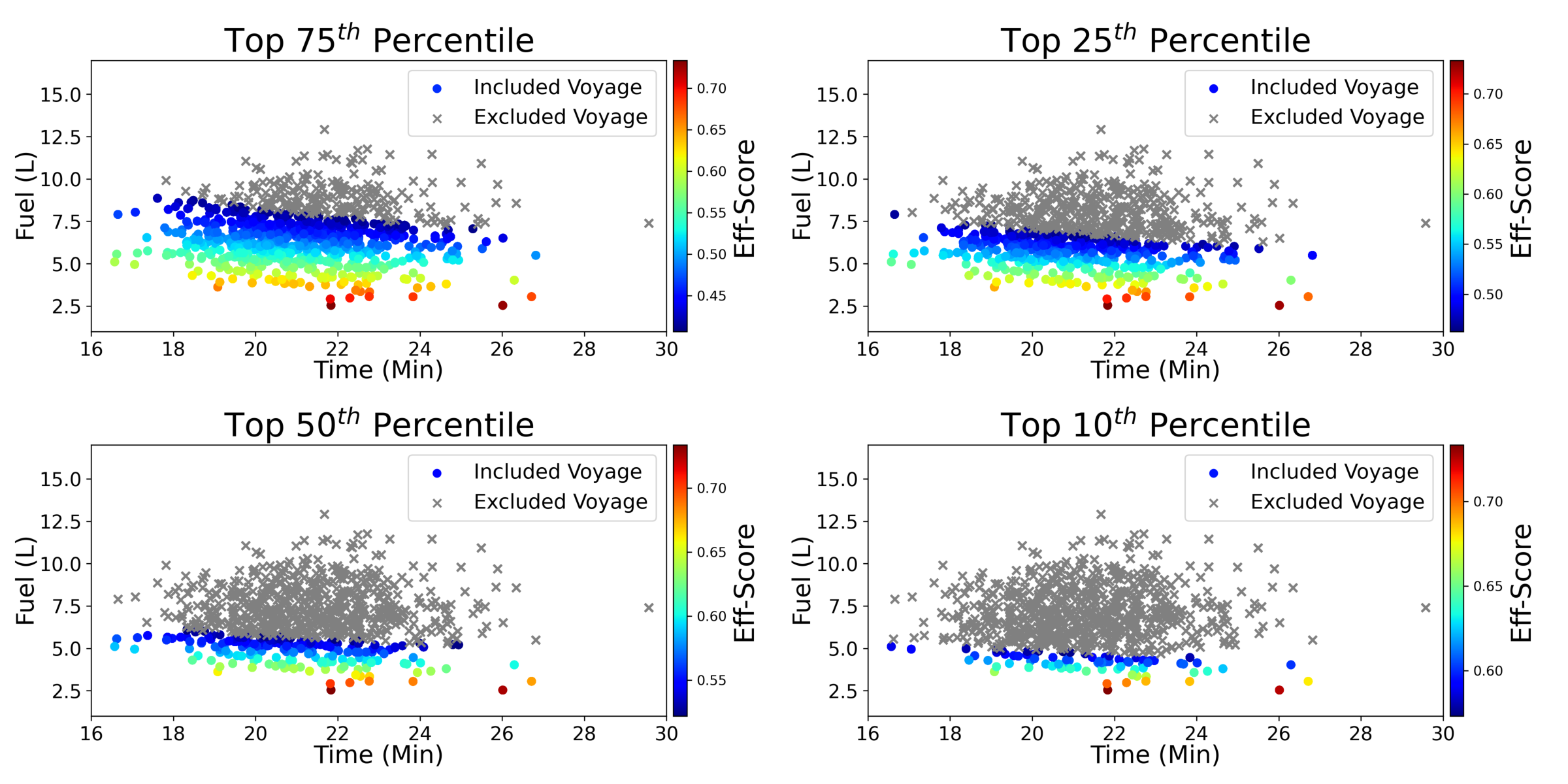}
\caption{Four clusters of voyages based on their efficiency.}
\label{vog_clusters}
\end{figure}

There are four voyage data clusters, namely $Top75Pr$, $Top50Pr$, $Top25Pr$, and $Top10Pr$, as shown in Figure~\ref{vog_clusters}, These clusters are categorized on their respective Eff-Score percentiles, enabling a structured analysis of voyage efficiency across various percentile groups.

\begin{figure*}[htb]
\centering
\includegraphics[width=\linewidth]{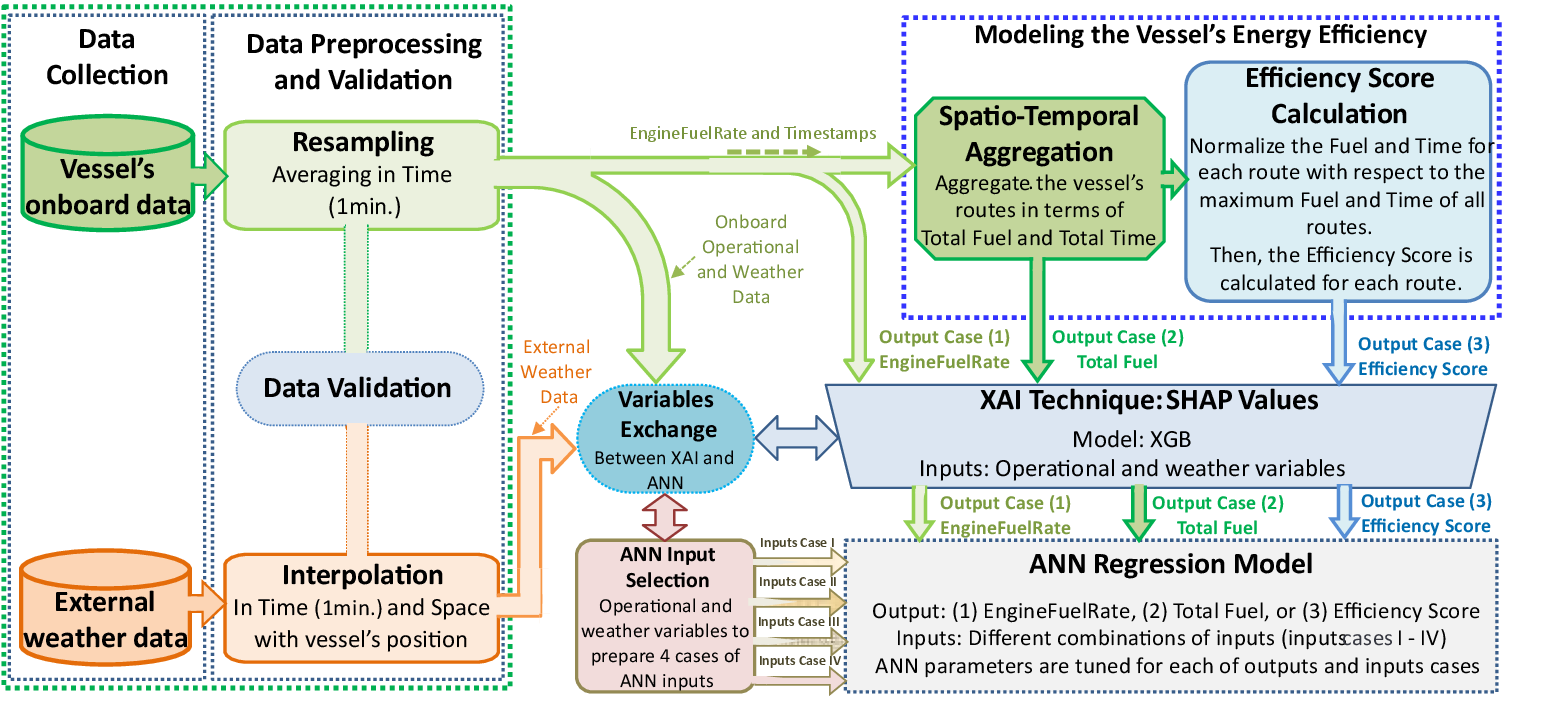}
\caption{Workflow of modeling and analysis of energy efficiency in short-sea shipping}
\label{Fig_workflow}
\end{figure*}

We introduced a data-driven approach for modeling of vessel energy efficiency, by integrating data from various sources and employing explainable artificial intelligence (XAI) and artificial neural network (ANN) to gain clear insights into a vessel's energy efficiency in SSS. For more details, refer to our publication~\cite{abuella2023xai}.\\
The workflow for modeling of vessel energy efficiency is shown in Figure~\ref{Fig_workflow}.


\subsection{Framework of Voyage Optimization}\label{sec_framework_EEimprov}
\begin{figure}[htb]
\centering
\includegraphics[width=0.5\linewidth]{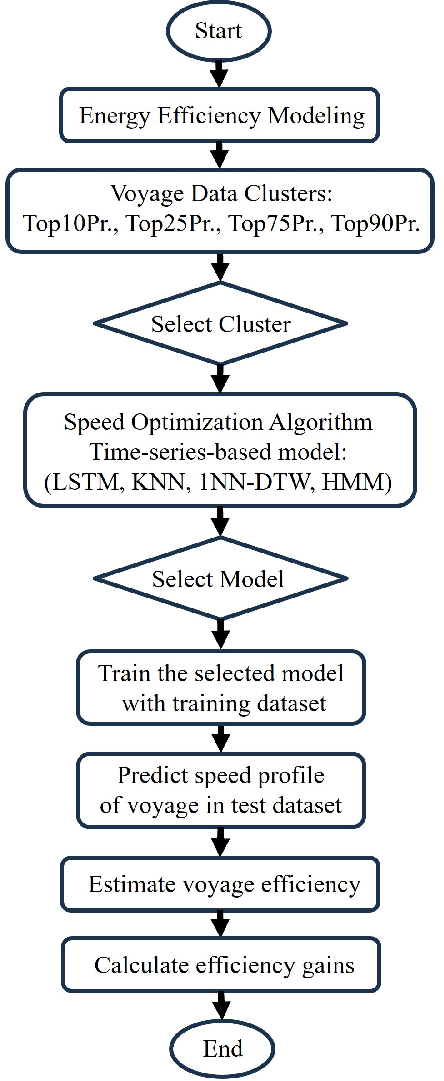}
\caption{Framework of vessel voyage optimization.}
\label{fig_Fuel_Min_problem}
\end{figure}

One of the main objective of this paper is to improve the vessel voyage by optimizing its speed to enhance the vessel's energy efficiency. In other words, reducing the vessel's fuel consumption within constrained arrival time. More details about the voyage optimization approach can be found in~\cite{abuella_TSAEE}.\\
The framework of the developed approach for improving the vessel voyages is depicted in Figure~\ref{fig_Fuel_Min_problem}.

For the purpose of vessel voyage optimization, we need a model to optimize the vessel's speed profile for improving the vessel's energy efficiency.
This ideal model should mainly be able to:
\begin{itemize}
    \item Model the temporal dependencies in vessel speed profiles.
    \item Incorporate external variables, such as weather conditions, into the modeling process.
    \item Adapt to changing conditions to provide real-time optimized speed profiles.
\end{itemize}

In order to meet these requirements, we adopt several time-series analysis models, specifically they are:
Long Short-Term Memory (LSTM), k-Nearest Neighbors (kNN), 1-Nearest Neighbor with Dynamic Time Warping (1NN-DTW), and Hidden Markov Model (HMM). These models are described with some details as follow.

\subsubsection{Long Short-Term Memory (LSTM)}\label{Sec_lstm}
It is a type of recurrent neural network (RNN) architecture designed for sequential data analysis. It is particularly useful for modeling and predicting time-series data, making it a valuable tool for improving energy efficiency in vessel operations~\cite{hochreiter1997long}.\\
LSTM networks are well-suited for tasks involving time-series sequences of varying lengths, capturing long-term dependencies, and handling irregular temporal patterns. The LSTM architecture deals specifically to address the problem of vanishing gradients that often occurs in other RNN structures~\cite{staudemeyer2019understanding}.

The LSTM's capability to learn the hidden insights from selected data, particularly in terms of vessel's efficiency scores, comes with a trade-off. The data clustering may result in a smaller dataset, which potentially limiting the LSTM's deep learning capabilities. Additionally, data clustering increases the susceptibility of LSTM to overfitting.\\
LSTM can be employed to model and predict the vessel's speed to improve its fuel efficiency, taking into account various factors such as weather conditions and operational parameters 
\subsubsection{k-Nearest Neighbors (kNN)}\label{Sec_knn}
The kNN algorithm is a non-parametric model that makes no assumptions and operates on the principle of determining an unknown observation's class by measuring distances to nearby observations, attributing the observation to the majority class of its nearest neighbors. The parameter k represents the number of neighboring observations considered when classifying a given observation, and its value is determined through a search for the optimal choice that maximizes accuracy in the training set.

The paper by Cover and Hart~\cite{cover1967nearest} is a foundational work in pattern recognition and machine learning, introduced the KNN algorithm for pattern classification by utilizing the nearest neighbors to classify data points based on majority vote.

KNN can be employed to analyze and predict energy consumption patterns based on similar historical data. By considering the nearest neighbors of a given operational scenario, KNN provides a straightforward approach to making energy-efficient decisions. The choice of the parameter 'k,' representing the number of neighbors, plays a crucial role in the accuracy of KNN-based models.
\subsubsection{1-Nearest Neighbor with Dynamic Time Warping (1NN-DTW) }\label{Sec_dtw}
This model is a distinctive variant of kNN with (k=1) and with a distance-wise measure that utilizes Dynamic Time Warping (DTW) algorithm for a comparison of two time-series sequences that may have different lengths, time shifts, and speed variations~\cite{berndt1994using, keogh2005exact, ding2008querying, rakthanmanon2012searching, bagnall2017great}.
The 1NN-DTW algorithm can be utilized to measure the similarity between the given speed profile and other measured vessel speed profiles with higher efficiency scores. 
1NN-DTW is capable to capture the temporal dependency of speed profiles, even in short-sea shipping where the options of control the ship to improve its energy efficiency are limited.

Thus, the 1NN-DTW identifies the efficient observed speed profile that most similar to the given speed profile to be chosen as suggested an optimized speed profile for the current journey of the vessel that can lead to improved energy efficiency.
\subsubsection{Hidden Markov Model (HMM)}\label{Sec_hmm}
This model is based on probabilistic modeling and use a variety of techniques from the statistical modeling, and they are widely used in time-series analysis and pattern recognition.\\
The work by Rabiner in 1989~\cite{rabiner1989tutorial} is an essential reference in the application of HMM. This tutorial delved into the formulation of a statistical method for representing speech, showcasing a successful HMM system implementation with a focus on discrete or continuous density parameter distributions.

In the context of improving energy efficiency in constrained environments, HMMs can be applied to model the underlying patterns and transitions in ship operational data, allowing for more informed decisions on optimizing energy consumption.\\
The HMM estimates three main weather states and dynamically adjusts the SpeedOverGround (SOG) predictions. During calm weather, it selects the maximum observed speed in the historical data at that condition. In moderate conditions, it uses the average speed, while in rough weather, it relies on the minimum speed profiles. 
For instance, Eq.~\ref{Eq_sog_rough} indicates that the predicted speed (SOG$_{\text{Pred}_i}$) by HMM is the minimum value of measured speed profiles in rough weather state, and likewise for other weather states.

\begin{equation}\label{Eq_sog_rough}
SOG_{\text{Pred.}_i} = \min(SOG_{\text{Meas.}} \, | \, \text{Rough weather})
\end{equation}
This adaptability to changing weather conditions enhances voyage optimization, making it a crucial component of the overall framework.\\
\subsection{Framework of Path Identification}\label{Sec:path_id_method}
\begin{figure}[htb]
\centering
    \begin{subfigure}[b]{0.49\linewidth}
        \centering
        \includegraphics[width=0.79\linewidth]{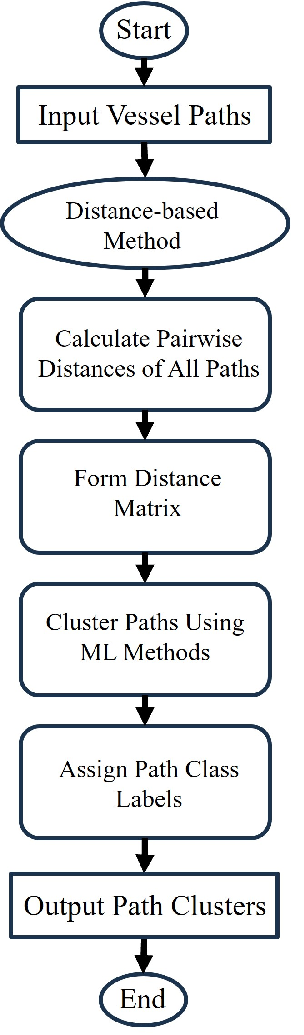}
        \caption{}
        \label{fig:fw_distance}
    \end{subfigure}
\hfill
\centering
     \begin{subfigure}[b]{0.49\linewidth}
        \centering
        \includegraphics[width=0.77\linewidth]{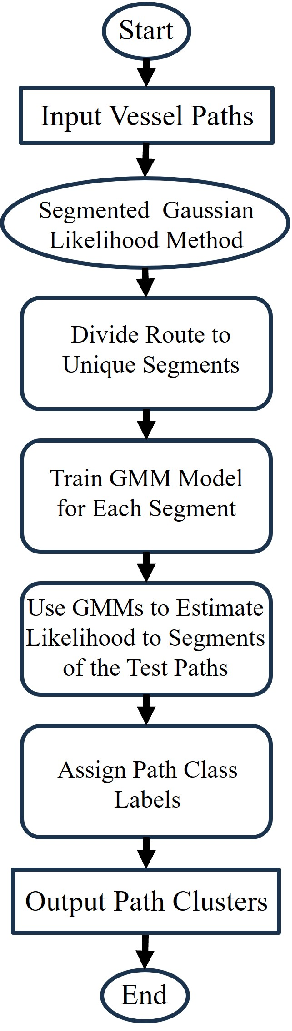}
        \caption{}
        \label{fig:fw_likelihood}
    \end{subfigure}
\caption{Framework of vessel path identification. (a) Flowchart of distance-based method. (b) Flowchart of segmented Gaussian likelihood method.}
\label{fig:framework_path_id}
\end{figure}

The theoretical background and description of the underling methodology of our proposed framework are covered within this section. Additional details about the framework of path identification can be found in~\cite{abuella_pathclustering}.\\
The framework of vessel path identification is depicted in Figure~\ref{fig:framework_path_id}.

\subsubsection{Problem Formulation}\label{Sec:prob_form}
~ \\
The equations~(\ref{eq:Path_Clust1}-\ref{eq:Path_Clust3}) serve as a mathematical representation to describe the clustering of vessel paths.\\
It is worth mentioning that the clustering process is conducted sequentially, point by point, while the labeling of path clusters is performed for the entire voyage. Therefore, each voyage has a single label of path cluster.

\begin{equation} \label{eq:Path_Clust1}
Voyages \in Path\ Clusters\\
\end{equation} 
\begin{equation} \label{eq:Path_Clust2}
Voyages=\{ts_1[p_1,\dots, p_n],\dots,ts_j[p_1,\dots,p_n] \} \\
\end{equation} 
\begin{equation} \label{eq:Path_Clust3}
Path\ Cluster\ Set = \{cluster_1,\dots, cluster_k\}
\end{equation}
where:\\
$Voyages$: a collection of time series, each representing a voyage of the vessel taken following a given path, with a predicted cluster. \\
\textit{$ts_j$}: a time series corresponding to voyage $j$, i.e., a sequence of $n$ data points, where each data point $p$ represents the vessel position and is defined by a pair of coordinates, namely latitude and longitude.\\
$n$: the number of time steps (duration) of each voyage, which can differ from one voyage to another.\\
$j$: a total number of voyages.\\
$Path\ Cluster\ Set$: a set of $k$ clusters into which the path of voyages are being clustered.

{\textit{Remark:} In the definition of $k$ clusters, it is important to clarify that this study is for labeling or identifying predefined fairways for vessels navigating in confined waters rather than open sea, resulting in a predetermined number of path clusters, $k$.}

\subsubsection{Distance-Based Method}\label{Sec:Distance_method}
~ \\
The similarity between two paths is measured by the average nearest neighbor distance (ANND), as shown in Eq.~(\ref{eq:nearest_distance}).\\
\begin{equation} \label{eq:nearest_distance}
ANND (i, j) = \frac{1}{n_i} \sum_{k=1}^{n_i} Distance(P_i^k,  NN(P_j^k))
\end{equation}
where:\\
$ANND(i, j)$: is the average nearest neighbor distance between path $i$ and path $j$, present in the distance matrix at row $i$ and column $j$. It is a symmetric, meaning that $ANND(i,j)$ is the same as $ANND(j,i)$\\
$Distance(P_i^k, NN(P_j^k))$ : The distance between the $k^{th}$ point in path $i$, denoted as $P_i^k$, and its corresponding nearest neighbor point in path $j$, indicated as $NN(P_j^k)$. $n_{i}$ is the total number of points in path $i$.\\ 
The measure $Distance$ is an Euclidean distance. However, for longer curved routes, Haversine or Great-circle distance would be more suitable.

\begin{figure}[htb]
  \centering
  \includegraphics[width=\linewidth]{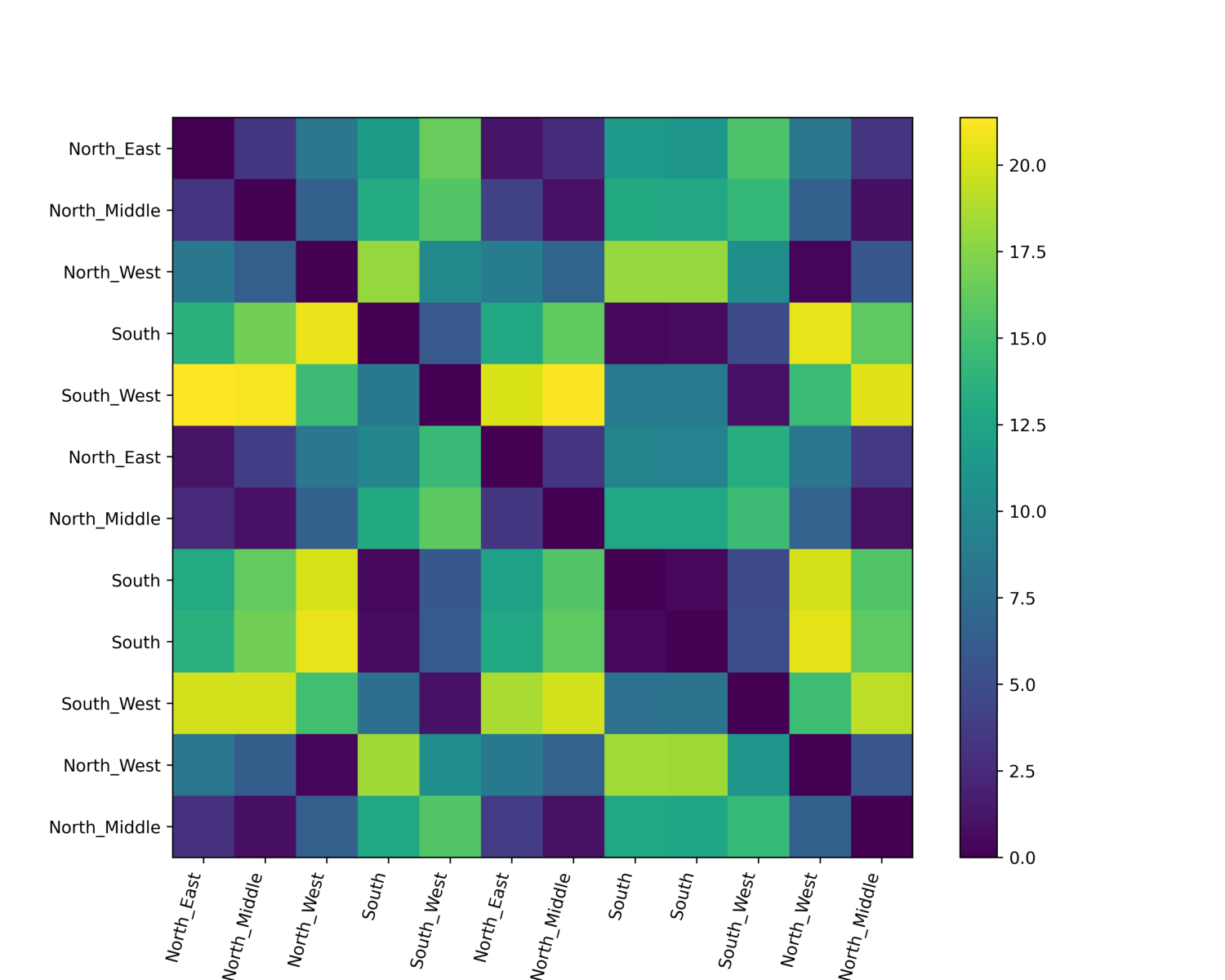}
  \caption{{Heatmap of the distance matrix for 12 paths of Cinderella II vessel.}}
  \label{fig:distance_matrix}
\end{figure}

The ANND, as expressed in Eq.~(\ref{eq:nearest_distance}), is computed by averaging the distances between each point in one path and its nearest neighbor in the other path.\\
Then, the similarity value (i.e., ANND) of this pair of paths is stored as an element in the distance matrix.\\
A lower ANND indicates that the paths within a cluster are more similar.
The distance matrix will have dimensions $(m \times m)$ , where $m$ is the number of paths.\\
For instance, the computed distance matrix for a set of 12 paths is illustrated in Figure~\ref{fig:distance_matrix}.

After the construction of the distance matrix, the machine learning (ML) technique is applied to cluster the paths based on their corresponding values in this distance matrix. The ML techniques that we used are k-means, Gaussian Mixture Model (GMM), and hierarchical clustering. \\
\subsubsection{Segmented Gaussian Likelihood Method}\label{Sec:Gauss_method}
~ \\
In addition to identifying the vessel's path, to better understand how the vessel changes its paths, we employ Gaussian distributions on several distinct segments of the vessel route.
This technique can be summarized as follows:
\begin{itemize}
    \item Utilize a training dataset comprising vessel position information that should adequately represent all potential paths of the vessel route.
    \item Divide the route into different distinct segments. 
    \item Train a single GMM model for each segment to find the Gaussian distributions of all segments of the route. 
    \item Estimate likelihoods of the segments by using the trained GMM models with their corresponding segments of each vessel voyage in the test dataset. 
    \item Label the path clusters based on the estimated likelihood at the unique segments of the route.
\end{itemize}

\section{Implementation of Frameworks}\label{sec:implemen}
In this section, we present specific case studies demonstrating how the frameworks of the paper, applied with real-world data, provide practical insights and solutions.

\subsection{Case Study of Voyage Efficiency Modeling and Optimization }\label{Sec:case_voyg_modoptz}

The implementation of our approach of a time-series analysis-based voyage optimization framework for a fixed-route vessel of our case study is depicted in Figure~\ref{fig_Fuel_Min_problem}, and the step-by-step process is described by algorithms~\ref{data_proc_Alg} and~\ref{TSA4EE_Alg}.\\
In this case study, we utilized a dataset collected from a passenger ship named Buro, which operates in Sweden. The dataset from the vessel Cinderella II was also available; however, it was not utilized for vessel voyage optimization. This decision was made due to the dataset's duration being limited to five months, encompassing only the summer and autumn seasons. As a result, it does not cover the full spectrum of seasonal variability in weather impacts.\\
For more detailed information about setting up the models and their specifications including various parameters, you may refer to the source codes, which are developed in Python 3.9.7 to produce the results of this study. 
These source codes are available at:~\url{https://github.com/MohamedAbuella/TSA4EESSS.}

\begin{algorithm}[htb]
\caption{Speed Optimization Models for Improving Voyage Efficiency}
\label{TSA4EE_Alg}

\KwData{Refer to \textbf{Algorithm~\ref{data_proc_Alg}} for data processing and clustering.\\ Add SOG$_{\text{Meas.}}$ and Weathers to Inputs.}

\ForEach{$C_{k}$ in CT (sorted by Eff-Score)}{
    \ForEach{\textbf{model} in [LSTM, kNN, 1NN-DTW, HMM]}{
        Train \textbf{model} with voyages in $C_k$\;
        \ForEach{Voyage $v_{i}$ in test dataset $\notin C_k$ }{
            \If{model is LSTM}{
                Predict SOG$_{\text{Pred.}_i}$ using trained LSTM\;
            }
            \ElseIf{model is kNN}{
                Predict SOG$_{\text{Pred.}_i}$ using trained kNN\;
            }
            \ElseIf{model is 1NN-DTW}{
                Predict SOG$_{\text{Pred.}_i}$ based on the most similar voyage in $C_{k}$\;
            }
            \ElseIf{model is HMM}{
                Predict SOG$_{\text{Pred.}_i}$ based on weather states (Calm, Moderate, Rough)\;
            }
            Estimate efficiency for SOG$_{\text{Meas.}_i}$ and SOG$_{\text{Pred.}_i}$\;
        }
        Evaluate efficiency gains for test voyages\;
    }
}
\end{algorithm}


\subsection{Case Study of Path Identification}\label{Sec:case_path_id}

In this section, we describe the case study, including the data collection, preprocessing, and analysis. 

\begin{figure}[htb]
\centering
\begin{subfigure}[b]{0.46\linewidth}
\centering
    \includegraphics[width=\linewidth]{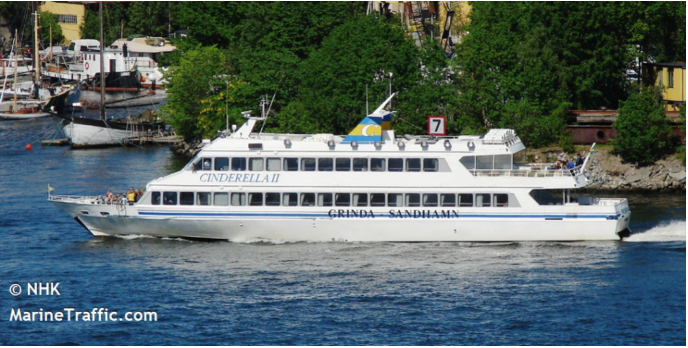}
    \caption{Cinderella II.}
    \label{fig:sship_cind_image}
\end{subfigure}
\hfill
\centering
\begin{subfigure}[b]{0.525\linewidth}
\centering
    \includegraphics[width=\linewidth]{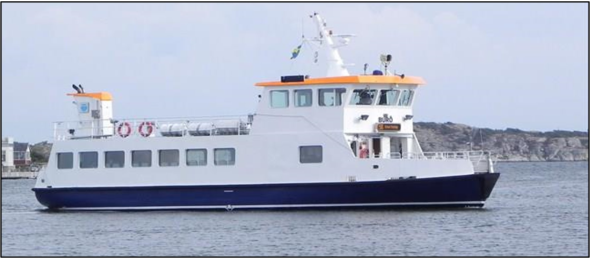}
    \caption{Buro.}
    \label{fig:ship_buro_image}
\end{subfigure}
\caption{{Images of two vessels that are used in the case study{~\cite{marinetrafffic_Cind}}.}}
\label{ship_image}
\end{figure}

In this study, we utilized datasets collected from two passenger ships, named Cinderella II and Buro, operating in Sweden.
The vessels are shown in Figure~{\ref{ship_image}}, additional information about vessels Cinderella II and Buro can be found in{~\cite{marinetrafffic_Cind}}. 

Cinderella II operates in Stockholm archipelago, east of Sweden. The data of Cinderella II spans over five months (July to November 2022). It comprises information on 124 voyages of this vessel, connecting the two main ports of Vaxholm in the east and Sodra in the west. 

While Buro works in Gothenburg, west of Sweden.
It dataset has been gathered over a period of 15 months (between January 2020 and March 2021). The data of Buro has 1755 voyages, between two main ports, Groto in the south and Ockero in the northwest.

\begin{figure}[htb]
\centering
\begin{subfigure}[b]{0.48\linewidth}
\centering
    \includegraphics[width=\linewidth]{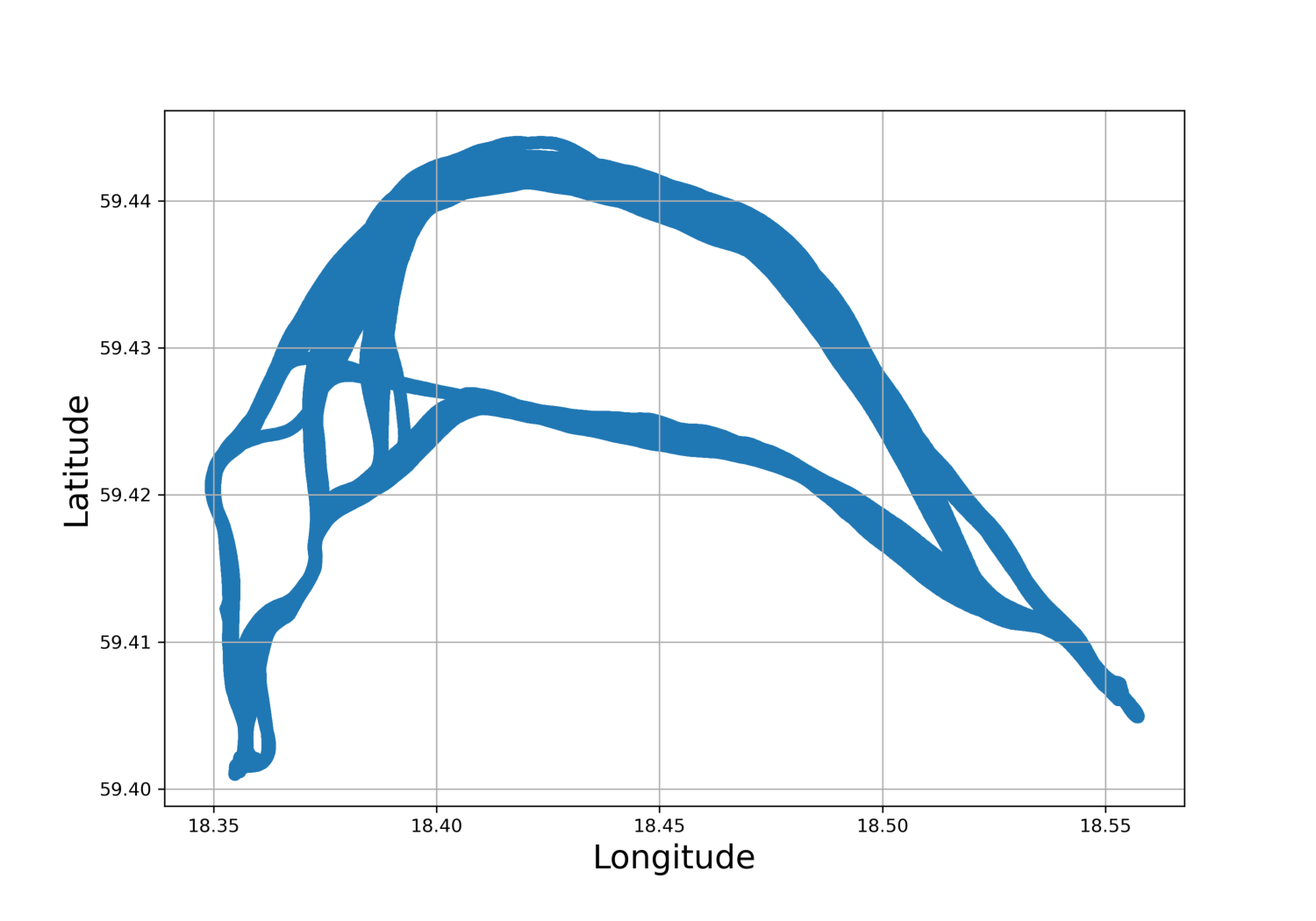}
    \caption{Cinderella II route.}
    \label{fig:ship_route}
\end{subfigure}
\hfill
\centering
\begin{subfigure}[b]{0.48\linewidth}
\centering
    \includegraphics[width=\linewidth]{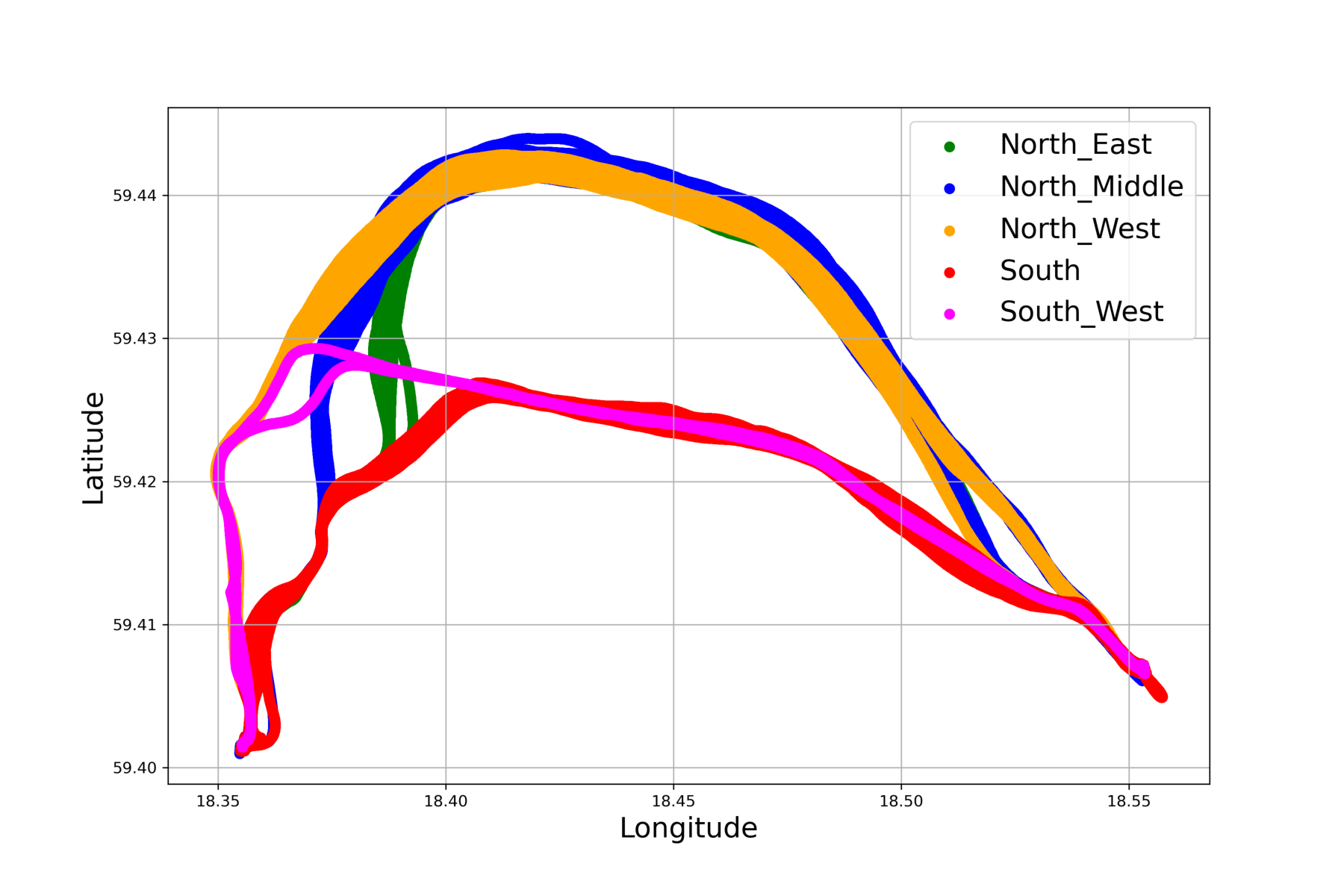}
    \caption{Five clusters of path.}
    \label{fig:1plt_path5cls}
\end{subfigure}
\begin{subfigure}[b]{0.48\linewidth}
\centering
    \includegraphics[width=\linewidth]{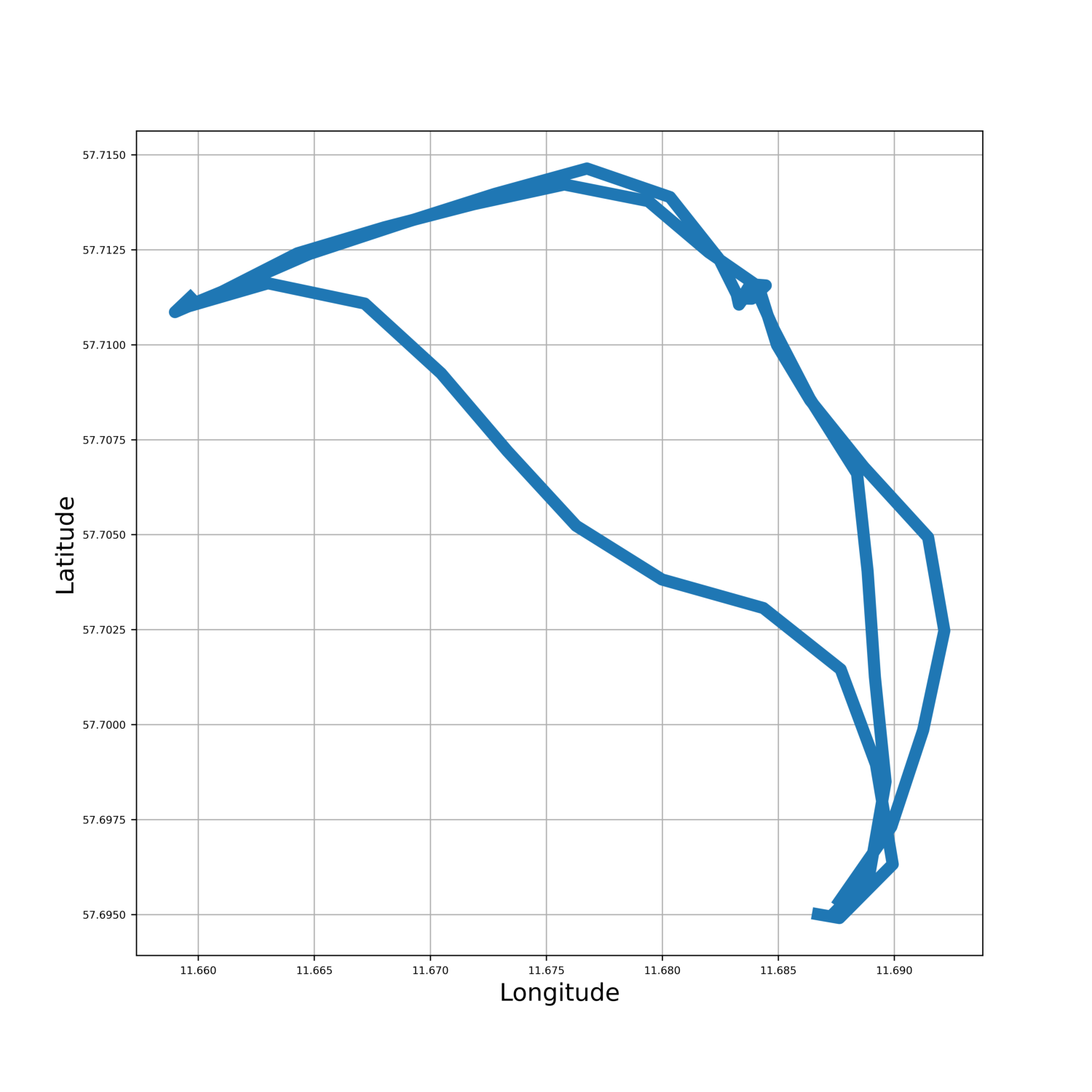}
    \caption{{Buro route.}}
    \label{fig:ship_Buro_route}
\end{subfigure}
\hfill
\centering
\begin{subfigure}[b]{0.48\linewidth}
\centering
    \includegraphics[width=\linewidth]{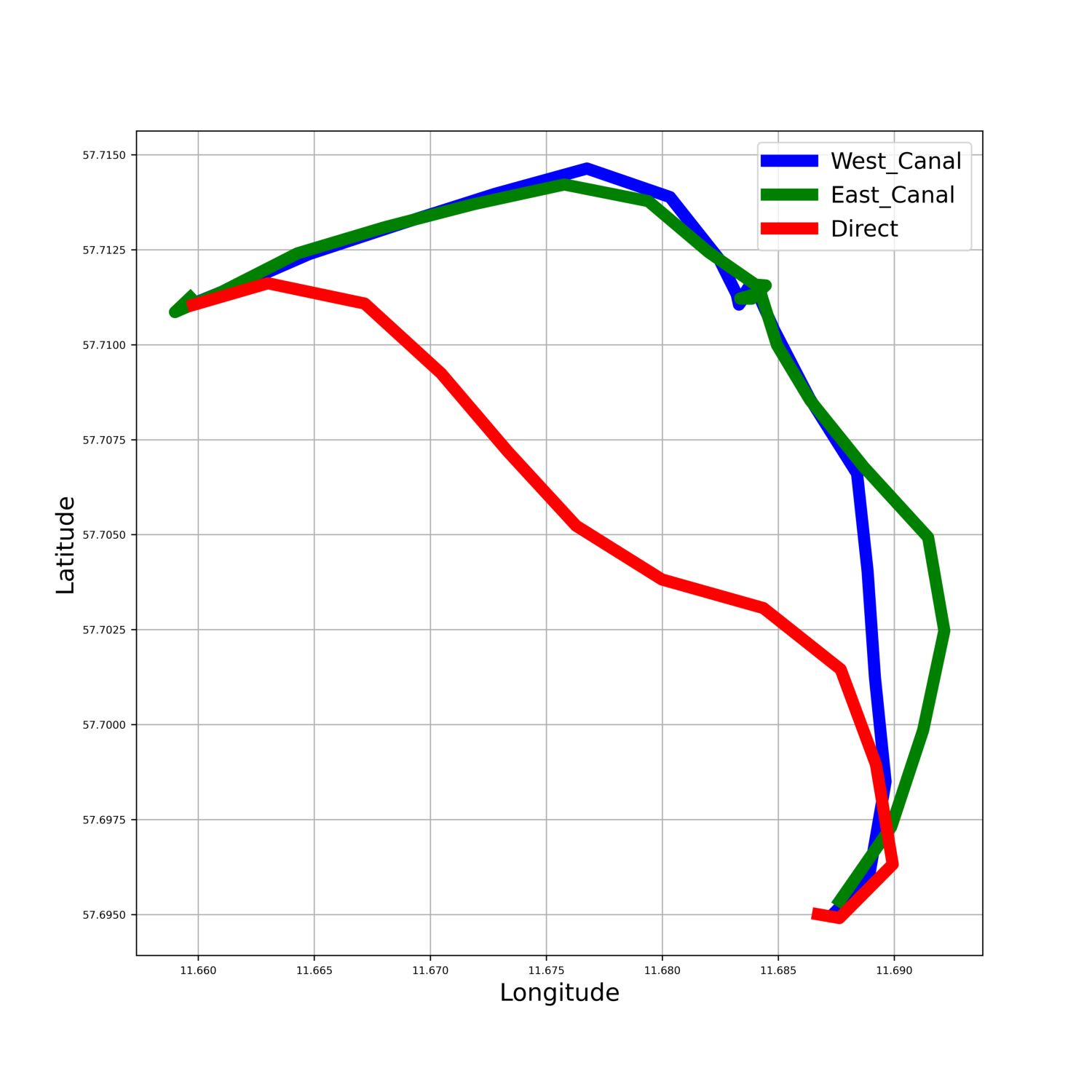}
    \caption{{Three clusters of path.}}
    \label{fig:1plt_path3cls}
\end{subfigure}
\caption{{Routes of the two vessels beside their all possible path clusters, which will be identified by  applying proposed framework.}}
\label{fig:Buro_Cind_route}
\end{figure}

\begin{figure}[htb]
    \centering
    \begin{subfigure}{1.1\linewidth}
      \centering
      \includegraphics[width=\linewidth]{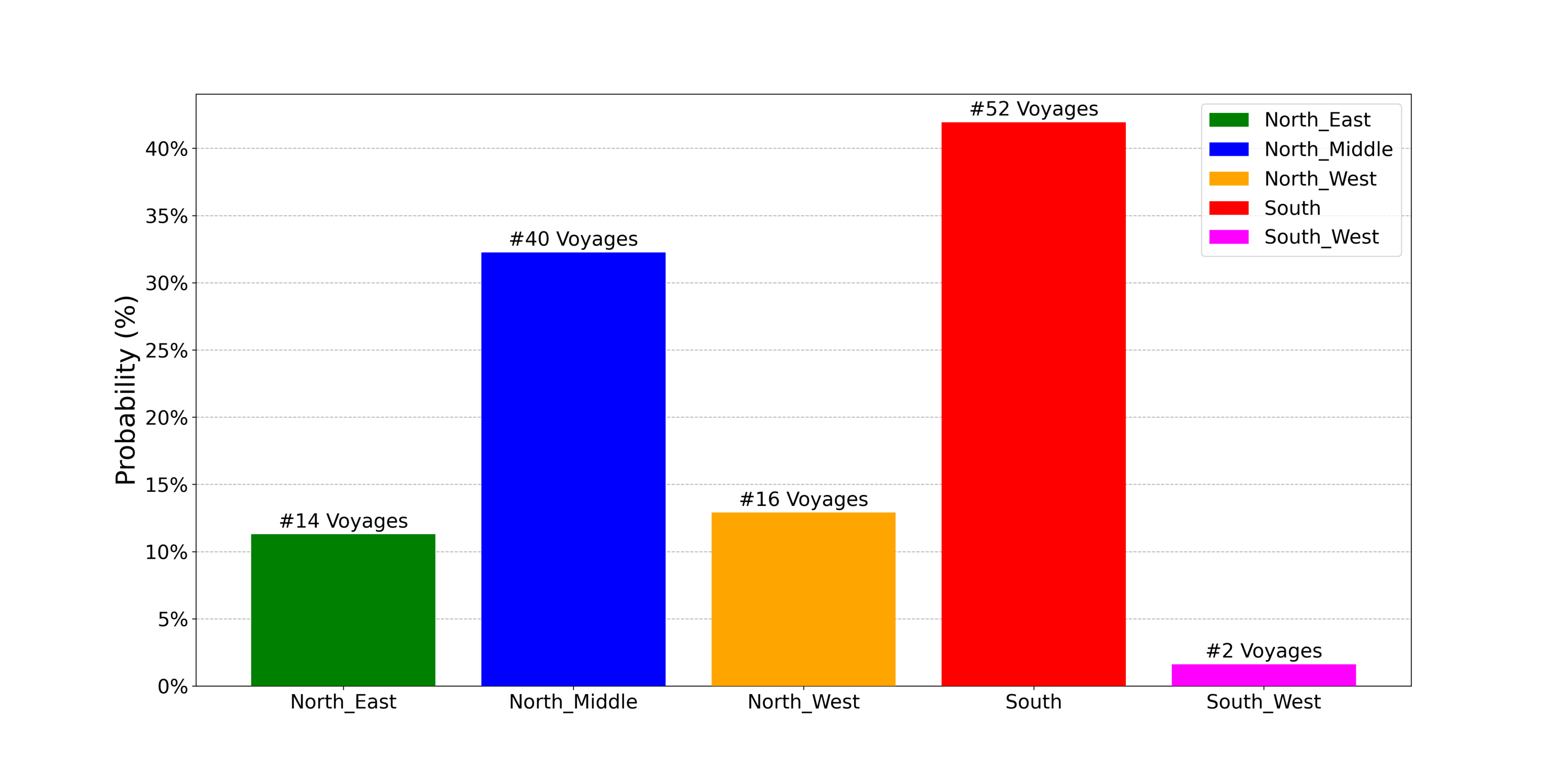}
      \caption{Distribution of Cinderella II voyages across the\\ five path clusters.}
      \label{fig:stats_path5cls}
    \end{subfigure}
    \begin{subfigure}{1.1\linewidth}
      \centering
      \includegraphics[width=\linewidth]{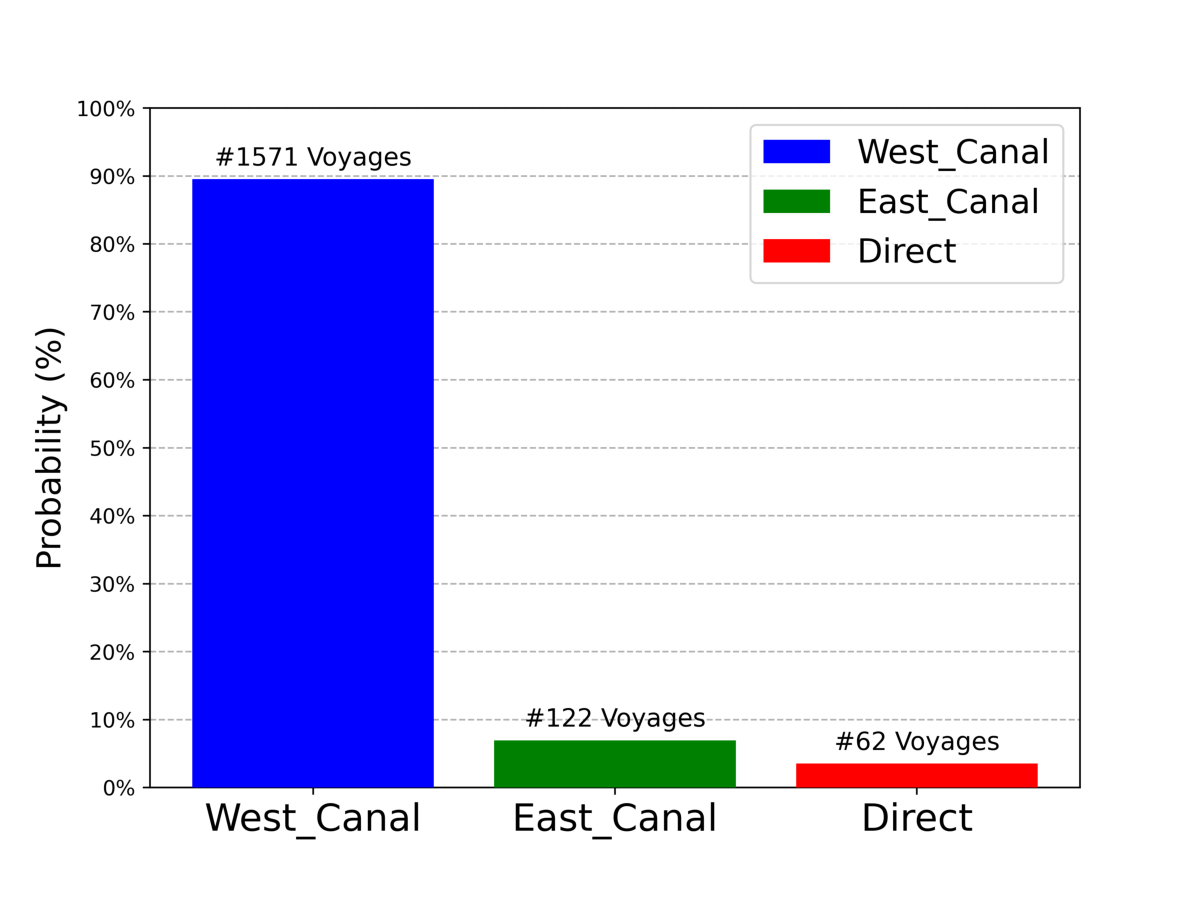}
      \caption{{Distribution of Buro voyages across the three\\  path clusters.}}
      \label{fig:stats_path3cls}
    \end{subfigure}
\caption{Distribution of voyages if both vessels across the the path clusters.}
\label{fig:stats_path_cls}
\end{figure}

In our approach, we emphasize the significance of data representation.
As a result, we group the path points based on their timestamps with a resolution of one second and store these grouped path points with distinctive Voyage IDs.\\
{The routes of Cinderella II and Buro, along their chosen paths, are depicted in Figure~{\ref{fig:Buro_Cind_route}}.
Path cluster distribution for both vessels are illustrated by the histograms in Figure~{\ref{fig:stats_path_cls}}.}

Afterward, these paths are ready to be processed by the path clustering approach to determine the overall path cluster.


For further details regarding the identification framework, the reader may refer to the source codes employed to produce the case study results. These codes are accessible at: \url{https://github.com/MohamedAbuella/Path_Clustering}.

\section{Results and Discussion}\label{Sec_results}
The vessel has a fixed-route which starts from the southern port to the northern port or vice versa.
This route can be divided into four segments, specifically North, Middle, South, and Direct, as depicted in Figure~\ref{Aggreg_rts}.\\
Cruising speeds are more common in North, South, and Direct segments of the vessel's route. Meanwhile, in the Middle segment, the vessel typically operates at maneuvering speeds, due to the presence of two ports located on west and east sides of the canal.

\subsection{Statistical Analysis }\label{Res_Stat_Analysis}
\begin{figure}[htb]
\centering
\includegraphics[width=\linewidth]{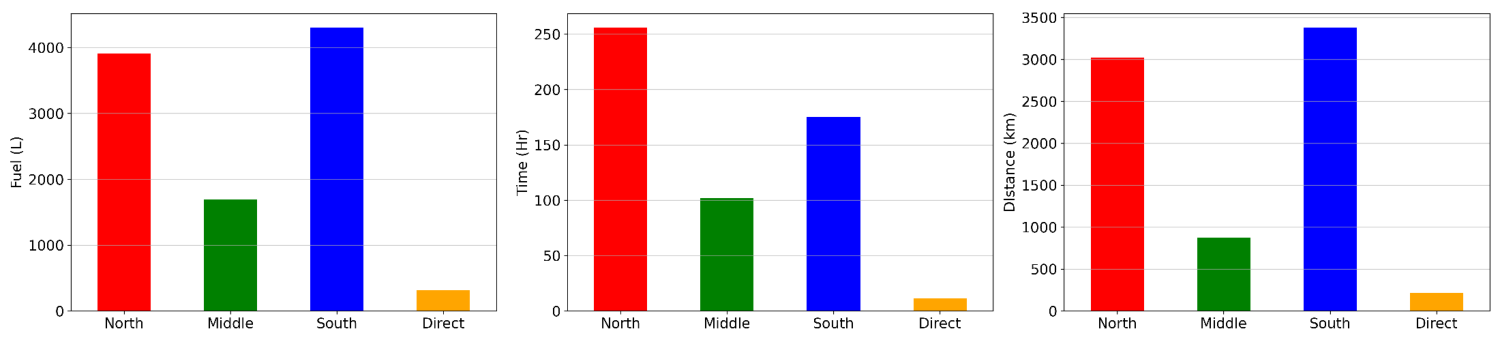}
\caption{Barplots for statistics of Fuel, Time, and Distance in different route segments,  for vessel Buro.}
\label{Stats_NMS_sections}
\end{figure}

We have first conducted a statistical analysis on the dataset.
Figure~\ref{Stats_NMS_sections} illustrates some important statistic for all aggregated voyage, with regard to the accumulated fuel, time, and distance at different route segments. 
These statistics represent the routes of vessel Buro, as depicted in Figure~\ref{fig:Buro_Cind_route}.

As it can be seen from Table~\ref{Stats_Speed_Modes}, the difference of fuel consumption of the and cruising speeds is 5.47\%, so that and also based on the recommendations from domain experts in compliance with maritime regulations including safety and traffic considerations, it might be more practical to primarily focus on optimizing cruising speed.
\begin{table}[htb]
\centering
\caption{Statistics of the dataset for different speed modes}
\label{Stats_Speed_Modes}
\begin{tabular}{|c|c|c|c|}
\hline
\multirow{2}{*}{Variable} & \multicolumn{2}{c|}{Speeds} & \multirow{2}{*}{Difference (\%)} \\
\cline{2-3}
& All & Cruising & \\
\hline
Fuel, total (Liter) & 1329.2 & 1256.62 & 5.47\% \\
Time, total (Hour) & 48.04 & 22.32 & 53.57\% \\
Distance, total (km) & 608.72 & 349.2 & 42.68\% \\
Speed, average (m/s) & 2.67 & 1.67 & 37.5\% \\
\hline
\end{tabular}
\end{table}

\subsection{Modeling of Energy Efficiency}\label{Res_Modeling}

The first step in optimizing the model is to identify the best set of input parameters. We consider four cases of ANN inputs, where each case consists of different combinations of operational and weather variables. Further details about these ANN input cases are provided in Tables~\ref{tab:scenarios_descrip} and \ref{vars_abbrev_tab}.

\begin{table}[htb]
\caption{Description of the four input cases of ANN.} \label{tab:scenarios_descrip}
\centering
\begin{tabular}{c|c|c|c}
\hline
\multirow{1}{*}{\begin{tabular}[c]{@{}c@{}} Inputs \\ Case \end{tabular}}  & \multirow{1}{*}{\begin{tabular}[c]{@{}c@{}} Operational \\ Variables \end{tabular}}  & \multicolumn{2}{c} {Weather Variables} \\
&& onboard data  &  external sources \\ \hline
I &\multirow{4}{*}{\begin{tabular}[c]{@{}c@{}} Vessel's location, \\ speed, and direction \\ are used for all cases  \end{tabular}}  &  wind   & ---  \\
II & & --- & wind, wave, and current  \\
III &  & wind  &  wave and current  \\
IV & & wind   &  wind, wave, and current  \\
\hline
\end{tabular}
\end{table}

\begin{figure}[htb]
\centering
\begin{subfigure}{\linewidth}
\centering
        \includegraphics[width=0.75\linewidth]{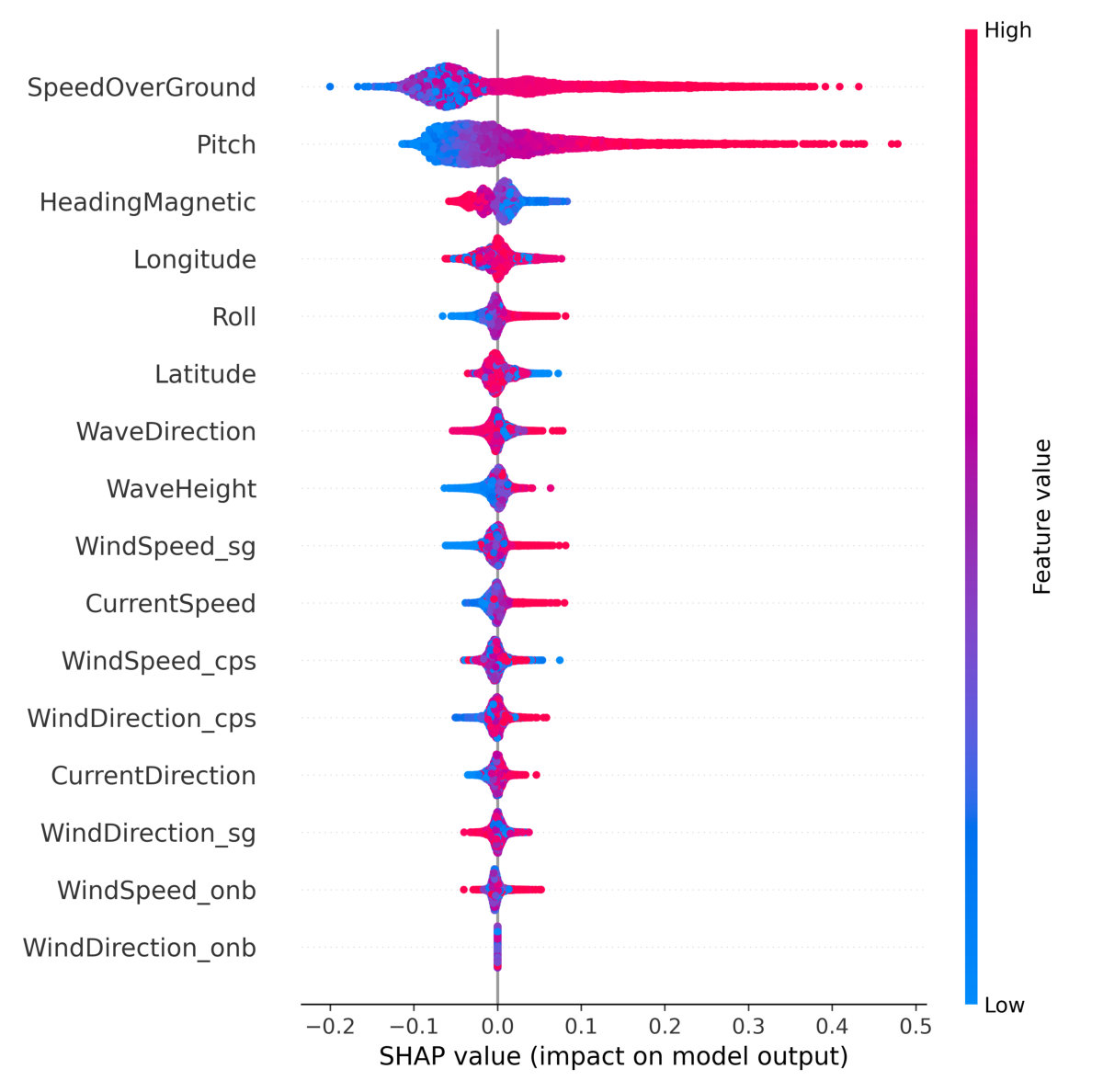}
    \caption{\centering{Model's output is EngineFuelRate, ($R^2$=0.7615)}}
    \label{Shap_Reg_EFR}
\end{subfigure}
\hfill
\begin{subfigure}{\linewidth}
\centering
        \includegraphics[width=0.75\linewidth]{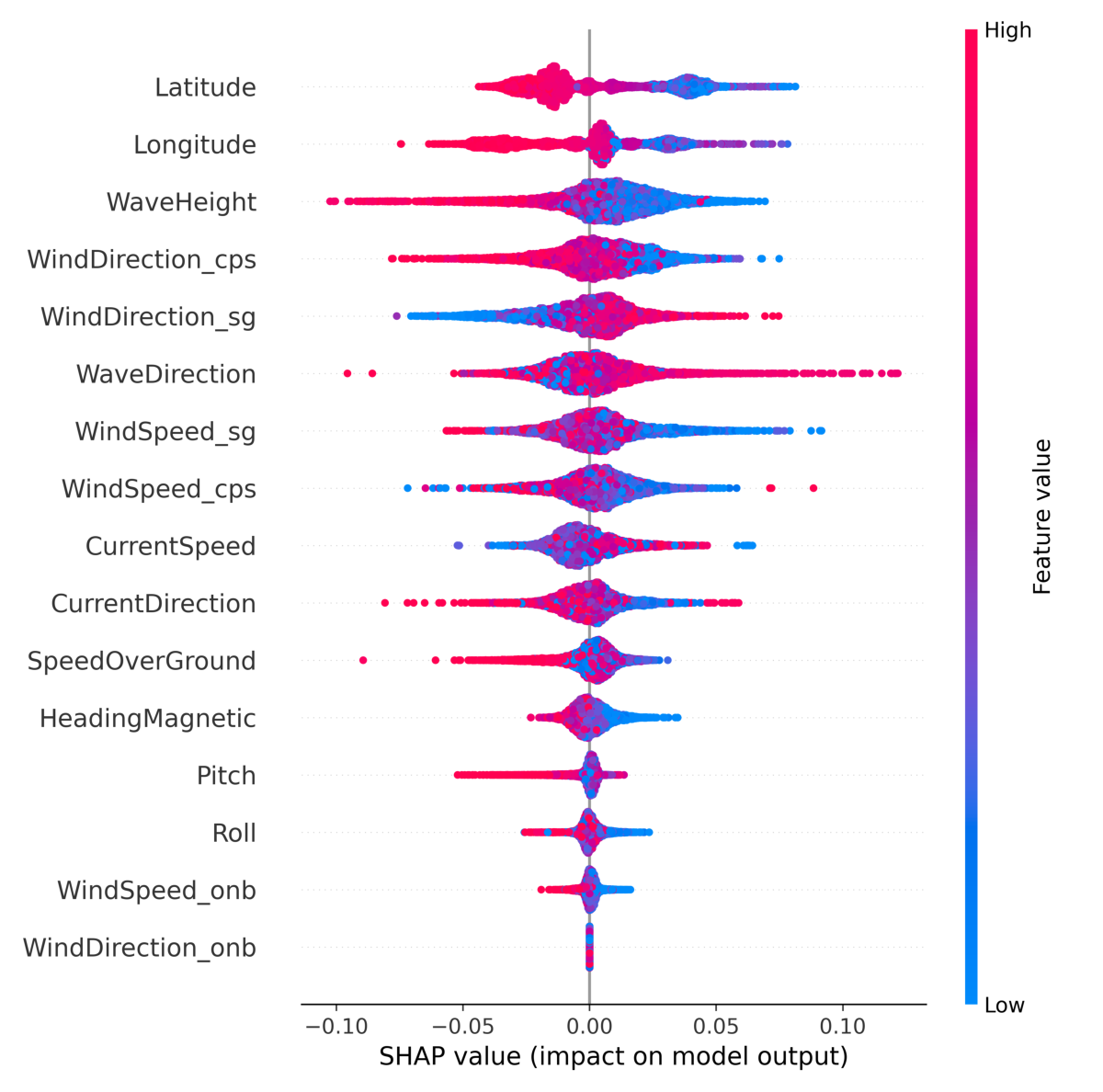}    
    \caption{\centering{Model's output is Efficiency Score,  ($R^2$=0.8324)}}
    \label{shap_global_eff}
\end{subfigure}
\caption{Beeswarm plots of SHAP values for XGBoost regression model with different outputs}
\label{Shap_Reg_EFR_Eff_Loc_allvars}
\end{figure}

The Beeswarm plot in Figure~\ref{shap_global_eff} indicates that the vessel's location has the most significant impact on the Efficiency Score. Therefore, a spatial analysis was conducted to identify the impact of various combinations of operational and weather variables on the Efficiency Score concerning the vessel's location.

\begin{figure}[htb]
 \centering
\begin{subfigure}[!ht]{\linewidth}
    \centering 
    \includegraphics[width=0.75\linewidth] 
{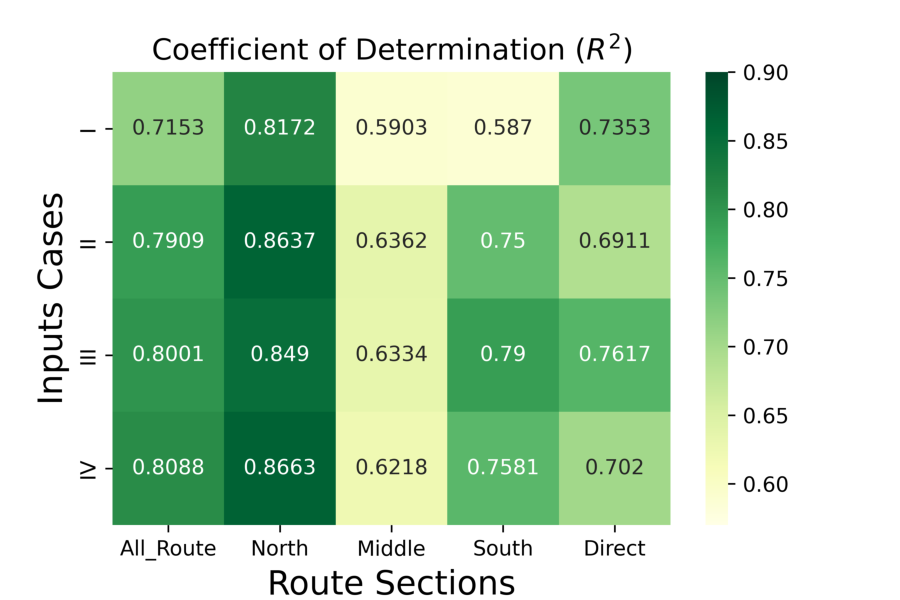}
    \caption{EngineFuelRate}
    \label{Heatmap_ANN_EFR_R2}
\end{subfigure}
\hfill
\begin{subfigure}[!ht]{\linewidth}
    \centering 
    \includegraphics[width=0.75\linewidth]{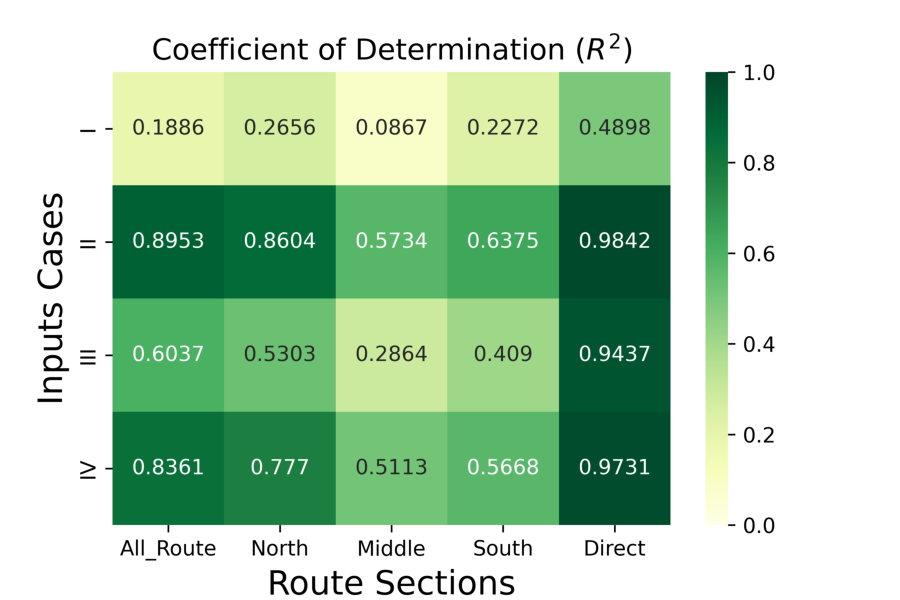}
    \caption{Efficiency Score} 
    \label{Heatmap_ANN_Eff_R2}
\end{subfigure}
\caption{Results ($R^{2}$) for ANN regression with EngineFuelRate and Efficiency Score as outputs across different input cases in relation to varying vessel's route sections.}
\label{Fig_Heatmap_ANN_Nav_Weath_OnbEx}
\end{figure}

The results are shown as heatmaps in Figure~\ref{Fig_Heatmap_ANN_Nav_Weath_OnbEx}, revealing that the direct route from south to north or vice versa, located on the open sea, is particularly susceptible to the impact of weather conditions. Thus, for this direct section of vessel routes, the estimation of Efficiency Score, as shown in Figure~\ref{Heatmap_ANN_Eff_R2}, has the highest accuracy with different input combinations.

Meanwhile, in the north section, where strong either head or tail wind is more frequent (in this area, west winds dominate) with respect to the vessel route, the Efficiency Score estimation has the second highest accuracy, as in Figure~\ref{Heatmap_ANN_Eff_R2}. 

In the other case, when it comes to estimating EngineFuelRate, as shown in Figure~\ref{Heatmap_ANN_EFR_R2}, the results are not accurate. For instance, the direct sections of the route are not achieving the highest accuracy, even though they are supposed to experience more weather conditions than other sections due to these sections being the most similar to an open sea.

\subsection{Improving Voyage Efficiency}\label{Res_voyag_Eff}

After that we have implemented the framework for improving the vessel's energy efficiency, as presented in Section~\ref{sec_framework_EEimprov}.
Then, we evaluate the results, and for a sake of a fair comparison, we are injected both the actual measured and optimized speed profiles into the same estimation model of energy efficiency to predict fuel and time before and after the improving framework of energy efficiency is being implemented. Once the fuel and time estimated, we compute the efficiency to determine how much energy has been saved.

One of our main metrics to evaluate the model performance of voyage efficiency optimization is the gain of efficiency scores, as represented in (\ref{eq:eff_gain}). 
\begin{equation}
Eff._{Gain} = \frac{Eff.Score_{{Pred.}} - Eff.Score_{Meas.}}{Eff.Score_{Meas.}} \times 100
\label{eq:eff_gain}
\end{equation}
Where $Eff.Score_{{Meas.}}$ and $Eff.Score_{{Pred.}}$ represent the voyage efficiency obtained with measured and predicted speed profiles, respectively.

\begin{table}[htb]
\centering
\caption{Average efficiency gains (Eff. Gains \%, see Eq.~\ref{eq:eff_gain}) and counts of improved voyages (IV Count \#) out of 162 voyages in the test dataset.}
\label{Table_Eff_gain_couts}
\begin{tabular}{c|c|cccc cccc}
\hline
\multirow{2}{*}{Cluster} & \multirow{2}{*}{Efficiency Score} & \multirow{2}{*}{LSTM} & \multirow{2}{*}{kNN} & \multirow{2}{*}{1NN-DTW} & \multirow{2}{*}{HMM}
\\ & & & & & \\
\hline
\multirow{2}{*}{Top10Pr} 
& Eff. Gains (\%) & 2.61 & 2.13 & 3.20 & 6.05 \\
& IV Count (\#) & 134 & 114 & 127 & 139 \\
\hline
\multirow{2}{*}{Top25Pr} 
& Eff. Gains (\%) & 2.38 & 1.58 & 3.23 & 1.30 \\
& IV Count (\#) & 129 & 107 & 128 & 107 \\
\hline
\multirow{2}{*}{Top50Pr} 
& Eff. Gains (\%) & 0.97 & 0.98 & 2.58 & 7.34 \\
& IV Count (\#) & 100 & 106 & 117 & 140 \\
\hline
\multirow{2}{*}{Top75Pr} 
& Eff. Gains (\%) & -0.84 & 0.50 & 2.28 & 9.31 \\
& IV Count (\#) & 60 & 93 & 119 & 141 \\
\hline
\multirow{2}{*}{Average} 
& Eff. Gains (\%) & 1.28 & 1.30 & 2.82 & 6.00 \\
& IV Count (\#) & 105.75 & 105.00 & 122.75 & 131.75 \\
\hline
\end{tabular}
\end{table}
\begin{table}[htb]
\centering
\caption{Average and deviation of gains (\%) of Eff-Scores in three weather states, when models are trained with the four data clusters}
\label{table_Stats_Effall_gains}
\begin{tabular}{|c|c c|c c|c c|c c|}
\hline
\multicolumn{1}{|c|}{Model} & \multicolumn{2}{|c|}{LSTM} & \multicolumn{2}{|c|}{kNN} & \multicolumn{2}{|c|}{1NN-DTW} & \multicolumn{2}{|c|}{HMM} \\
\hline
Weather & Avg & Std & Avg & Std & Avg & Std & Avg & Std \\
\hline
Calm & 0.17 & 4.72 & 1.17 & 3.29 & 2.21 & 4.30 & 3.96 & 6.19 \\
Moderate & 1.53 & 4.12 & 1.26 & 4.19 & 3.4 & 4.47 & 5.33 & 6.56 \\
Rough & 1.94 & 3.48 & 1.42 & 3.65 & 2.84 & 4.90 & 8.17 & 9.08 \\
\hline
Average & 1.21 & 4.11 & 1.28 & 3.71 & 2.82 & 4.56 & 5.82 & 7.28 \\
\end{tabular}
\end{table}

As shown in Table~\ref{Table_Eff_gain_couts}, the HMM model achieves the highest average efficiency gain of 6.00\%, followed by the 1NN-DTW model (2.82\%), the kNN model (1.30\%), and the LSTM model (1.28\%). In terms of the number improved voyages out of 162 voyages in test dataset,  the HMM model also improves the energy efficiency of the most average number of improved voyages (131.75 out of 162 voyages).

The HMM model achieves its best performance when trained on the Top75Pr cluster, which includes voyages with lower Eff-Scores and frequently encountered adverse weather conditions. Such performance underscores the HMM model's capability to learn the hidden patterns between the vessel speed and weather states, ultimately facilitates for developing more efficient speed profiles.


\begin{figure}[htb]
\centering
\begin{subfigure}{0.48\linewidth}
    \includegraphics[width=\linewidth]{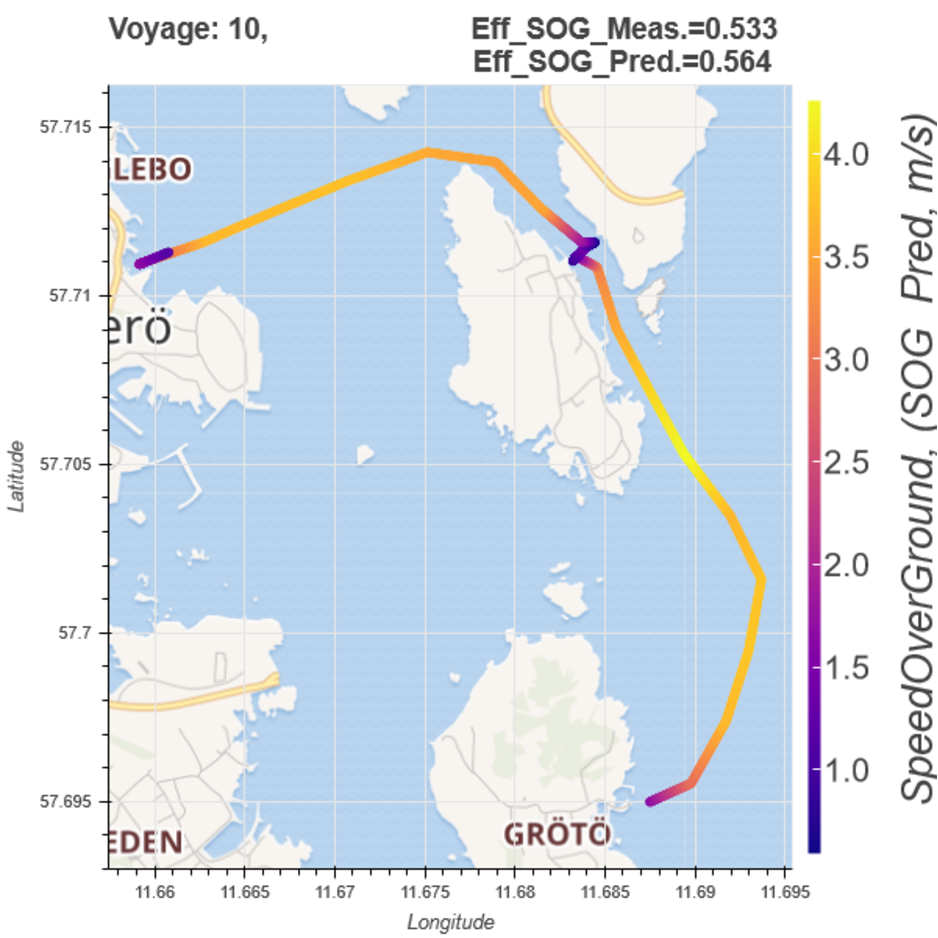}
    \caption{LSTM-based model}
    \label{voyage_LSTM}
\end{subfigure}
\hfill
\begin{subfigure}{0.48\linewidth}
    \includegraphics[width=\linewidth]{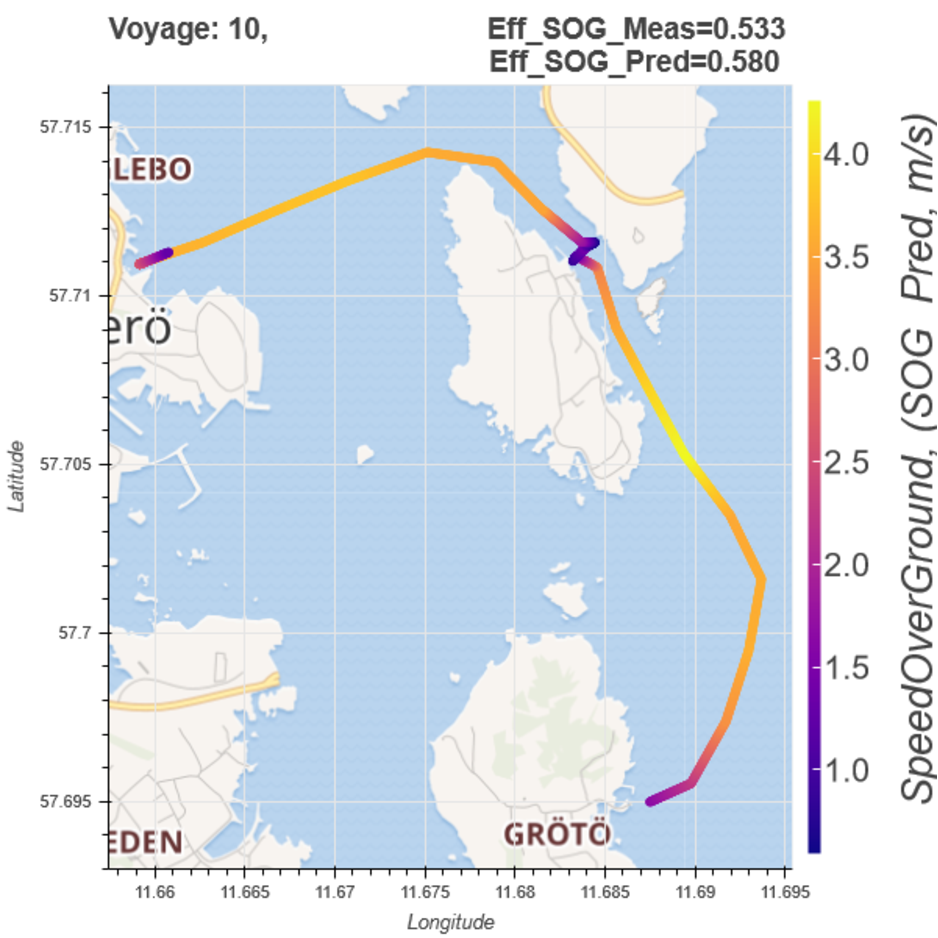}
    \caption{kNN-based model}
    \label{voyage_kNN}
\end{subfigure}
\hfill
\begin{subfigure}{0.48\linewidth}
    \includegraphics[width=\linewidth]{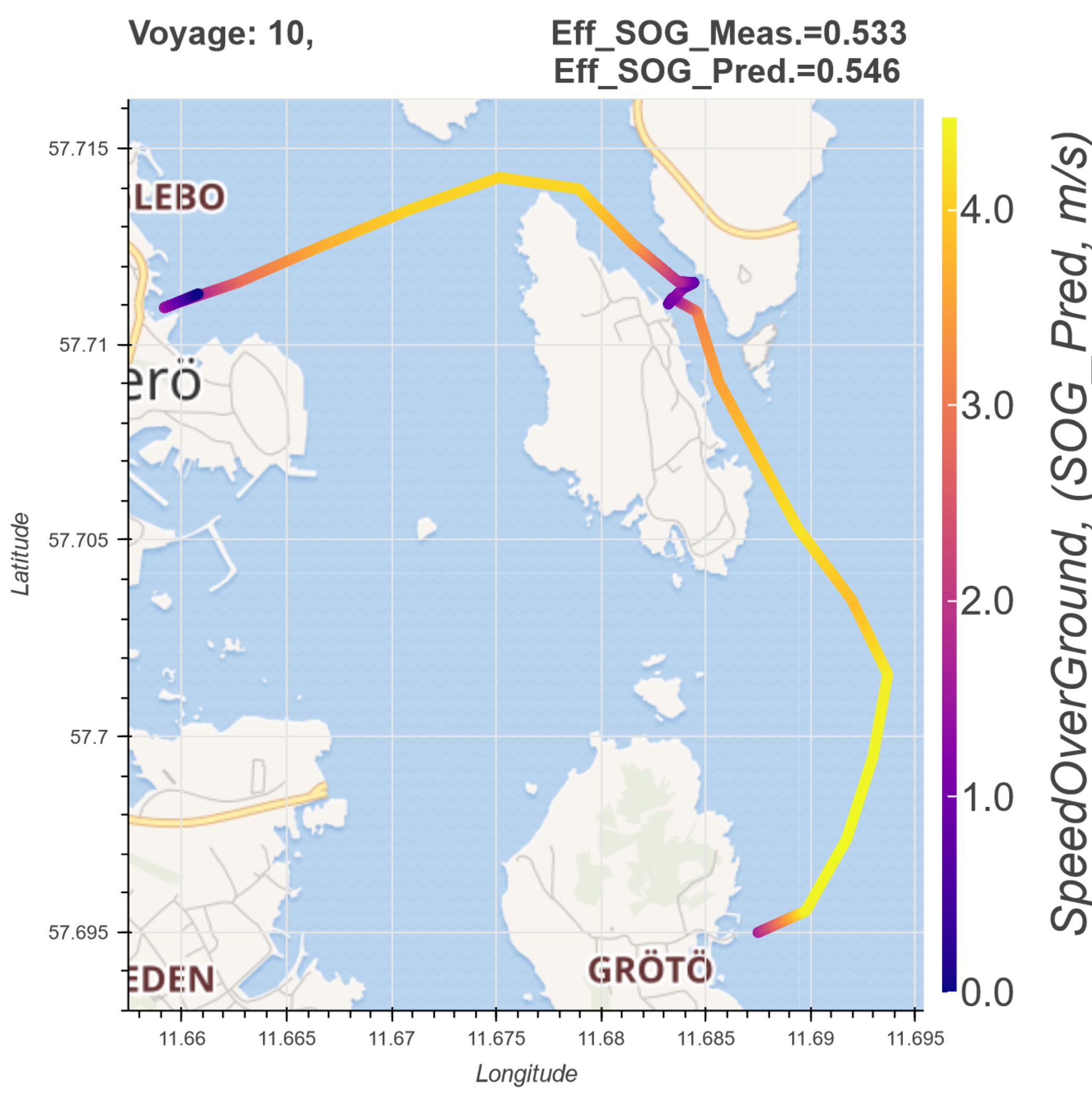}
    \caption{1NN-DTW-based model}
    \label{voyage_DTW}
\end{subfigure}
\hfill
\begin{subfigure}{0.48\linewidth}
    \includegraphics[width=\linewidth]{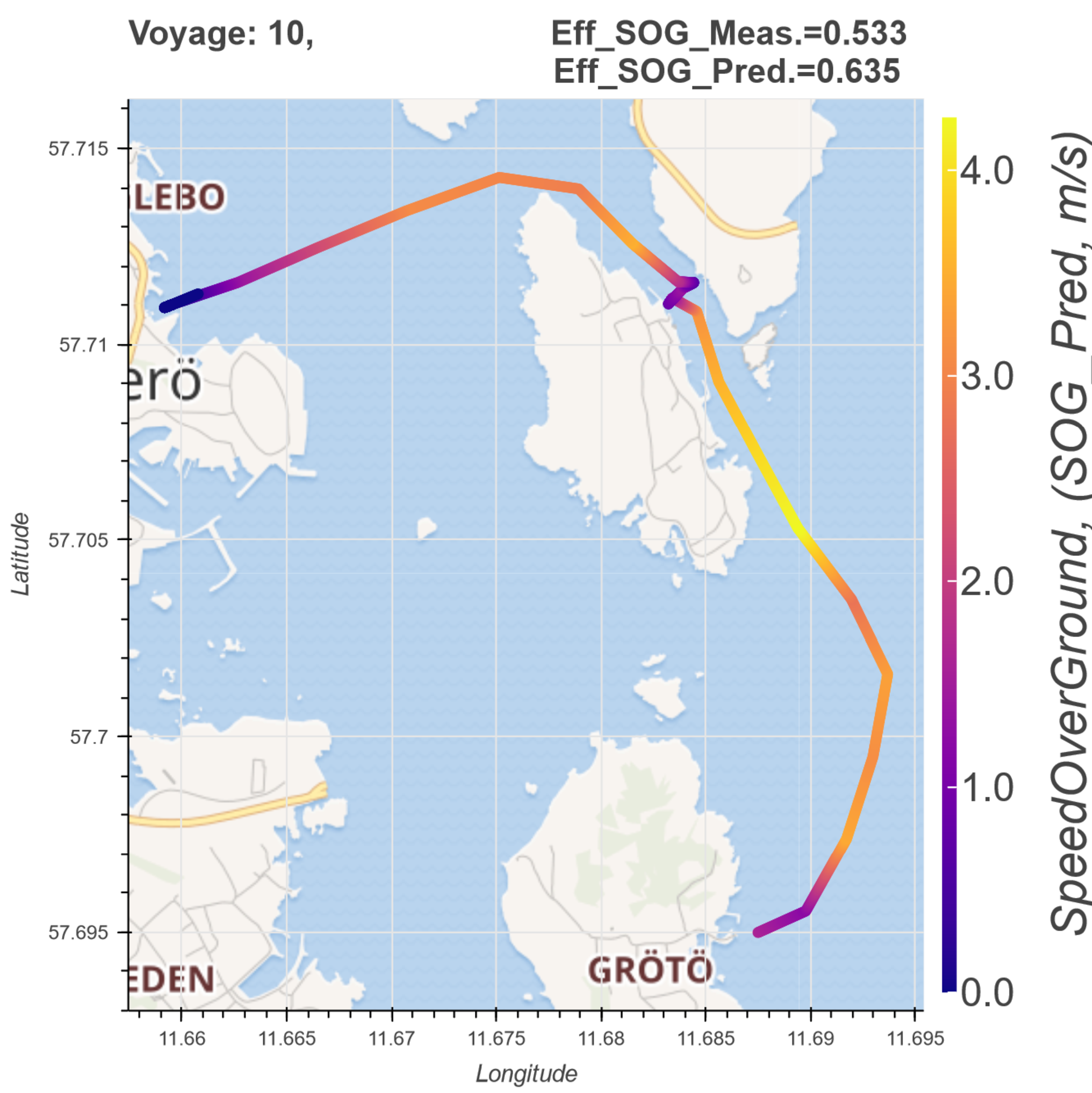}
    \caption{HMM-based model}
    \label{voyage_HMM}
\end{subfigure}
\caption{Predicted speed profile for a test voyage. From four time-series based models incorporate weather data as inputs and are trained by Top10Pr cluster.}
\label{Plots_sog_eff_seq}
\end{figure}

\begin{figure}[htb]
\centering
\begin{subfigure}{0.49\linewidth}
    \includegraphics[width=\linewidth]{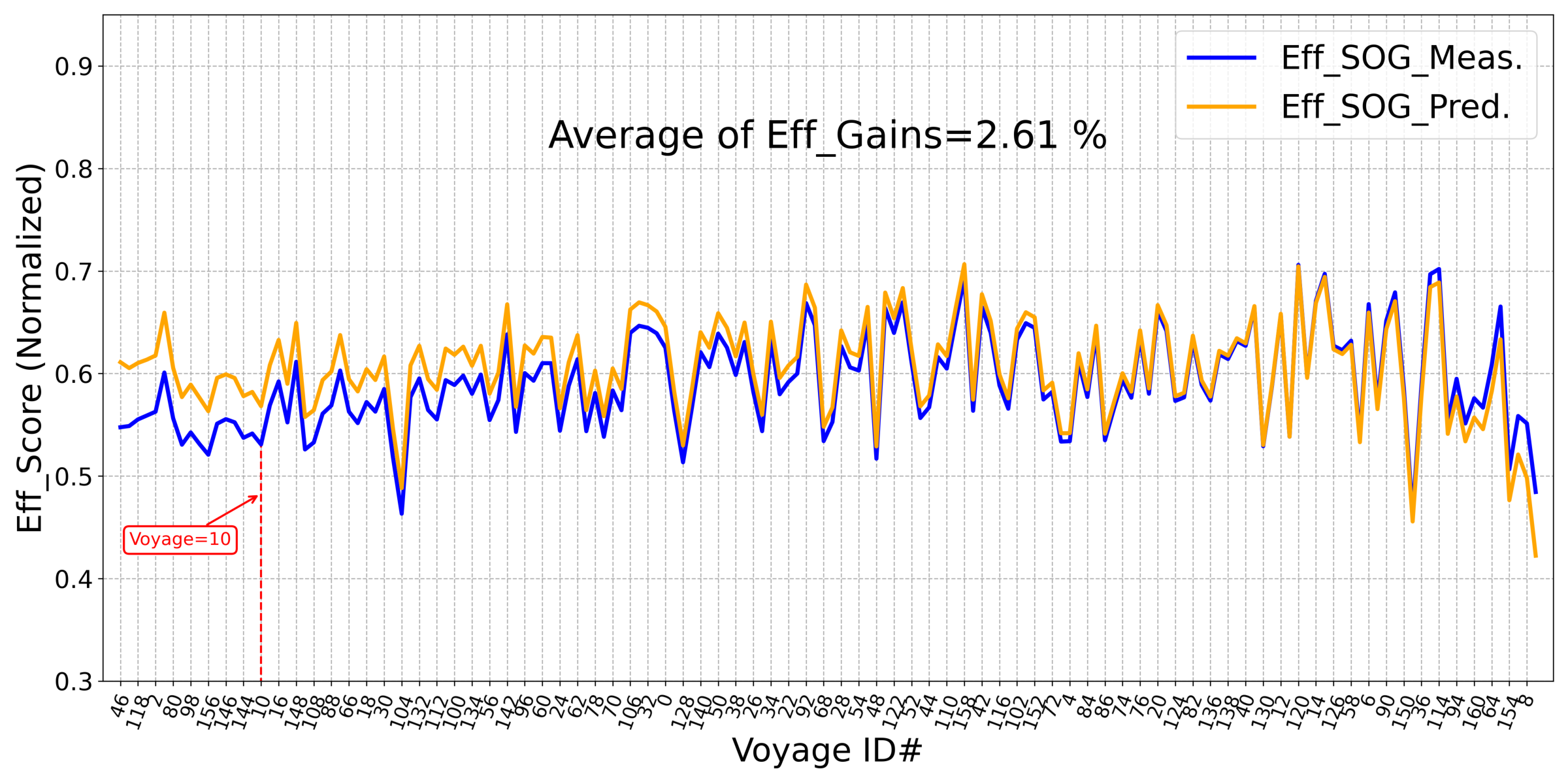}   
    \caption{LSTM-based model}
    \label{Sort_Effs_graph_LSTM}
\end{subfigure}
\hfill
\begin{subfigure}{0.49\linewidth}
    \includegraphics[width=\linewidth]{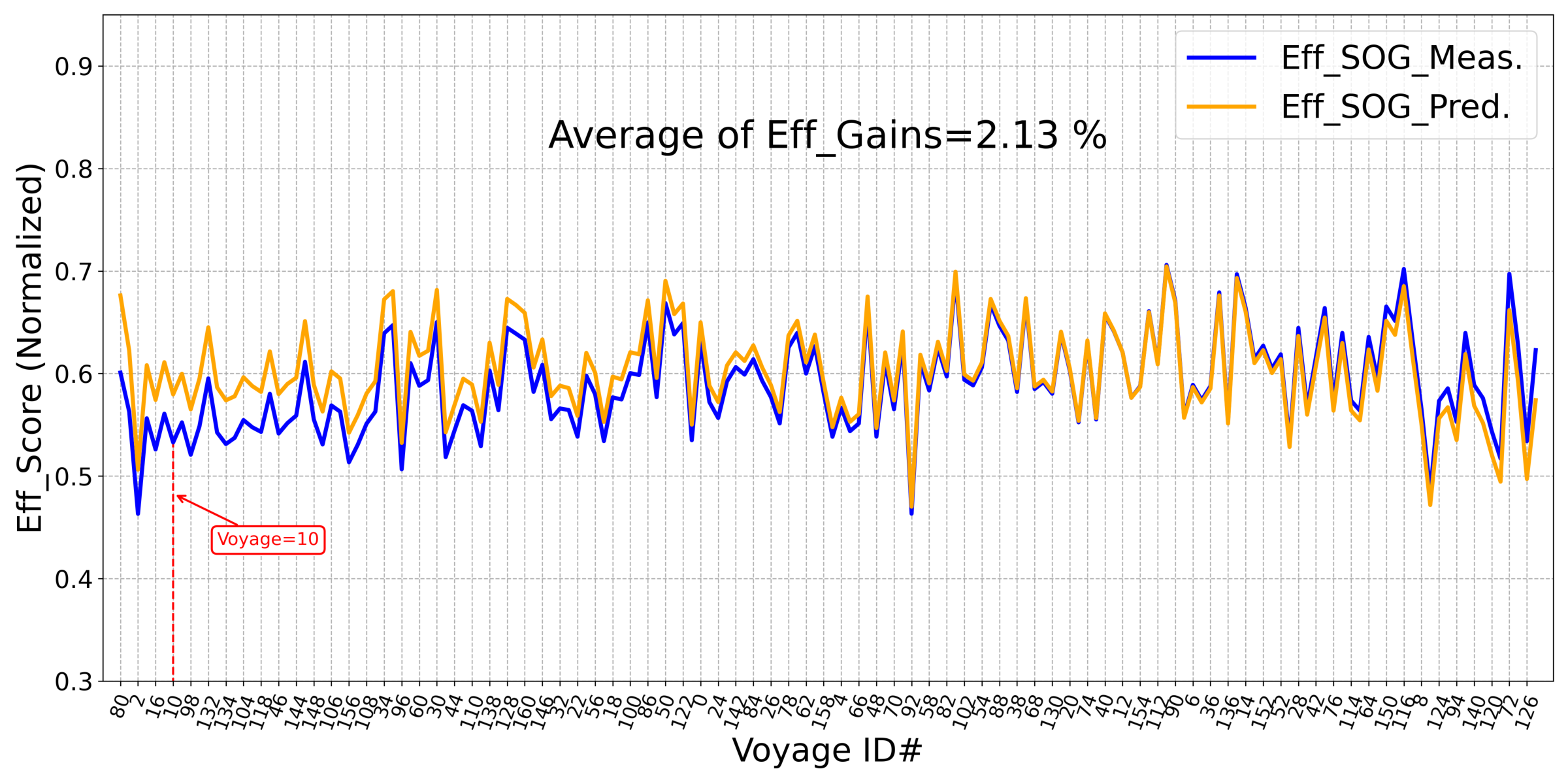}
    \caption{kNN-based model}
    \label{Sort_Effs_graph_kNN}
\end{subfigure}
\hfill
\begin{subfigure}{0.49\linewidth}
    \includegraphics[width=\linewidth]{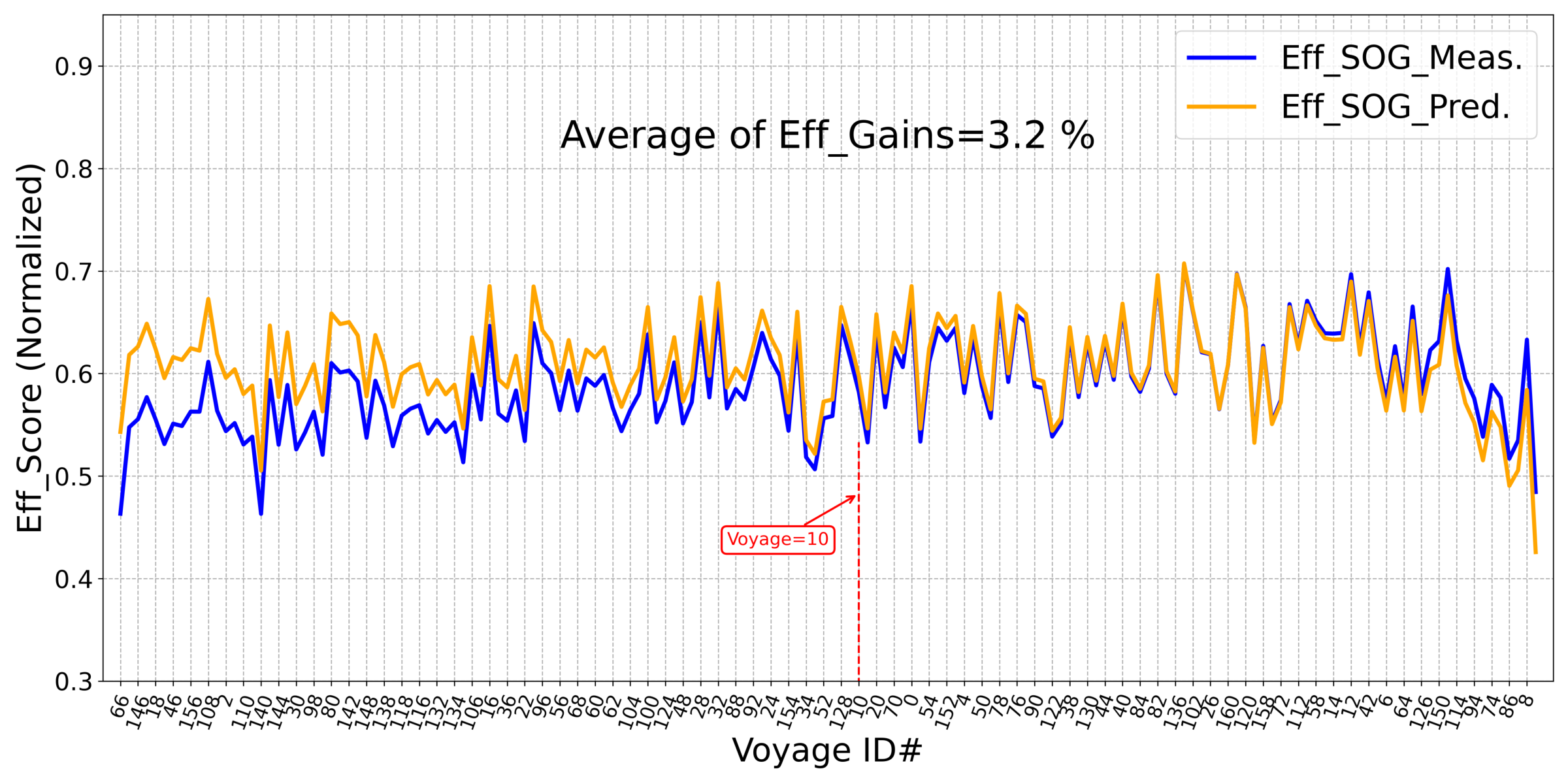}
    \caption{1NN-DTW-based model}
    \label{Sort_Effs_graph_DTW}
\end{subfigure}
\hfill
\begin{subfigure}{0.49\linewidth}
    \includegraphics[width=\linewidth]{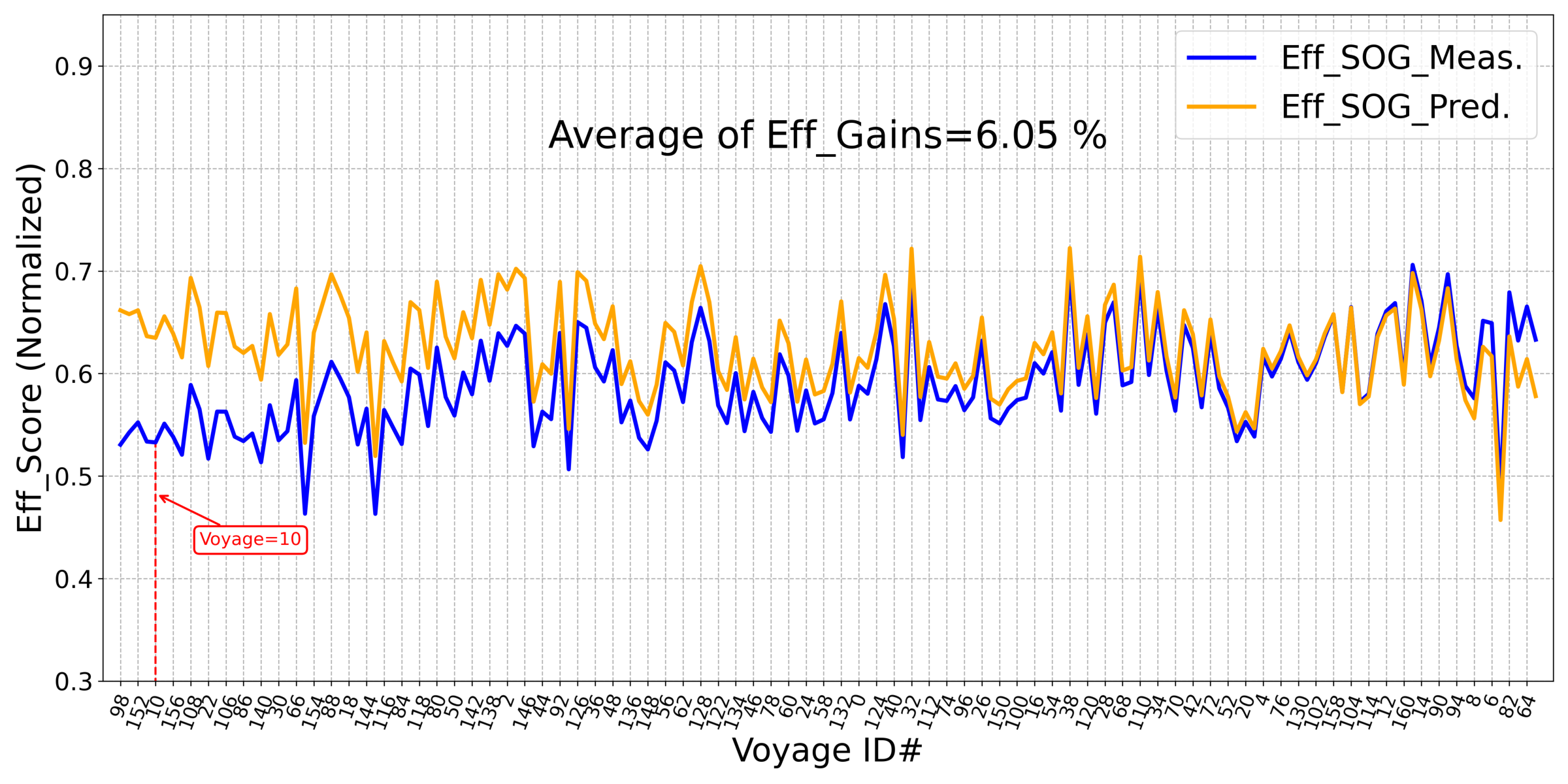}
    \caption{HMM-based model}
    \label{Sort_Effs_graph_HMM}
\end{subfigure}
\caption{Sorted Eff-Scores based on their gain, for 162 test voyages. From four time-series based models incorporate weather data as inputs and are trained by Top10Pr cluster,  
$Eff._{\text{Gain}}$ as in Eq.~\ref{eq:eff_gain}.}
\label{Plots_sort_effs}
\end{figure}

We also present the results by plots illustrating the predicted speed profiles for a test voyage, which are generated by four time-series based models that incorporate weather data as inputs. These models were trained using data from the Top10Pr cluster. 

In summary, the HMM-based model is the most effective model for improving energy efficiency for a vessel voyage in short sea. The HMM model is able to learn the complex relationships between the input features (e.g., speed and weather) and the output feature (Eff-Score), even in different weather conditions.

\subsection{Path Identification}\label{Res_path_id}


In this section, we present the results of our spatial clustering approach for vessel path identification and discuss the implications of these findings. The approach involved the utilization of position information and various clustering techniques, specifically k-means, hierarchical clustering, and Gaussian distributions clustering to the paths of two vessels, Buro and Cinderella II. These vessels operate in distinct locations, leading to diverse datasets.

\subsubsection{Evaluation Metrics for Path Clustering}\label{eval_metrics}
~ \\
The results of vessel path identification are evaluated through visual inspection and tabulation using metrics such as confusion matrix, precision, recall, and F1-score~\cite{yan2021artificial}. 

The hits and messes of path clustering are presented by the confusion matrix.
For our results of path clustering, the confusion matrix is a one-vs-one type matrix.
Then, the confusion matrix is converted into a one-vs-all type matrix (binary-class confusion matrix)~as shown in Eq.~(\ref{eq:cofus_mat_2class}), for calculating class-wise metrics like precision, recall, and F1-score. 

\begin{equation} \label{eq:cofus_mat_2class}
\begin{array}{c|cc}
& \text{Pred. Pos.} & \text{Pred. Neg.} \\
\hline
\text{Act. Pos.} & \text{TP } & \text{FN } \\
\text{Act. Neg.} & \text{FP } & \text{TN }
\end{array}
\end{equation}
where True Positives (TP), False Positives (FP), True Negatives (TN), and False Negatives (FN) are determined by comparing the predicted (Pred.) and actual (Act.) path clusters.

{The confusion matrix transformation involves considering one class as positive at a time, while combining all other classes as negative. This process is repeated iteratively for each class, resulting in multiple binary-class confusion matrices.}

The following performance metrics were used:
\begin{itemize}
    \item Precision: the ratio of true positives to the total number of predicted positives.
    \item Recall: the ratio of true positives to the total number of actual positives in the data.
    \item F1-score: the harmonic mean of precision and recall. 
\end{itemize}

The equations for precision, recall, and F1-score are shown in Eqs.~(\ref{eq:Precision}), (\ref{eq:Recall}), and (\ref{eq:F1score}).

\begin{equation}\label{eq:Precision}
Precision = \frac{TP}{TP + FP}
\end{equation} 

\begin{equation}\label{eq:Recall}
Recall = \frac{TP}{TP + FN}
\end{equation}

\begin{equation}\label{eq:F1score}
F1\text{-}score= 2\times\frac{(Precision \times Recall)}{(Precision + Recall)}
\end{equation}
\subsubsection{Results of Path Clustering for Cinderella II Vessel}\label{cind_case}
~ \\
Table~\ref{Table:Result_Kmean_GMM_Summary} shows the results of applying k-means or Gaussian Mixture Model (GMM) models to identify the vessel paths from the distance matrix in the distance-based method of the path identification approach.
Notably, the paths with clusters of North-West, South, and South-West achieved an F1-score of 1.0, indicating that the approach correctly identified all the paths of these clusters.

\begin{figure}[htb]
  \centering
  \includegraphics[width=\linewidth]{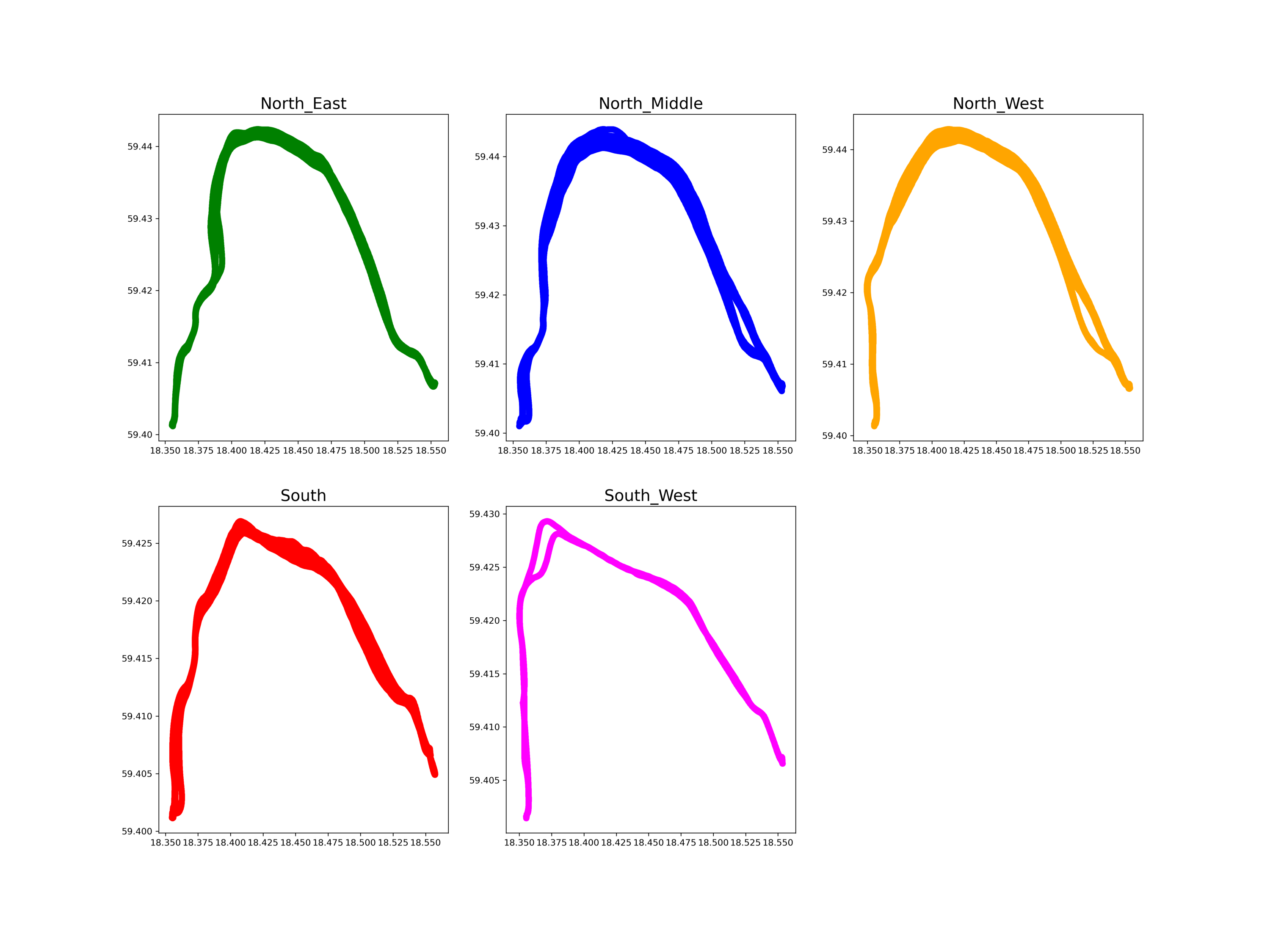}
  \caption{Display of the five clusters of path for Cinderella II vessel, (ground truth).}
  \label{fig:5plts_path5cls}
\end{figure}

\begin{figure}[htb]
  \centering
    \includegraphics[width=1.1\linewidth]{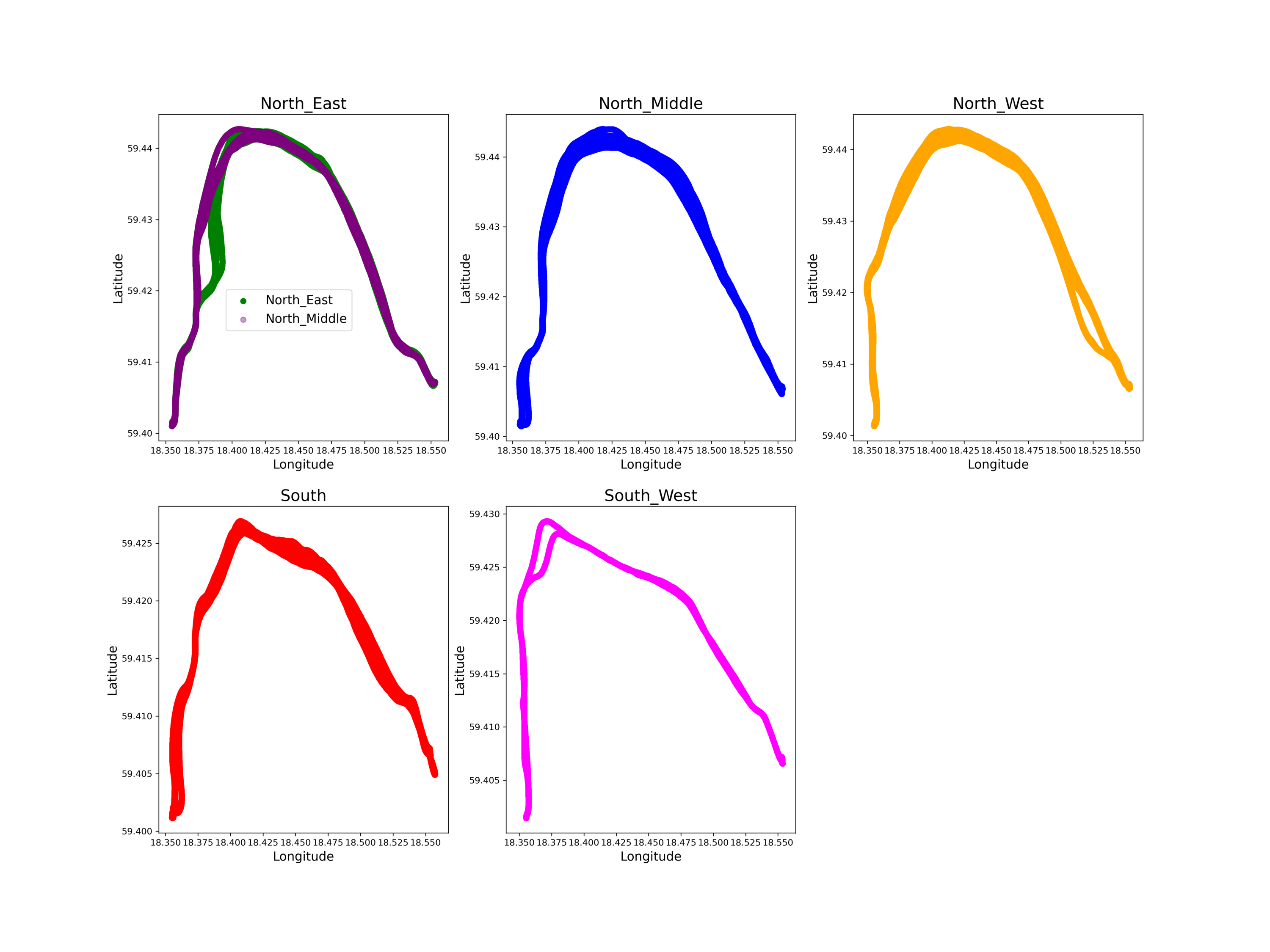}
  \caption{Results of both k-means and GMM clustering to five path clusters for Cinderella II.}
  \label{fig:Kmean_GMM_clus_path5cls}
\end{figure}

\begin{figure}[htb]
  \centering
  \includegraphics[width=1.1\linewidth]{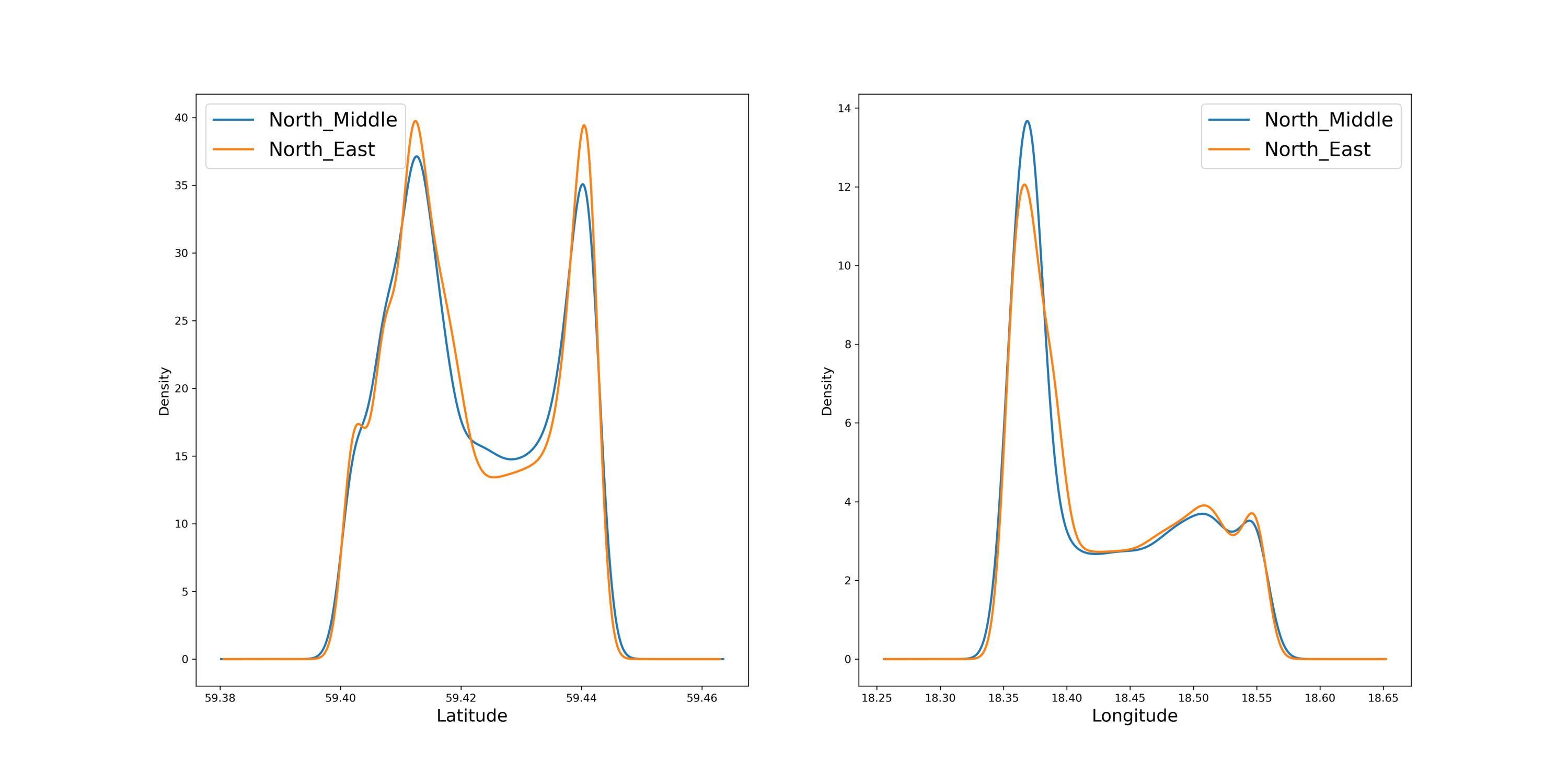}
  \caption{Probability distribution of location coordinates for mis-clustered paths of Cinderella II by both k-means and GMM.}
  \label{fig:Fig_Cind_5paths_PDF_Misclustered}
\end{figure}

In contrast, the North-East and North-Middle paths exhibited lower F1 scores compared to other clusters. 
The path cluster of North-Middle is the most challenging path to identify since six such paths have been clustered as North-East, as can be seen by comparing Figures~\ref{fig:5plts_path5cls} and~\ref{fig:Kmean_GMM_clus_path5cls}, which are the visualization for all the paths, color-marked based on their ground truth clusters.
Figure~\ref{fig:Fig_Cind_5paths_PDF_Misclustered} illustrates the probability distribution of mis-clustered paths with respect to latitude and longitude coordinates. It is obvious that these paths have nearly identical coordinates, which makes them challenging paths to cluster with k-means or GMM.
This suggests that there is still room for improvement by using other ML clustering methods.

\begin{figure}[htb]
  \centering
  \includegraphics[width=1.1\linewidth]{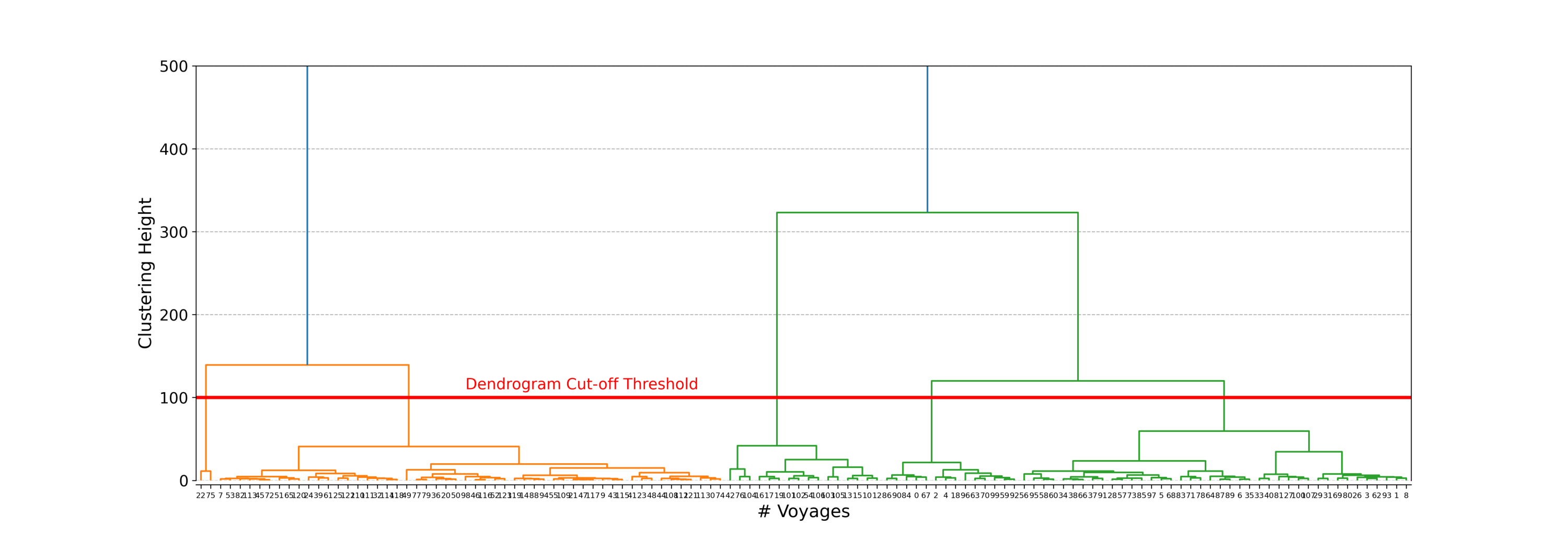}
  \caption{Results of hierarchical clustering the paths of Cinderella vessel to five clusters.}
  \label{fig:Hierch_clus_path5cls}
\end{figure}

Table~\ref{Table:Result_Hierch_Summary} presents the results of employing hierarchical clustering to the distance matrix in the distance-based method for clustering the vessel paths. In hierarchical clustering, there is a parameter called "Dendrogram Cut-off threshold," and its value should be selected depending on the number of path clusters. Hence, as illustrated in Figure~\ref{fig:Hierch_clus_path5cls}, this parameter is denoted by the Y-axis as a clustering height, and its value is set to 100 for clustering the vessel path into five clusters.

\begin{figure}[htb]
  \centering
  \includegraphics[width=\linewidth]{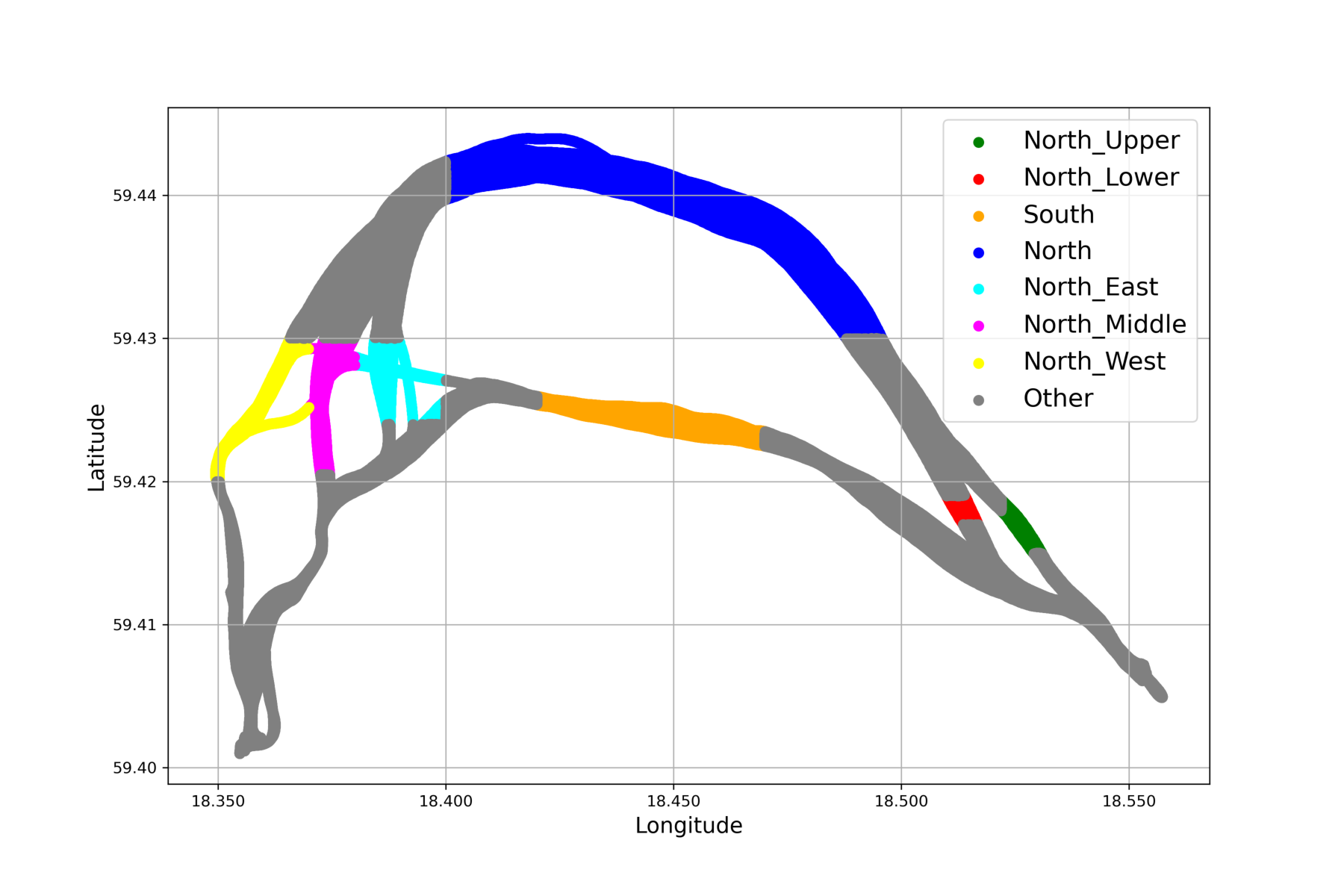}
  \caption{Distribution of the route of Cinderella II into eight segments.}
  \label{fig:Distribution_path8cls}
\end{figure}

\begin{figure}[htb]
  \centering
  \includegraphics[width=\linewidth]{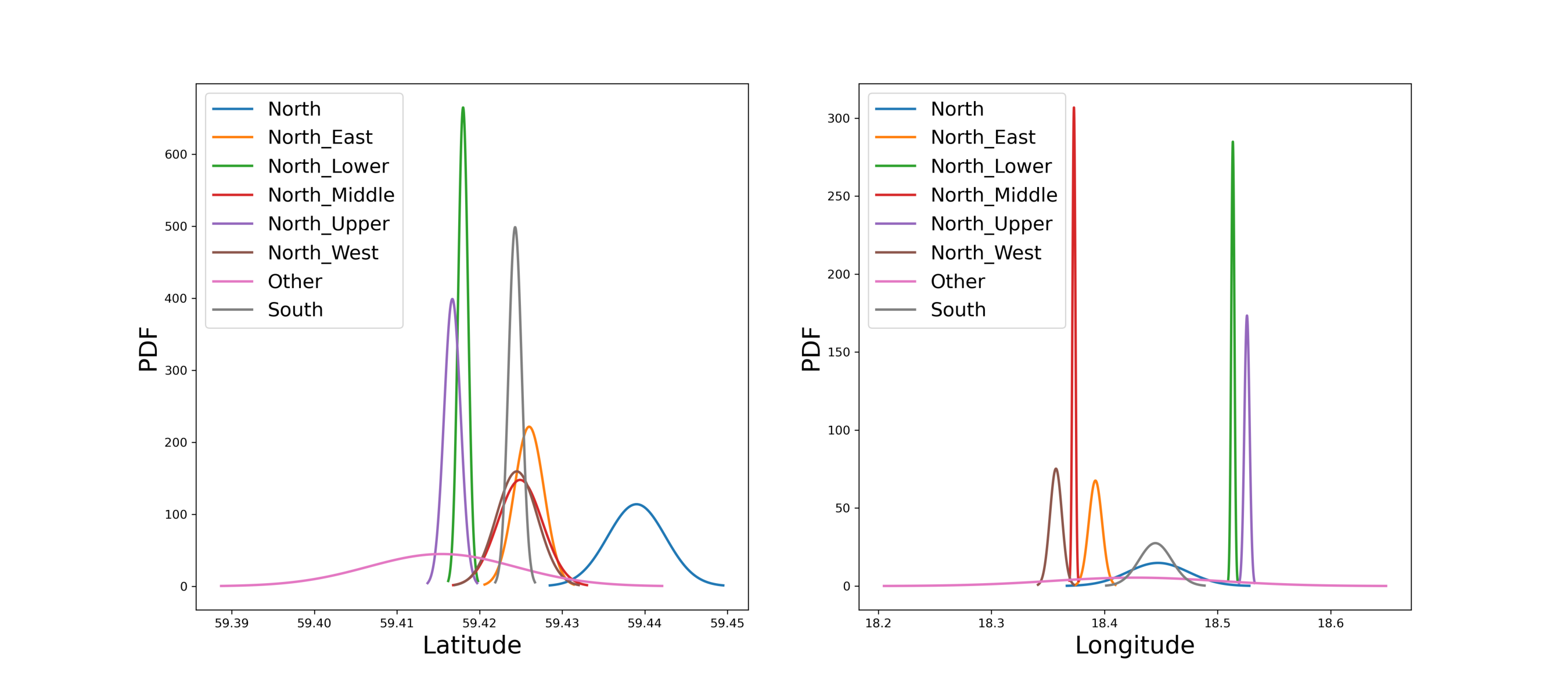}
  \caption{Probability distributions of location coordinates for the eight segments of the route of Cinderella II.}
  \label{fig:Prob_path8cls_latlon}
\end{figure}

\begin{figure}[htb]
  \centering
  \includegraphics[width=\linewidth]{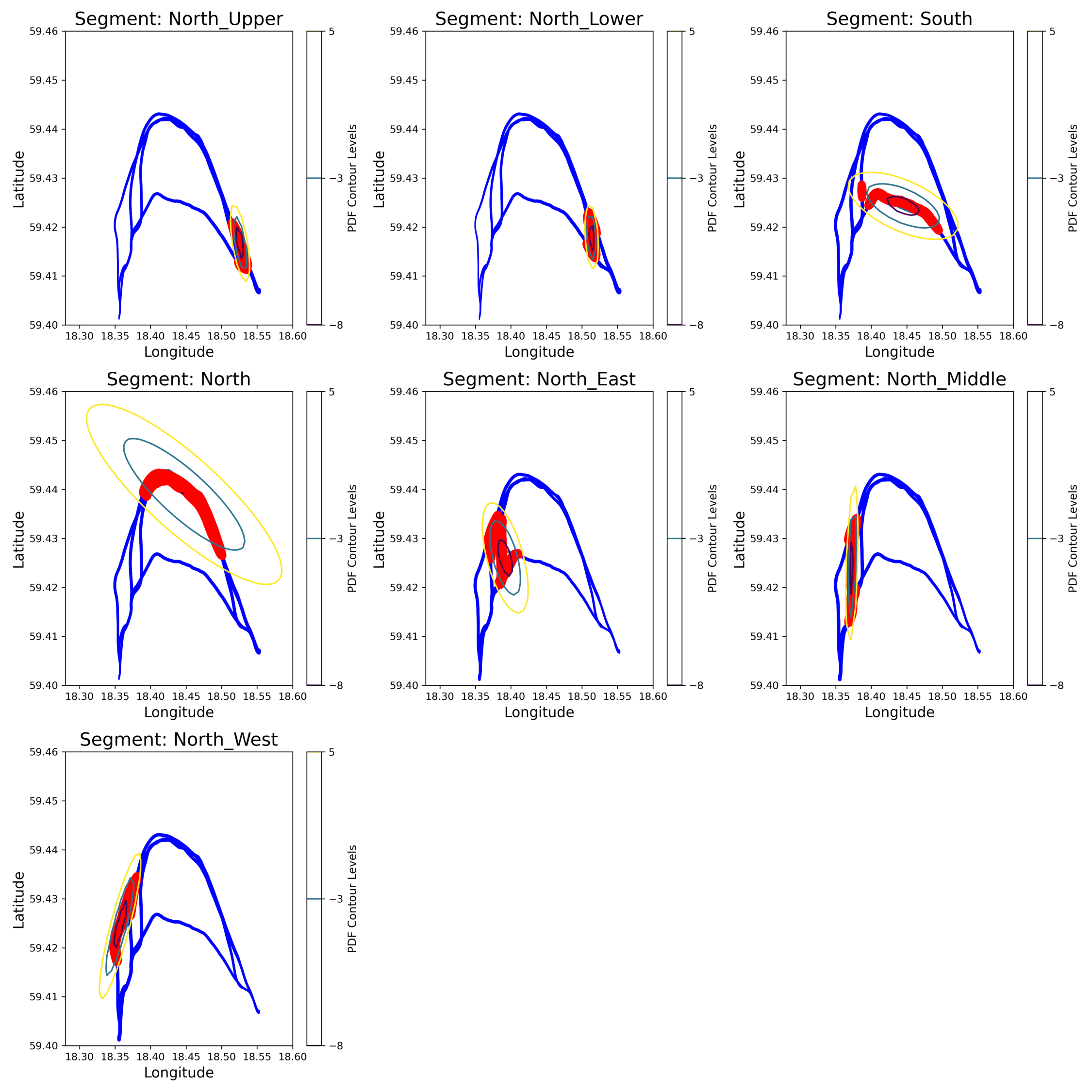}
  \caption{Gaussian distributions for seven segments of Cinderella II vessel route.}
  \label{fig:Gauss_path8cls}
\end{figure}

Remarkably, all path clusters achieved an F1-score of 1, indicating that hierarchical clustering successfully identified all paths with high accuracy from the distance matrix using the distance-based method.
This suggests that the choice of ML clustering technique with the distance matrix can influence the accuracy of path identification.
Table~\ref{Table:Result_Gauss_Summary} displays the outcomes of clustering, now by applying the segmented likelihood Gaussian method. This method achieved perfect precision, recall, and F1-score for all path clusters.\\
Figures~\ref{fig:Distribution_path8cls}, \ref{fig:Prob_path8cls_latlon}, and \ref{fig:Gauss_path8cls} present visualizations for the segmented Gaussian likelihood method.

The accuracy in results by hierarchical and segmented Gaussian likelihood clustering for path clusters indicates the efficacy of the developed approach of spatial clustering for vessel path identification.
\begin{table}[htb]
    \centering
    \caption{Result of implementing of both k-means and GMM clustering of the paths of Cinderella II vessel to five clusters.}
    \label{Table:Result_Kmean_GMM_Summary}
    
    \begin{subtable}{\linewidth}
        \centering
        \caption{Precision, Recall, and F1-score}
        \begin{tabular}{cccc}
            \toprule
            \textbf{Paths} & \textbf{Precision} & \textbf{Recall} & \textbf{F1-score} \\
            \midrule
            North-East (NE)    & 0.7 & 1   & 0.824 \\
            North-Middle (NM)  & 1   & 0.85 & 0.919 \\
            North-West (NW)   & 1   & 1   & 1     \\
            South (S)      & 1   & 1   & 1     \\
            South-West (SW)   & 1   & 1   & 1     \\
            \bottomrule
        \end{tabular}
    \end{subtable}
    
    \vspace{1em} 
    
    \begin{subtable}{\linewidth}
        \centering
        \caption{Confusion Matrix}
        \begin{tabular}{ccccccc}
            \toprule
            \multirow{2}{*}{\textbf{Actual}} & \multicolumn{5}{c}{\textbf{Predicted}} & \multirow{2}{*}{\textbf{Total}} \\
            \cline{2-6}
            & NE & NM & NW & S & SW & \\
            \midrule
            NE & 14 & 0 & 0 & 0 & 0 & 14 \\
            NM & 6  & 34 & 0 & 0 & 0 & 40 \\
            NW & 0  & 0  & 16 & 0 & 0 & 16 \\
            S  & 0  & 0  & 0 & 52 & 0 & 52 \\
            SW & 0  & 0  & 0 & 0 & 2 & 2 \\
            \textbf{Total} & 20 & 34 & 16 & 52 & 2 & 124 \\
            \bottomrule
        \end{tabular}
    \end{subtable}
\end{table}

\begin{table}[htb]
    \centering
    \caption{Result of implementing both hierarchical clustering and segmented likelihood Gaussian clustering of paths for Cinderella II vessel into five clusters.}
    \label{Table:Result_Hierch_Summary}
    \label{Table:Result_Gauss_Summary}
    
    \begin{subtable}{\linewidth}
        \centering
        \caption{Precision, Recall, and F1-score}
        \begin{tabular}{cccc}
            \toprule
            \textbf{Paths} & \textbf{Precision} & \textbf{Recall} & \textbf{F1-score} \\
            \midrule
            North-East (NE)    & 1 & 1   & 1 \\
            North-Middle (NM)  & 1 & 1 & 1 \\
            North-West (NW)   & 1   & 1   & 1     \\
            South (S)      & 1   & 1   & 1     \\
            South-West (SW)   & 1   & 1   & 1     \\
            \bottomrule
        \end{tabular}
    \end{subtable}
    
    \vspace{1em} 
    
    \begin{subtable}{\linewidth}
        \centering
        \caption{Confusion Matrix}
        \begin{tabular}{ccccccc}
            \toprule
            \multirow{2}{*}{\textbf{Actual}} & \multicolumn{5}{c}{\textbf{Predicted}} & \multirow{2}{*}{\textbf{Total}} \\
            \cline{2-6}
            & NE & NM & NW & S & SW & \\
            \midrule
            NE & 14 & 0 & 0 & 0 & 0 & 14 \\
            NM & 0  & 40 & 0 & 0 & 0 & 40 \\
            NW & 0  & 0  & 16 & 0 & 0 & 16 \\
            S  & 0  & 0  & 0 & 52 & 0 & 52 \\
            SW & 0  & 0  & 0 & 0 & 2 & 2 \\
            \textbf{Total} & 14 & 40 & 16 & 52 & 2 & 124 \\
            \bottomrule
        \end{tabular}
    \end{subtable}
\end{table}


\subsubsection{Results of Path Clustering for Buro Vessel}\label{buro_case}
~ \\
In this part the results from applying the framework of both hierarchical clustering and segmented likelihood Gaussian clustering of Buro vessel paths to three clusters. Following a similar evaluation procedure as in the case of Cinderella II vessel.

\begin{figure}[htb]
  \centering
  \includegraphics[width=1.1\linewidth]{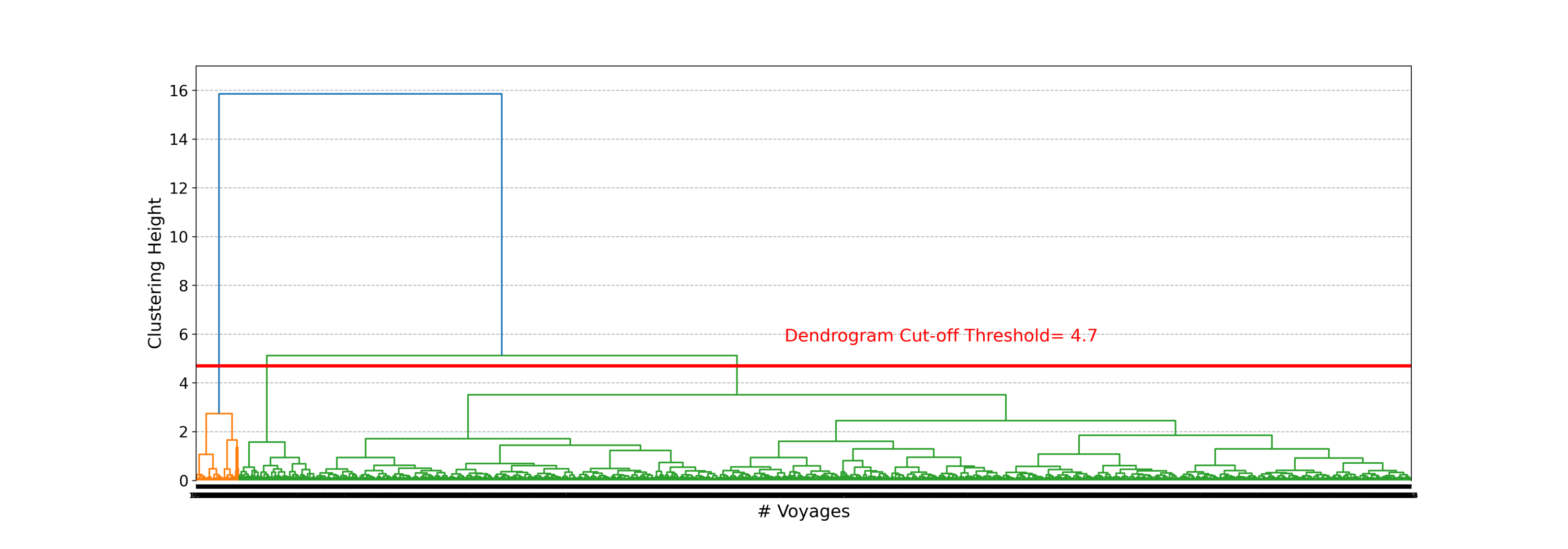}
  \caption{{Results of hierarchical clustering to three path clusters of Buro vessel.}}
  \label{fig:buro_Hierch_clus_path5cls}
\end{figure}

\begin{table}[htb]
    \centering
    \caption{{Result of Buro vessel by implementing hierarchical clustering to three path clusters}}
    \label{Table:Buro_Result_Hierch_Summary}    
    \begin{subtable}{\linewidth}
        \centering
        \caption{Precision, Recall, and F1-score}
        \begin{tabular}{cccc}
            \toprule
            \textbf{Paths} & \textbf{Precision} & \textbf{Recall} & \textbf{F1-score} \\
            \midrule
            Direct          & 1     & 1     & 1     \\
            East\_Canal      & 0.917 & 1     & 0.957 \\
            West\_Canal      & 1     & 0.993 & 0.996 \\
            \bottomrule
        \end{tabular}
    \end{subtable}
    
    \vspace{1em} 
    
    \begin{subtable}{\linewidth}
        \centering
        \caption{Confusion Matrix}
        \begin{tabular}{ccccc}
            \toprule
            \multirow{2}{*}{\textbf{Actual}} & \multicolumn{3}{c}{\textbf{Predicted}} & \multirow{2}{*}{\textbf{Total}} \\
            \cline{2-4}
            & Direct & E-C & W-C & \\
            \midrule
            Direct & 62 & 0 & 0 & 62 \\
            E-C & 0 & 122 & 0 & 122 \\
            W-C & 0 & 11 & 1560 & 1571 \\
            \textbf{Total} & 62 & 133 & 1560 & 1755 \\
            \bottomrule
        \end{tabular}
    \end{subtable}
\end{table}
\vfill

\begin{figure}[htb]
  \centering
  \includegraphics[width=\linewidth]{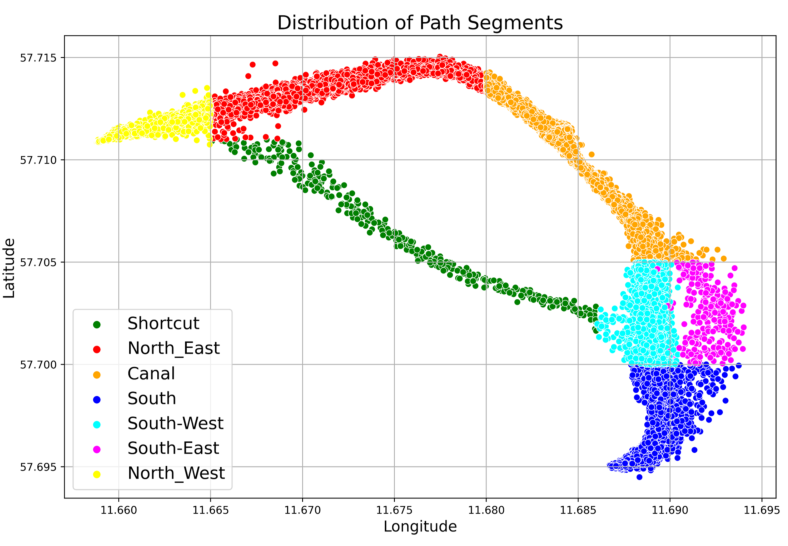}
  \caption{{Distribution of the route of Buro vessel into seven segments.}}
  \label{fig:Buro_Distribution_path7cls}
\end{figure}

\begin{figure}[htb]
  \centering
  \includegraphics[width=\linewidth]{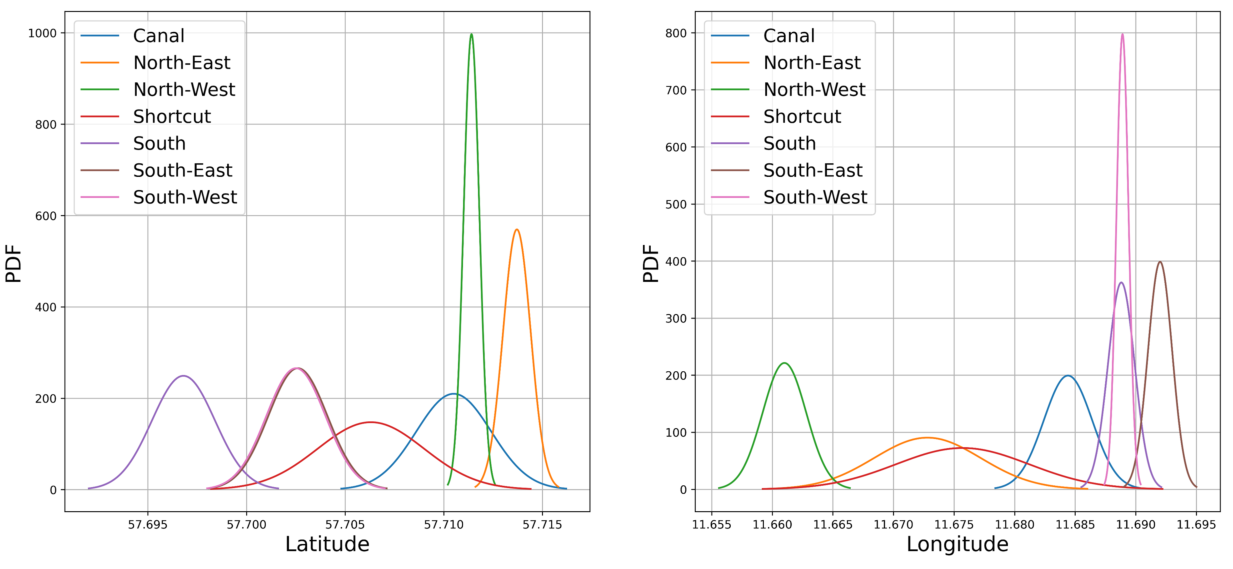}
  \caption{{Probability distributions of location coordinates for the seven segments of the route of Buro vessel.}}
  \label{fig:Buro_Prob_path7cls_latlon}
\end{figure}

\begin{figure}[htb]
  \centering
  \includegraphics[width=\linewidth]{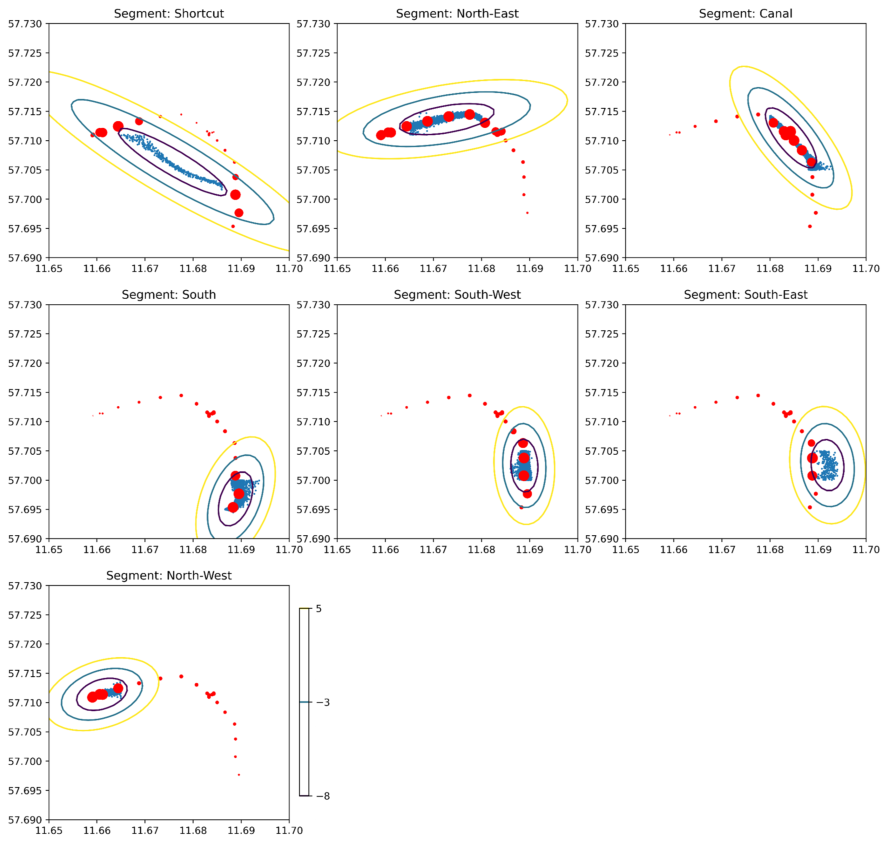}
  \caption{{Gaussian distribution for seven segments of the route of Buro vessel.}}
  \label{fig:Buro_Gauss_path7cls}
\end{figure}

\begin{figure}[htb]
  \centering
  \includegraphics[width=\linewidth]{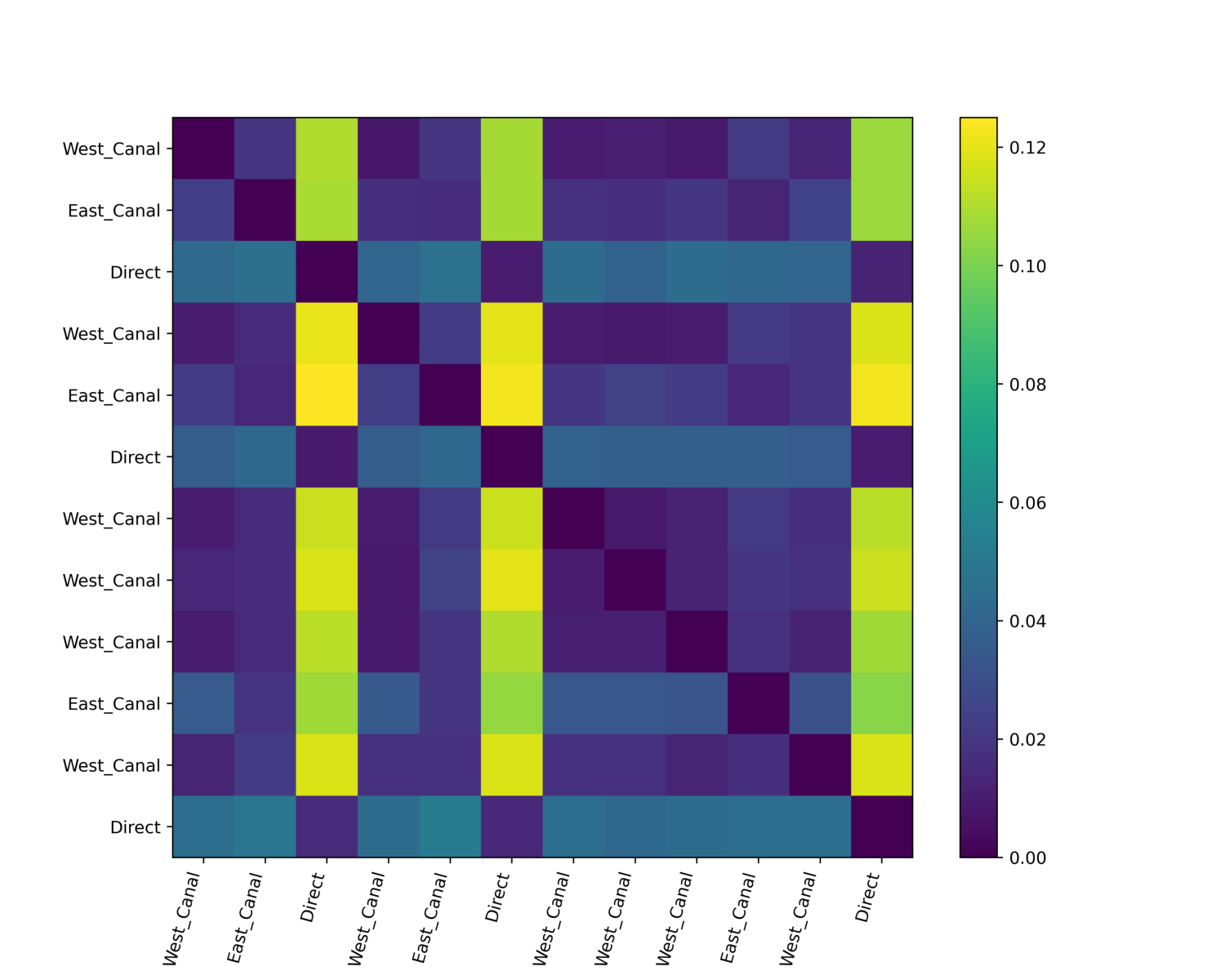}
  \caption{{Heatmap of the distance matrix for 12 paths of Buro vessel.}}
  \label{fig:buro_distance_matrix}
\end{figure}

\begin{table}[htb]
    \centering
    \caption{{Result of Buro vessel by implementing segmented likelihood Gaussian clustering to three path clusters}}
    \label{Table:Buro_Result_Hierch_Summary}    
    \begin{subtable}{\linewidth}
        \centering
        \caption{Precision, Recall, and F1-score}
        \begin{tabular}{cccc}
            \toprule
            \textbf{Paths} & \textbf{Precision} & \textbf{Recall} & \textbf{F1-score} \\
            \midrule
            Direct          & 1     & 1     & 1     \\
            East\_Canal      & 1  & 1     & 1 \\
            West\_Canal      & 1     & 1 & 1 \\
            \bottomrule
        \end{tabular}
    \end{subtable}
    
    \vspace{1em} 
    
    \begin{subtable}{\linewidth}
        \centering
        \caption{Confusion Matrix}
        \begin{tabular}{ccccc}
            \toprule
            \multirow{2}{*}{\textbf{Actual}} & \multicolumn{3}{c}{\textbf{Predicted}} & \multirow{2}{*}{\textbf{Total}} \\
            \cline{2-4}
            & Direct & E-C & W-C & \\
            \midrule
            Direct & 62 & 0 & 0 & 62 \\
            E-C & 0 & 122 & 0 & 122 \\
            W-C & 0 & 0 & 1571 & 1571 \\
            \textbf{Total} & 62 & 122 & 1571 & 1755 \\
            \bottomrule
        \end{tabular}
    \end{subtable}
\end{table}
{As it can be seen, the results of likelihood Gaussian clustering also achieved F1-score of 1, as in Cinderella vessel.
But for the hierarchical clustering in Buro vessel case has F1-scores of 0.957 and 0.996 for clustering East\_Canal and West\_Canal paths respectively. This results can be considered remarkable when take into account that these two paths, East\_Canal and West\_Canal, are challenging to be clustered, since they have several paths that are slightly different form each other. Figure{~\ref{fig:buro_distance_matrix}} shows the heatmap of the distance matrix of 12 sample paths for Buro vessel. Notably, the East\_Canal and West\_Canal paths exhibit high similarity, which make not easy clustering task.

Choosing a proper value of the agglomerative threshold for the hierarchical clustering to get the three clusters, as shown in Figure~{\ref{fig:buro_Hierch_clus_path5cls}}, with an agglomerative threshold = 4.7.}

\section{Conclusion}\label{Sec_conclusion}
In this section, we present briefly the key findings of this paper and provide recommendations for future research or practical implications for the maritime industry.

\subsection{Modeling and Improving Voyage Energy Efficiency} 
\begin{itemize}
\item The modeling approach by using the Efficiency Score, instead of directly working with the EngineFuelRate onboard signal, is more effective in facilitating decision-making. 
\item The resulting model is based on a more comprehensive understanding of the critical factors that impact fuel consumption, both temporally and spatially, resulting in more dependable counterfactual predictions. 
\item The quantitative evaluation indicates that estimating the Efficiency Score produces more precise and less biased outcomes than estimating the measured EngineFuelRate.
\item The study employs four distinct models: LSTM, kNN, 1NN-DTW, and HMM, to optimize vessel speed profiles with the objective of enhancing energy efficiency in short sea voyages. The key observation is that model performance varies significantly across these algorithms.
However, the performance of the models varies depending on the data cluster used to train the model and the weather conditions.
\item We developed a data-driven framework for optimizing vessel speed profiles to improve energy efficiency in SSS. The framework integrates a data-driven modeling approach to energy efficiency with the 1NN-DTW algorithm. We evaluated the added value of the framework using a real-world dataset and found that it can effectively improve vessel energy efficiency, especially with limited options, which are common in short-sea shipping.
\item 1NN-DTW exhibits the ability to capture temporal dependencies within speed profiles, especially within the constraints of short-sea shipping where opportunities for actively controlling the vessel to enhance its energy efficiency are restricted
\item Although the kNN can handle multivariate data and incorporating additional features like weather conditions, in this case study, the 1NN-DTW performs better due to its specialized handling of time-dependent data and inherent patterns.
\item The result findings emphasize that in terms of searching the best behavior of vessel from the observed data, the 1NN-DTW exhibits superior performance compared to LSTM and kNN, since the 1NN-DTW selects the best measured speed profiles. On the other hand, the HMM is the most effective approach in our study. where the HMM optimizes these measured speed profiles further by offering strategies to them informed by their weather states.
\item The study also reveals that the HMM model exhibits notable adaptability to different weather states (Calm, Moderate, and Rough). In each weather state, the HMM consistently delivers efficiency gains, indicating its ability to adapt speed profiles according to varying environmental conditions. This adaptability is crucial for real-world maritime applications, where weather can change rapidly.
\end{itemize}
\subsection{Vessel Path Identification} 
\begin{itemize}
\item The approach is able to identify the vessel paths with partially defined or unknown paths.
\item In the distance-based method, the hierarchical clustering used in the approach outperforms k-means and GMM clustering techniques.
\item The approach of hierarchical clustering includes a user-customized, a cut-off threshold, which allows desired control for the number of path clusters, enhancing the flexibility and adaptability of the proposed approach.
\item In the distance-based method, adopting ANND as a measure of similarity makes path clustering less affected by noise or outliers and provides a more intuitive interpretation of path similarity, ultimately enhancing the robustness and interpretability of our approach.
\item The segmented Gaussian likelihood method is particularly useful for identifying and analyzing the vessel path alterations at different segments of the vessel route.
\item The proposed approach is computationally efficient and has the potential to be a valuable tool for planning vessel paths. Accurate path identification can contribute to safer and more efficient maritime transportation practices, aiding in route planning, collision avoidance, and navigation optimization.
\item Nevertheless, the framework has some potential limitations, such as the segmented Gaussian likelihood method exhibiting sensitivity to segment definition which could affect its salable performance, particularly in complex maritime scenarios.
Moreover, while the study case demonstrates that the framework is computationally efficient, it is essential to discuss any potential scalability issues, especially when dealing with large datasets, since the computational efficiency may vary depending on the dataset size and the nature of paths.
\end{itemize}
\subsection{Future work and Recommendations} 
Future work could include the following:
\begin{itemize}
\item For modeling of vessel energy efficiency, considering spatial dimension, such as distance variable, might be worth of investigation.
\item Including the use of heuristic algorithms such as genetic programming, for optimizing voyage efficiency.
\item Exploring the scalability and real-world applicability of the proposed path clustering approach, as well as its integration with related systems of maritime transportation.
\item Applying incremental map-matching algorithms for real-time vessel path identification.
\item Future research in path detection may involve the study of graph theory and the application of evolutionary algorithms, including ant colony optimization.
\end{itemize}

\section*{Acknowledgment}
This research work has been funded and supported by Vinnova.
We would like to thank CetaSol AB for their support and for providing the resources necessary to conduct this research.\\
We also wish to thank the diverse group at the Center for Applied Intelligent Systems Research (CAISR), Halmstad University, for helpful discussions.

\section*{Supplementary Materials}
The source codes that are implemented on Python 3.9.7 to produce the results are available at:~\url{https://github.com/MohamedAbuella/TSA4EESSS} and~\url{https://github.com/MohamedAbuella/Path_Clustering}.

\bibliographystyle{IEEEtran}
\bibliography{ihelm_bib}

\begin{IEEEbiography}[{\includegraphics[width=1in,height=1.25in,clip,keepaspectratio]{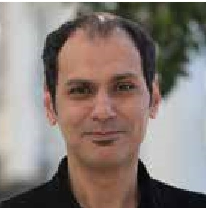}}]
{Mohamed Abuella} received his M.S. and PhD degrees in Electrical and Computer Engineering from Southern Illinois University at Carbondale and University of North Carolina at Charlotte, in 2012 and 2018 respectively. He is a postdoctoral researcher at Halmstad University since 2022.
His research interests include energy analytics and AI for sustainability.
\end{IEEEbiography}

\begin{IEEEbiography}[{\includegraphics[width=1in,height=1.25in,clip,keepaspectratio]{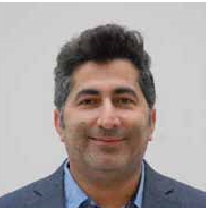}}]
{Hadi Fanaee-T} is an Associate Professor at Halmstad University, Sweden. He completed his PhD (with distinction) in Computer Science, under supervision of Professor Joao Gama at the Faculty of Science of University of Porto, Portugal in November 2015. He was the finalist in ERCIM Cor Baayen Young Researcher Award 2017. Prior to this position he was a postdoctoral fellow at the department of biostatistics, University of Oslo, Norway, and also worked as a postdoctoral researcher in European FP7 Project "MAESTRA" at INESC TEC research institute, Portugal. His main research interests are interdisciplinary applications of tensor decompositions, data fusion, anomaly/event detection and spatiotemporal data mining. He is the first-author of several journal and conference papers. He has served as a PC member to over 20 prestigious conferences (e.g. IJCAI, AAAI, ECML-PKDD, ISMIS, ACM SAC, IEEE DSAA, etc.) and also reviewer to several high-impact journals (e.g. TDKE, TKDD, DAMI, ML, KAIS, KBS, CSUR.
\end{IEEEbiography}

\begin{IEEEbiography}[{\includegraphics[width=1in,height=1.25in,clip,keepaspectratio]{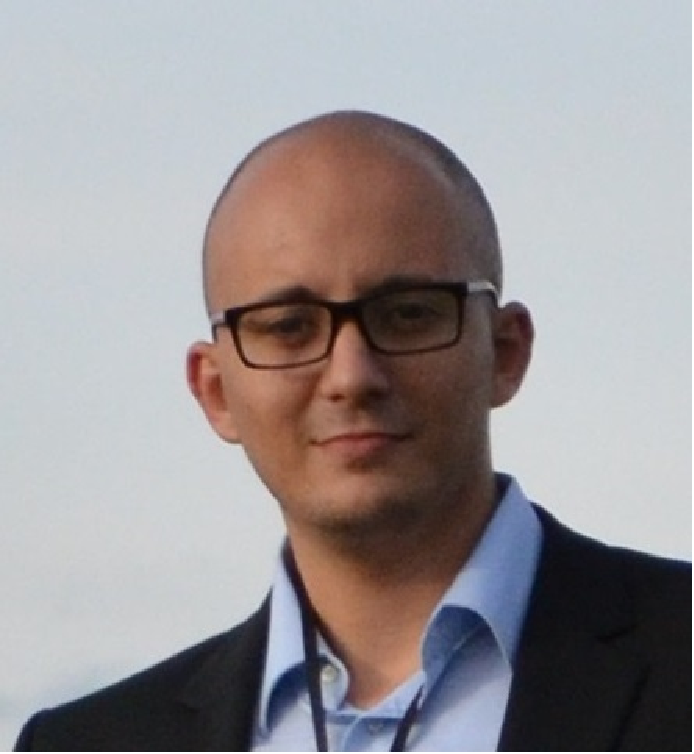}}]
{M. AMINE ATOUI} obtained his Ph.D. degree from LARIS, Polytech' Angers, France, in 2015. Currently, he is affiliated with the Center for Applied Intelligent Systems Research at Halmstad University, Sweden. His research interests encompass probabilistic and explainable Machine Learning, causal and Bayesian Inference, transmission/communication, and automatic control.
\end{IEEEbiography}

\begin{IEEEbiography}[{\includegraphics[width=1in,height=1.25in,clip,keepaspectratio]{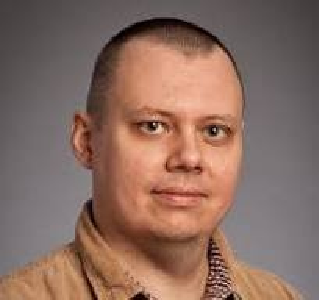}}]
{Slawomir Nowaczyk} is a Professor in Machine Learning at the Center for Applied Intelligent Systems Research, Halmstad University, Sweden. He received his MSc degree from Poznan University of Technology in 2002 and his PhD from the Lund University of Technology in 2008. During the last decades, his research has focused on machine learning, knowledge representation, and self-organising systems. The majority of his work concerns industrial data streams, often with predictive maintenance as the goal. Given that accurate and relevant labels are usually impossible to obtain in such settings, Slawomir’s contributions primarily take advantage of proxy labels, such as transfer learning and multi-task learning, or concern semi-supervised and unsupervised modelling. He is a board member of the Swedish AI Society and a research leader for the School of Information Technology at Halmstad University. Slawomir has led multiple research projects on applying Artificial Intelligence and Machine Learning in different domains, such as transport and automotive, energy, smart cities, and healthcare. In most cases, this research was done in collaboration with industry and public administration organisations – inspired by practical challenges and leading to tangible results and deployed solutions.
\end{IEEEbiography}

\begin{IEEEbiography}[{\includegraphics[width=1in,height=1.25in,clip,keepaspectratio]{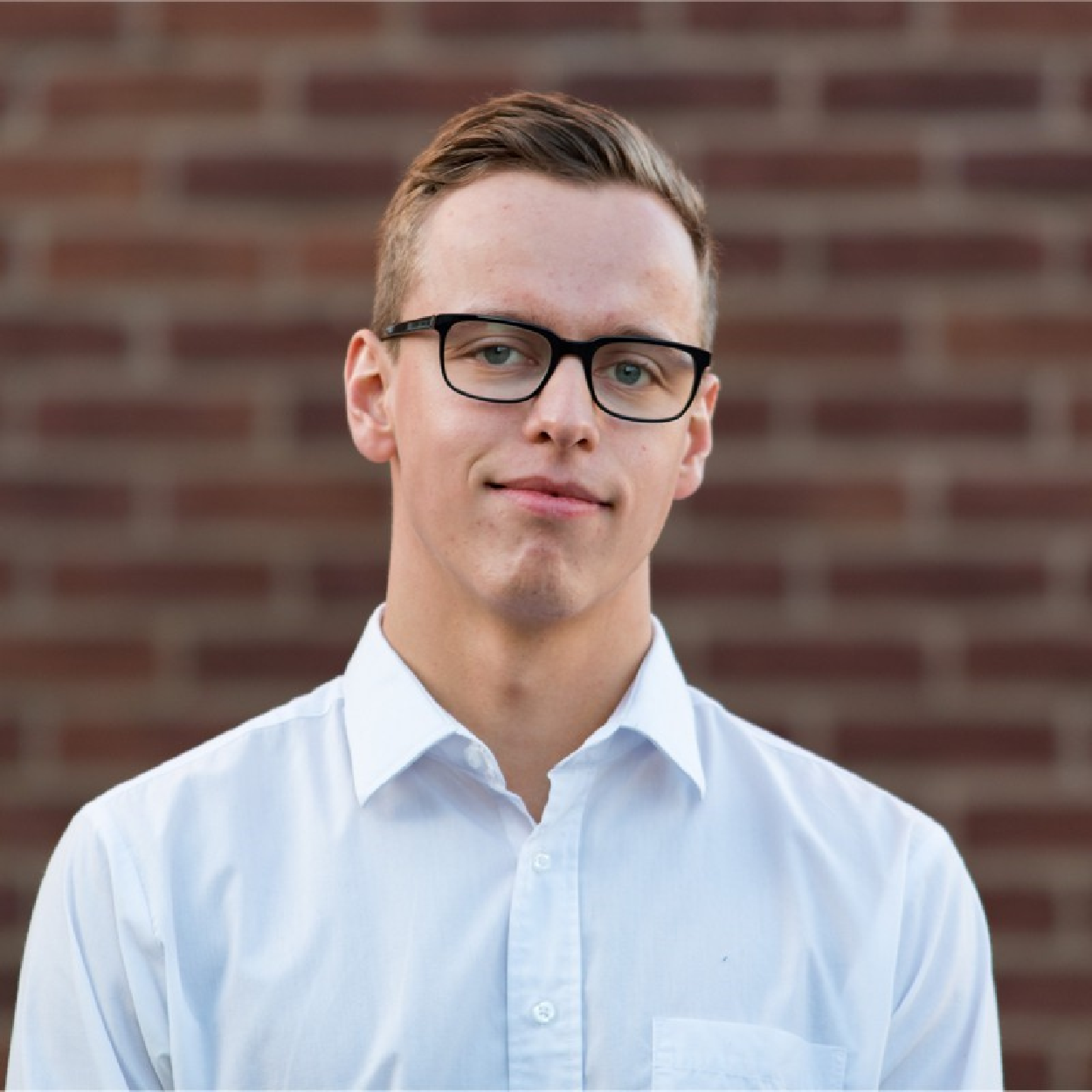}}]
{Simon Johansson} is a MSc graduate of Chalmers University of Technology's program in Engineering Mathematics and Computational Science in 2020, currently works in Cetasol, a marine company specialising in CO2 reduction and energy optimisation for vessels. His practical application of computational methods and dedication to environmental sustainability align with his role, contributing to global efforts to mitigate climate change. Simon's commitment to advancing eco-friendly solutions in the marine industry reflects a seamless integration of academic excellence and real-world impact.
\end{IEEEbiography}

\newpage
\begin{IEEEbiography}[{\includegraphics[width=1in,height=1.25in,clip,keepaspectratio]{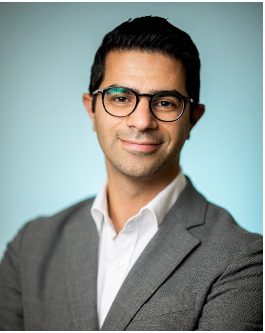}}]
{Ethan Faghani} is the CEO and founder of Cetasol. Before Cetasol, Ethan was Chief Engineer of Automation and AI at Volvo Penta. Ethan has experience working with cutting-edge technologies in other transportation segments in both big enterprises and his own founded startup. Ethan obtained his Ph.D. in mechatronics from UBC and Innovation and Entrepreneurship from Stanford Business School.
\end{IEEEbiography}

\vfill

\end{document}